\documentclass[reprint, aps, nofootinbib, prb, nobibnotes]{revtex4-1}
\usepackage{amsmath} \usepackage{graphicx} \usepackage{dcolumn}
\usepackage{bm} 
\usepackage{amssymb}
\usepackage{pstricks}
\usepackage{float}
\usepackage{subfigure}
\usepackage{braket}
\usepackage{wasysym}
\usepackage{xr}
\usepackage{bibunits}

\begin{document}
\newlength{\figurewidth}
\setlength{\figurewidth}{\columnwidth}

\newcommand{\prtl}{\partial}
\newcommand{\la}{\left\langle}
\newcommand{\ra}{\right\rangle}
\newcommand{\dla}{\la \! \! \! \la}
\newcommand{\dra}{\ra \! \! \! \ra}
\newcommand{\we}{\widetilde}
\newcommand{\smfp}{{\mbox{\scriptsize mfp}}}
\newcommand{\smp}{{\mbox{\scriptsize mp}}}
\newcommand{\sph}{{\mbox{\scriptsize ph}}}
\newcommand{\sinhom}{{\mbox{\scriptsize inhom}}}
\newcommand{\sneigh}{{\mbox{\scriptsize neigh}}}
\newcommand{\srlxn}{{\mbox{\scriptsize rlxn}}}
\newcommand{\svibr}{{\mbox{\scriptsize vibr}}}
\newcommand{\smicro}{{\mbox{\scriptsize micro}}}
\newcommand{\scoll}{{\mbox{\scriptsize coll}}}
\newcommand{\sattr}{{\mbox{\scriptsize attr}}}
\newcommand{\sth}{{\mbox{\scriptsize th}}}
\newcommand{\sauto}{{\mbox{\scriptsize auto}}}
\newcommand{\seq}{{\mbox{\scriptsize eq}}}
\newcommand{\teq}{{\mbox{\tiny eq}}}
\newcommand{\sinn}{{\mbox{\scriptsize in}}}
\newcommand{\suni}{{\mbox{\scriptsize uni}}}
\newcommand{\tin}{{\mbox{\tiny in}}}
\newcommand{\scr}{{\mbox{\scriptsize cr}}}
\newcommand{\tstring}{{\mbox{\tiny string}}}
\newcommand{\sperc}{{\mbox{\scriptsize perc}}}
\newcommand{\tperc}{{\mbox{\tiny perc}}}
\newcommand{\sstring}{{\mbox{\scriptsize string}}}
\newcommand{\stheor}{{\mbox{\scriptsize theor}}}
\newcommand{\sGS}{{\mbox{\scriptsize GS}}}
\newcommand{\sBP}{{\mbox{\scriptsize BP}}}
\newcommand{\sNMT}{{\mbox{\scriptsize NMT}}}
\newcommand{\sbulk}{{\mbox{\scriptsize bulk}}}
\newcommand{\tbulk}{{\mbox{\tiny bulk}}}
\newcommand{\sXtal}{{\mbox{\scriptsize Xtal}}}
\newcommand{\sliq}{{\text{\tiny liq}}}

\newcommand{\smin}{\text{min}}
\newcommand{\smax}{\text{max}}

\newcommand{\saX}{\text{\tiny aX}}
\newcommand{\slaX}{\text{l,{\tiny aX}}}

\newcommand{\svap}{{\mbox{\scriptsize vap}}}
\newcommand{\sjam}{J}
\newcommand{\Tm}{T_m}
\newcommand{\sTS}{{\mbox{\scriptsize TS}}}
\newcommand{\sDW}{{\mbox{\tiny DW}}}
\newcommand{\cN}{{\cal N}}
\newcommand{\cB}{{\cal B}}
\newcommand{\br}{\bm r}
\newcommand{\cH}{{\cal H}}
\newcommand{\cHlt}{\cH_{\mbox{\scriptsize lat}}}
\newcommand{\sthermo}{{\mbox{\scriptsize thermo}}}

\newcommand{\bu}{\bm u}
\newcommand{\bk}{\bm k}
\newcommand{\bX}{\bm X}
\newcommand{\bY}{\bm Y}
\newcommand{\bA}{\bm A}
\newcommand{\bb}{\bm b}

\newcommand{\lintf}{l_\text{intf}}

\newcommand{\tXW}{{\text{\tiny XW}}}
\newcommand{\tRL}{{\text{\tiny RL}}}

\def\Xint#1{\mathchoice
   {\XXint\displaystyle\textstyle{#1}}%
   {\XXint\textstyle\scriptstyle{#1}}%
   {\XXint\scriptstyle\scriptscriptstyle{#1}}%
   {\XXint\scriptscriptstyle\scriptscriptstyle{#1}}%
   \!\int}
\def\XXint#1#2#3{{\setbox0=\hbox{$#1{#2#3}{\int}$}
     \vcenter{\hbox{$#2#3$}}\kern-.5\wd0}}
\def\ddashint{\Xint=}
\def\dashint{\Xint-}
\title{The chemical bond as an emergent phenomenon}

\author{Jon C. Golden} \affiliation{Department of Physics, University
  of Houston, Houston, TX 77204-5005}

\author{Vinh Ho$^\dagger$}
%\footnote{Present address: UT Southwestern Medical
%  School; Dallas, TX 75390-9003} 
\affiliation{Department of Chemistry,
  University of Houston, Houston, TX 77204-5003}

% \author{Joseph Hutchins} \affiliation{Department of Chemistry,
%   University of Houston, Houston, TX 77204-5003}

\author{Vassiliy Lubchenko} \email{vas@uh.edu} \affiliation{Department
  of Chemistry, University of Houston, Houston, TX 77204-5003}
\affiliation{Department of Physics, University of Houston, Houston, TX
  77204-5005}

\date{\today}

%\SectionsOn
%\SectionNumbersOn

%\begin{document}

\begin{abstract}

  We first argue that the covalent bond and the various closed-shell
  interactions can be thought of as symmetry broken versions of one
  and the same interaction, viz., the multi-center bond. We use
  specially chosen molecular units to show that the symmetry breaking
  is controlled by density and electronegativity variation.
  % Depending on the precise combination of these factors, the
  % symmetry breaking transition can be either continuous or
  % discontinuous. When present, the discontinuity is weak so that the
  % transition region is still approximately subject to a law of
  % corresponding states.
  We show that the bond order changes with bond deformation but in a
  step-like fashion, regions of near constancy separated by electronic
  localization transitions.  These will often cause displacive
  transitions as well so that the bond strength, order, and length are
  established self-consistently.  We further argue on the inherent
  relation of the covalent, closed-shell, and multi-center
  interactions with ionic and metallic bonding. All of these
  interactions can be viewed as distinct sectors on a phase diagram
  with density and electronegativity variation as control variables;
  the ionic and covalent/secondary sectors are associated with on-site
  and bond-order charge density wave respectively, the metallic sector
  with an electronic fluid.  While displaying a contiguity at low
  densities, the metallic and ionic interactions represent distinct
  phases separated by discontinuous transitions at sufficiently high
  densities. Multi-center interactions emerge as a hybrid of the
  metallic and ionic bond that results from spatial coexistence of
  delocalized and localized electrons. In the present description, the
  issue of the stability of a compound is that of mutual miscibility
  of electronic fluids with distinct degrees of electron localization,
  supra-atomic ordering in complex inorganic compounds comes about
  naturally. The notions of electronic localization advanced hereby
  suggest a high throughput, automated procedure for screening
  candidate compounds and structures with regard to stability, without
  the need for computationally costly geometric optimization.

\end{abstract}

\maketitle

%\tableofcontents
\begin{bibunit}[unsrt]

\section{Motivation}
\label{motivation}

Chemical bonding is traditionally discussed in terms of the covalent,
ionic, and metallic bond~\cite{doi:10.1021/ja02199a004}, and weaker,
closed-shell interactions such as secondary, donor-acceptor, hydrogen,
and van der Waals.~\cite{Pyykko, Alcock1972} The distinction between
these canonical bond types is not always clear-cut. For instance, a
directional, multi-center~\cite{PapoianHoffmann2000} bond holding
together identical atoms has an inherent ionic feature: In a
three-center, linear $pp\sigma$ bond,~\cite{doi:10.1021/ja00884a026,
  pimentel:446} the central atom contributes only a half orbital to
each of the individual $pp\sigma$ bonds.~\cite{MusherAngew1969} The
terminal atoms, on the other hand, each contribute one full orbital,
implying a non-uniform charge distribution over the bond. At the same
time, the three-center $pp\sigma$ bond can be thought of as a limiting
case of the metallic bond, since the appropriate electron count for an
infinite chain corresponds to a half-filled
band.~\cite{PapoianHoffmann2000} This identification is consistent
with the metallic luster of compounds in which covalent and secondary
bonds are comparable in length.~\cite{Alcock1972}.  In solid-state
contexts, interplay between ionic and covalent interactions is often
discussed using the van Arkel-Ketelaar triangle for binary
compounds;~\cite{doi:10.1021/ed012p53, vanArkel, Ketelaar,
  Burdett1995} or revealed, for instance, in the taxonomies of
classical valence compounds and Zintl phases.~\cite{GMiller1996,
  C5DT04140F}. Contiguity between the metallic and ionic extremes is
exemplified by metal-ammonia solutions,~\cite{SWmayo} whose electric
conductance ranges between largely electronic and ionic, depending on
the concentration of the metal.

Despite their distinct phenomenologies, canonical types of bonding are
not always easy to distinguish on formal grounds, the case of the
metallic bond being particularly subtle. Indeed, cohesive interactions
in periodic solids are usually discussed either in terms of molecular
orbitals (MO) or in terms of plane-wave, Bloch electronic
states.~\cite{ABW} The two approaches are not equivalent: The majority
of {\em aperiodic} condensed phases, such as liquids, frozen glasses,
or amorphous films, are expressly not Bloch solids. (Within the
Born-Oppenheimer approximation, liquids are solids as far as the
electrons are concerned.) Yet the Bloch and MO approaches are often
perceived as equivalent.  In fact, there have been calls to do away
with the concept of the metallic bond altogether, by including it
under the broader rubric of covalent interactions.~\cite{AllenBurdett}
Electron delocalization in aperiodic solids is of much subtler nature
than in crystals;~\cite{AndersonLoc, Mott1993} one no longer speaks of
allowed energy bands but, instead, of {\em mobility} bands.  The
latter are defined in a relatively narrow sense that individual
molecular orbitals (MOs) extend further than the mean-free path of a
charge carrier, a notion that harks back to the original motivation
for introducing the metallic bond via the mobility of the
electrons.~\cite{doi:10.1021/ja02199a004} The most important feature
of the metallic bond is that the electrons do indeed comprise a fluid;
an arbitrarily weak field results in a particle flow. No such flow can
take place in insulators, where the electrons are fully localized.

In its ideal form---even if unachievable in actual compounds---the
metallic bond is isotropic, not directional. This may directly reveal
itself in the material's being close-packed and/or malleable. Metallic
substances tend to be poor glass-formers and decrease in volume upon
crystallization, as perfect hard spheres would. In contrast, bond
directionality characteristic of covalent interaction may even lead to
expansion during freezing, as in water, silicon, or germanium.  From
the MO viewpoint, metals are special in that they possess a bulk
density of (formally non-bonding) states that can be easily occupied
at arbitrarily low temperatures. In contrast, the canonical covalent
bond is defined most unambiguously as being due to filled states.
From a thermodynamics vantage point, metallic and covalent
interactions can be thought of as distinct since the metallic and
insulating phases are separated by a phase transition.~\cite{Mott1993}
The latter viewpoint will prove most instructive in the present
context, the pertinent order parameter having to do with the degree of
localization of the electronic liquid.

\begin{figure}[t]
  \centering
  \includegraphics[width=0.95 \figurewidth]{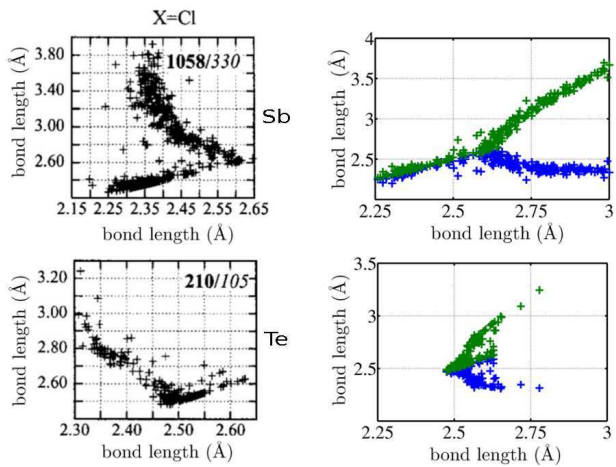}
  \caption{\label{LH} {\em Left}: The lengths of the two bonds in
    three-center bonding configurations compiled by Landrum and
    Hoffman.~\cite{LandrumHoffmann1998} {\em Right}: The
    Landrum-Hoffmann data replotted to show the two bond lengths as
    functions of the trimer length per bond. }
\end{figure}

The lack of clear dichotomies between canonically distinct chemical
interactions is brought home by a systematic study by Landrum and
Hoffmann,\cite{LandrumHoffmann1998} who have screened thousands of
compounds in the Oxford Structural Database for near linear trimeric
motifs XQX, where Q stands for Sb or Te and X for F, Cl, Br, or I.
Select parametric plots of the lengths $b_1$ and $b_2$ of the two
individual bonds, from Ref.\cite{LandrumHoffmann1998}, are reproduced
here on the l.h.s. of Fig.~\ref{LH}. The three-center bond in
Fig.~\ref{LH} is directional and exhibits
back-bonding.\cite{LandrumHoffmann1998} Indeed, the negative slopes of
the $b_2 \ne b_1$ portions of the parametric curves on the l.h.s. of
Fig.~\ref{LH} suggest that when shortened, the weaker bond supplies
electrons into the anti-bonding orbital of the stronger bond,
resulting in weakening of the latter. The shorter bond in the $b_2 \ne
b_1$ portions can be short enough to be considered covalent, while the
longer, weaker bond varies in strength from what one expects for a
relatively strong, directional secondary bond to a very weak van der
Waals interaction, which is essentially non-directional.  In the
symmetric region, $b_1 \approx b_2$, the bond strength can approach
that of the covalent bond, for sufficiently small $b_{1,2}$. Thus
empirical data suggest that the covalent, secondary, and three-center
bonds form a true {\em continuum} of interactions.  Despite some
scatter in the data, one notices a great deal of universality in the
$b_1$ vs. $b_2$ relation. While the synergic relation between
complementary covalent and secondary bonds is expected based on
straightforward molecular orbital
considerations,\cite{LandrumHoffmann1998} as just discussed, its
apparent near universality is less obvious.  This near universality
suggests a {\em law of corresponding states} may apply.

Laws of corresponding states, such as the familiar van der Waals
equation of state,\cite{LLstat} often arise on a systematic basis in
the context of critical points. If rescaled in terms of the critical
temperature, pressure, and density, equations of state for chemically
distinct substances will look identical near the critical point so
long as those substances belong in the same universality
class,~\cite{Goldenfeld} as defined by the symmetry and range of the
interaction; the detailed form of interaction becomes immaterial
because the molecular lengthscale is much shorter than the correlation
length. In those common situations when a critical point is a limiting
case of a set of discontinuous transitions---as is the case for the
liquid-to-vapor transition, for instance---laws of corresponding
states are still expected to hold approximately not too far from the
critical point.  When it holds, a law of corresponding states offers a
systematic way to reduce the complexity of the problem by allowing one
to use the simplest possible model from the universality class in
question.

\begin{figure}[t]
  \centering
  \includegraphics[width = 0.9 \figurewidth]{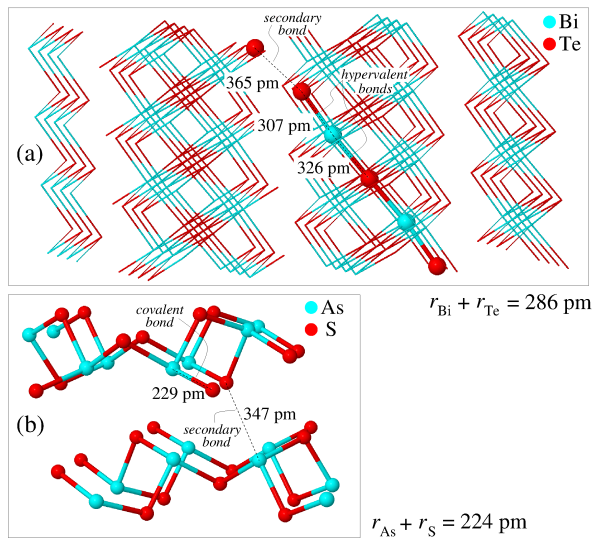}
  \caption{{\bf (a)} Structure of Bi$_2$Te$_3$ illustrating
    coexistence of multicenter, covalent and secondary bonding. {\bf
      (b)} Structure of As$_2$S$_3$ illustrating that the bonding is
    essentially covalent within the double layers. Sums of pertinent
    covalent radii are provided for the reader's
    reference. \label{coex}}
\end{figure}

A revealing way to view the Landrum-Hoffmann data is to graph the two
bond lengths as functions of the overall trimer length per bond $(b_1
+ b_2)/2$, see the r.h.s. of Fig.~\ref{LH}. The graphs explicitly show
an apparent {\em symmetry breaking} that takes place as the system
expands from a high density state, in which $b_1 \approx b_2$, to a
state where the central atom chooses to make a strong bond with a
specific neighbor, while settling on a weaker interaction with the
other neighbor. The antimony and tellurium cases are distinct in that
the broken-symmetry regime $b_2 \ne b_1$ can coexist with the
symmetric regime $b_2 = b_1$ in the latter case, but not in the
former. In the context of bulk phase transitions, lack of such
coexistence implies the transition is continuous, a critical point.
Incidentally, the density dependence of the bond lengths for the
antimony compounds in Fig.~\ref{LH} parallels that for the continuous
transition between the rhombohedral and simple-cubic
arsenic.~\cite{silas:174101} In contrast, a macroscopic phase
coexistence implies the transition is discontinuous, common example
being the liquid-to-solid transition.~\cite{LLstat, EhrenfestBook}
{\em Spatial} coexistence of distinct types of bonding can be directly
seen in the crystal of Bi$_2$Te$_3$, where multicenter, covalent, and
secondary bonding patterns form extended layers, see
Fig.\ref{coex}(a). This can be contrasted with As$_2$S$_3$,
Fig.~\ref{coex}(b), a compound made of covalently bonded double layers
that interact relatively weakly via secondary
interactions;~\cite{ZLMicro1} each atom is surrounded by a Lewis
octet.
 
Here we argue that not only are the covalent bond and secondary
interaction intimately related to each other and, in turn, to the
multi-center bond, but that the covalent and secondary interaction can
be thought of as {\em originating} from the multi-center bond as a
result of a symmetry-breaking transition driven by a delicate
interplay between steric repulsion and cohesive interaction.  For
sufficiently high electron count, the bond strength, order, and length
establish self-consistently as a result of the transition. The ensuing
differentiation in the bond length amounts to a breaking of spatial
symmetry in the nuclear arrangement, as in Fig.~\ref{LH}. The
transition could be either continuous or discontinuous.  In the latter
case, the discontinuity is weak and so a law of corresponding states
still holds approximately.

We also observe two types of symmetry breaking of purely electronic
origin, which occur even if the aforementioned differentiation in bond
length does not take place or is artificially prevented by constraining
the geometry. One type of transition occurs if one insists that the
trial electronic wave-function be a single Slater determinant. As a
result, the ground state as determined by solving the self-consistent
Hartree-Fock (SCHF) equations could become degenerate.  Each such
solution represents a charge-density wave~\cite{gruner2009density,
  CanadellWhangbo1991} (CDW) formed as a result of electron-electron
interactions. The true, quantum-mechanical ground state is unique; it
is a superposition of those distinct ``classical'' solutions of the
Hartree-Fock (HF) problem.~\cite{doi:10.1063/1.4832476} Still, the
degeneracy is physically consequential and may signal a structural
instability, if the overlap between those ``classical'' solutions is
small. The other, subtler type of electronic transition takes place
even as the molecular orbitals vary smoothly with geometry changes
while the electronic density is redistributed among bonding orbitals
and lone pairs.  This type of transition manifests itself as a
symmetry change of the effective wavefunctions of the bonding
electrons, which we determine using the {\em localized molecular
  orbital}~\cite{RevModPhys.32.296, RevModPhys.35.457,
  doi:10.1063/1.1677859, PerkinsStewart1980, Levine09} (LMO)
framework.  (We will also find the LMOs to be particulary convenient
in detecting the electronic symmetry breaking at the Hartree-Fock
level.)  Importantly we establish that away from narrow transition
regions, the bond order and related quantities are found to be robust
with respect to bond deformation, despite charge transfer accompanying
nuclear motions.

We thus find that the order of a chemical bond is established as a
result of transitions between states with distinct states of electron
localization, often accompanied by displacive transitions of the
nuclei. The bond order is apparently robust with respect to typical
geometry fluctuations in a way that is analogous to how a phase of
matter is stable while fluctuating within the corresponding free
energy minimum. It may seem surprising, at a first glance, that finite
systems such as molecules should exhibit phase transition-like
phenomena that are normally associated with bulk, macroscopic
phases. Yet a $D$-dimensional {\em quantum} system can be thought of a
classical system in $(D+1)$ dimensions, the extent of the extra
dimension proportional to the inverse
temperature.~\cite{RevModPhys.69.315} In this sense, finding the
ground state of even a small collection of atoms amounts to solving
for a partition function of a classical system extending indefinitely
along one spatial dimension; such systems do exhibit zero temperature
fixed points.~\cite{Goldenfeld} Note the latter are associated with
bound states.~\cite{Lfutile} Incidentally, a connection can be made
with earlier work of Kais, Herschbach, and others,~\cite{KaisSerraACP,
  doi:10.1063/1.1637581} who have viewed ionization and dissociation
as phase transitions with the nuclear charge being the control
parameter.

We next extend the above results to bulk systems proper to argue that
not only can distinct values of the bond order, but distinct {\em
  types} of chemical bond can be thought of distinct phases. The axes
on the corresponding ``phase diagram'' of chemical interactions are
density and electronegativity variation, respectively. The ionic and
metallic bond can be viewed as fully distinct phases in that they are
separated by one or more discontinuous phase transitions, above a
certain threshold density. Such transitions were described by Kohn
some 50 years ago.~\cite{PhysRevLett.19.789, PhysRevLett.19.439}
Multi-center bonding is viewed as a coexistence, or hybrid, of the
metallic and ionic bonding. The discontinuity of the transition stems
from poor mutual miscibility of localized and delocalized electrons.
Below the aforementioned threshold density, the metallic and ionic
bond form a continuous spectrum of interactions. At sufficiently low
values of mass density and electronegativity variation, the nuclear
arrangement undergoes a symmetry-lowering displacive transition so
that the multicenter bonding turns into a coexistence of covalent and
secondary bonding, as it did for small molecular fragments.  The
venerable density-functional theory~\cite{PhysRev.136.B864} provides a
formal foundation for and, at the same time, a convenient way to think
about the phase diagram of chemical interactions: The
covalent-secondary sector corresponds with the electrons forming a
bond-order wave, while the ionic sector to an on-site charge density
wave (CDW).~\cite{CanadellWhangbo1991} In the metallic sector, the
itinerant electrons---despite being subject to the field due to the
ionic cores that lowers the translational symmetry---could be thought
of as a uniform liquid in the continuum limit.

In small molecules and bulk systems alike, the structural
instabilities can be traced down to electronic instabilites arising
from the formation of a bond-order charge density wave, the troughs
and crests of the wave corresponding to weaker and stronger bonds in
the eventual distorted structure. We implement this notion to show
that an ambiguity in assigning of bonding electrons to effective
two-center bonds signals structural instabilities and suggest a novel
algoritm that can be used to speed up prediction of new compounds and
structures.

In thinking of interactions as sectors on a phase diagram, we borrow
the language from the renormalization group (RG) theory of phase
transitions,~\cite{Goldenfeld, RevModPhys.70.653, RevModPhys.55.583}
which operates on a space formed by coupling constants and, in
general, by Hamiltonians. In the RG language, interactions and phases
are interchangeable concepts. For instance, the paramagnetic and
polarized states of a ferromagnet are viewed as (attractive) fixed
points in a space formed by spin-spin couplings and the magnetic
field. In the present framework, the role of the order parameter is
played by the CDW type and strength; the description is
coarse-grained, in full analogy with the RG framework.

The possibility of making a compound thus can be viewed as a question
of coexistence of distinct types of charge density waves. For
instance, while the Heusler and half-Heusler compounds are readily
synthesized,~\cite{Graf20111} intermediate stoichiometries are not. In
fact, the full- and half-Heusler phases exhibit poor mutual
miscibility.~\cite{Romaka201345} Within the present formalism, those
intermediate stoichiometries formally correspond to structures that
interpolate between an insulating and metallic phase that are
separated by a discontinuous transition and thus are automatically
less stable than either of the two phases.  One can likewise
rationalize the variety of bonding preferences and structures in the
semi-metallic region of the periodic table. For instance, the present
notions provide a general understanding of why the structures of
di-pnictogen tri-chalcogenides, such as those depicted in
Fig.~\ref{coex}, show bonding ranging from essentially covalent in
lighter elements to multi-center,
hypervalent~\cite{PapoianHoffmann2000} interactions for heavier
species. In addition, the latter compounds are organized,
structure-wise, into stripes and ribbons despite the relatively
uniform spatial distribution of the two constituent elements. Here we
observe that the latter organization is analogous to lamellar
ordering, which is common during phase coexistence.~\cite{SWmayo} Last
but not least, the present results indicate that complications arising
from those ambiguous bonding preferences are subject to universal
relations in the form of laws of corresponding states. This justifies
the use of semi-empirical, meanfield treatments exemplified by
density-functional and tight-binding approximations, and even
implicit-electron treatments such as the {\em classical} DFT,
Landau-Ginzburg treatments of displacive transitions and multiferroic
phenomena.~\cite{PhysRevLett.3.412, Dove1997}

% For instance, the classic covalent bond can be seen as a low-density
% relic of many-body forces that arise in the solid state.
The article is organized as follows: In Section~\ref{trimer}, we
quantitatively analyse a substantial number of small molecular motifs
to elucidate the mechanism of the symmetry breaking that leads to the
emergence of the covalent and secondary bond from the multi-center
bond, and its interplay with the ionic interaction.
Section~\ref{localization} discusses the electronic symmetry breaking
underlying structural instabilities and the robustness of the concept
of the bond order in molecular systems.  In Section~\ref{solids}, we
extend those arguments to the solid state context and build a phase
diagram of chemical interactions. We summarize and discuss the present
results in Section~\ref{discussion}.

\section{Interplay of covalent, secondary, multi-center, and ionic
  interactions: Small molecules}

\label{trimer}

\begin{figure}[t]
  \centering
  \includegraphics[width = .85 \figurewidth]{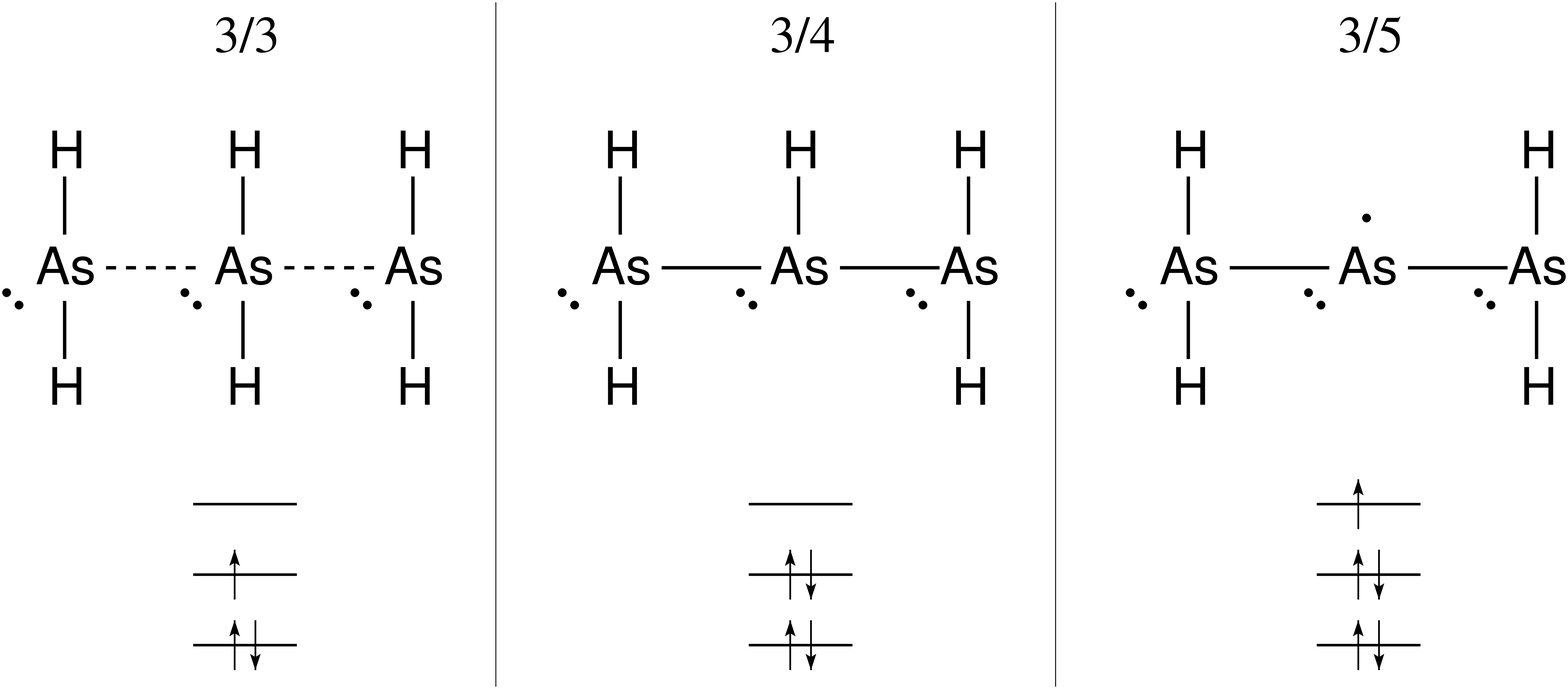}
  \caption{\label{Lewis} Lewis diagrams and the electronic
    configurations for the three-center $pp\sigma$ bond in the
    AsH$_2$-AsH$_n$-AsH$_2$ trimer, $n = 2, 1, 0$ left to right. The
    two As-As bond lengths are assumed to be equal. The dashed lines
    for the $n=2$, 3/3 case indicate there are formally only 1.5
    electrons per bond.}
\end{figure}

To investigate spatial symmetry breaking in small molecules, we
analyse a number of $pp\sigma$ bonded trimeric units suitably
passivated by hydrogens to achieve a desired number of electrons in
the $pp\sigma$ bond.  We begin from the AsH$_2$-AsH$_n$-AsH$_2$
trimer, $n = 0, 1, 2$.  According to the Lewis diagram in
Fig.~\ref{Lewis}, this formally corresponds to a three-center bond
that contains 5, 4, and 3 electrons, respectively, assuming each
arsenic carries a lone pair.  Our main focus is on the $n=1$ case,
AsH$_2$-AsH$_1$-AsH$_2$, which corresponds to the classic
3-center/4-electron bond and, at the same time, obeys the Lewis octet
rule.  Consistent with the latter notion, it is the only molecule of
the three that happens to be stable; the 3 and 5-electron molecules
dissociate into a (passivated) dimer and monomer. The two As-As bond
lengths, $b_1$ and $b_2$ respectively, are equal in the ground state
of the 3/4 molecule.

\begin{figure}[t]
  \centering
  \includegraphics[width = 0.9
  \figurewidth]{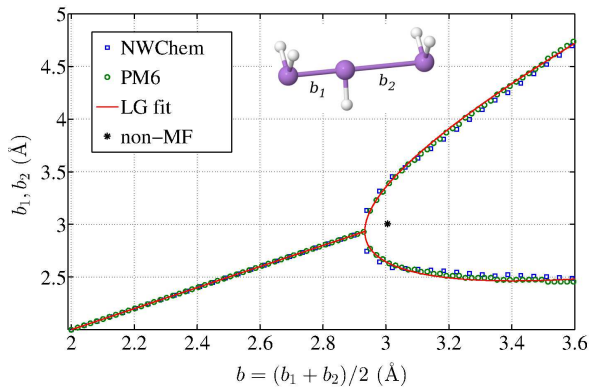}
  \caption{\label{b1b2} The equilibrium As-As bond lengths for the 3/4
    case as functions of the overall trimer length, per bond,
    c.f. right top panel of Fig.~\ref{LH}. The energies are calculated
    using MOPAC and NWChem, see text. The arsenic atoms are
    constrained to lie on a straight line, while the As-H bonds lie in
    planes perpendicular to that line. The As-H bonds are fixed at
    length 1.5~\AA~; mutual angle $90^\circ$ for the terminal bonds,
    see the inset. The molecule has two reflection planes. The
    asterisk denotes the location of the symmetry breaking corrected
    for zero-point vibrations of the molecule.}
\end{figure}

If stretched beyond a certain critical length, the molecule will
dissociate and thus break the $b_1 = b_2$ symmetry. This is shown in
Fig.~\ref{b1b2}, where we plot $b_1$ and $b_2$ as functions of the
overall trimer length per bond, $b \equiv (b_1 + b_2)/2$, which is
externally imposed. The full potential energy surface of the molecule,
as a function of the As-As bond lengths $b_1$ and $b_2$, is provided
in the Supplementary Material.  One set of curves in Fig.~\ref{b1b2}
corresponds to the semi-empirical approximation implemented in the
package MOPAC with PM6 parametrization.\cite{MOPAC2012,Stewart2007}
MOPAC treats explicitly only the valence electrons, while using only a
single, Slater-type basis function per atomic orbital and neglecting
overlap between wave-functions on different centers during the
self-consistent solution of the Hartree-Fock (HF)
problem.~\cite{doi:10.1021/ja00457a004} (The corresponding matrix
elements of the Hamiltonian are generally non-zero, of course.) The
other set of curves is produced by a more accurate, ab initio
approximation as implemented in the package
NWChem~\cite{Valiev20101477} using the aug-cc-pVTZ-pp basis and a
small, ten-electron effective core potential.  MOPAC-optimized
geometry compares well with the more accurate method, despite the
relatively crude level of approximation. There are several motivations
behind our use of MOPAC, to be discussed in due time.

The symmetry breaking transition in Fig.~\ref{b1b2},
c.f. Fig.~\ref{LH}, is continuous and apparently similar to classic
examples of symmetry breaking such as the Curie point or the critical
point in liquids.  The transition can be formally described, at a
meanfield level, using the Landau-Ginzburg expansion of the free
energy~\cite{Goldenfeld} as a function of an order parameter
reflecting the extent of symmetry breaking. In the present context, a
convenient order parameter is the displacement $\Delta b \equiv (b_1 -
b_2)/2$ of the central arsenic off the midpoint between the terminal
arsenics. The corresponding Landau-Ginzburg expansion then reads:
\begin{equation} \label{F} F(\Delta b) = \frac{a_2}{2} (b_c - b) \,
  {(\Delta b)^2} + \frac{a_4}{4} {(\Delta b)^4},
\end{equation}
where $b_c$ stands for the critical value of the trimer length $b$ per
bond. The equilibrium value of the order parameter is determined by
optimizing the ``free energy'' (\ref{F}). The quantities $a_2$ and
$a_4$ are system-dependent parameters. In the symmetry broken region,
the displacement of the middle arsenic is thus given by a simple
formula:
\begin{equation} \label{lcs} \Delta \tilde{b} = \pm ( \tilde{b} -1
  )^{1/2},
\end{equation}
where we have rescaled the control parameter by its critical value: $
\tilde{b} \equiv b/b_c$, and the order parameter by an appropriate
combination of the critical length $b_c$ and the expansion
coefficients $a_i$: $\Delta \tilde{b} \equiv \Delta b /[a_2
b_c/a_4]^{1/2}$. Eq.~(\ref{lcs}) is, of course, a system-independent,
universal relation and thus constitutes a law of corresponding states.
The best fit of the functional form (\ref{lcs}) to the $b_1$ vs. $b_2$
dependences in Fig.~\ref{b1b2} is shown as the solid red line in the
same figure, 

The apparently excellent fit of the bond lengths $b_1$ and $b_2$ to
the meanfield expression (\ref{lcs}) suggests correlations do not
significantly affect the symmetry breaking. Still, the location of the
bifurcation point in Fig.~\ref{b1b2} is only a lower bound on the
value of the critical length at which the actual symmetry breaking
would occur. This is because already zero-point, let alone
finite-temperature vibrations within either of the two individual
minima on the symmetry-broken energy surface will allow the system to
cross the barrier separating the minima, if the latter barrier is
sufficiently low.  Accordingly, one may assess the fluctuation-induced
lowering of the critical point~\cite{Goldenfeld} semi-quantitatively,
by requiring that those zero-point vibrations do not exceed the
half-width of the barrier, see the graphical illustration in the
Supplementary Material. The so estimated location of the critical
point is shown by the asterisk in Fig.~\ref{b1b2}; it differs
meaningfully from its meanfield value.

\begin{figure}[t]
  \centering
  \includegraphics[width = 0.95
  \figurewidth]{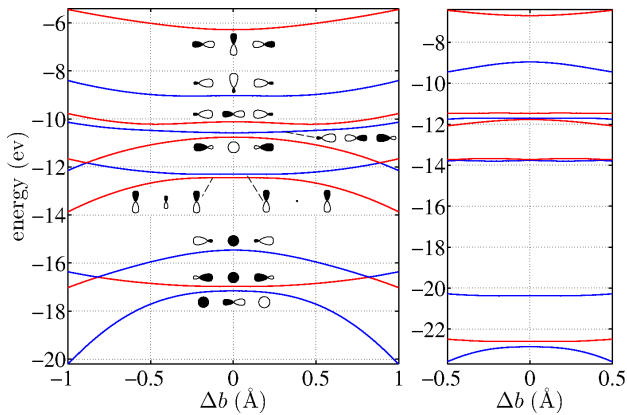}
  \caption{\label{Walsh} Walsh diagrams showing the energies of
    individual occupied MO as functions of the displacement $\Delta b$
    of the central arsenic for the 3/4 trimer
    AsH${_2}$-AsH-AsH${_2}$. The distance $b_1 + b_2$ between the
    terminal arsenics is 5.98~\AA, which is just beyond the meanfield
    symmetry breaking point. The left and right panels show MOPAC PM6 and
    NWChem data, respectively; note the difference in the horizontal
    ranges.}
\end{figure}

In the aforementioned examples of the Curie point and the continuous
vapor-to-liquid transition, the symmetric state is favored by entropic
forces, while cohesive interactions favor the symmetry broken
state.~\cite{L_AP} At the critical point, the two forces are in
balance so that fluctuations of the order parameter incur zero cost.
Likewise, we inquire what competing factors could drive the transition
in Fig.~\ref{b1b2}. It should be immediately clear that in the $b
\equiv (b_1+b_2)/2 \to \infty$ limit, the lowest energy state is
asymetric. Informally speaking, a bond is better than no bond. In a
more formal vein, symmetry lowering transitions in small molecules are
often associated with Jahn-Teller instabilities.~\cite{ABW,
  Bersuker2006, Hargittai2008} (Such instability would have to be
second order at least, in this case, because of symmetry.)  In
contrast, we observe in Fig.~\ref{Walsh} that the HOMO is actually
{\em stabilized} in the symmetric configuration. At the same time, the
molecular terms behave all but generically near the symmetry breaking
point.  Fig.~\ref{Walsh} displays the terms for a broad range of the
displacement of the central arsenic as calculated using MOPAC PM6 and
NWChem; the two calculations produce qualitatively similar results.

\begin{figure}[t]
  \centering
  \includegraphics[width = \figurewidth]{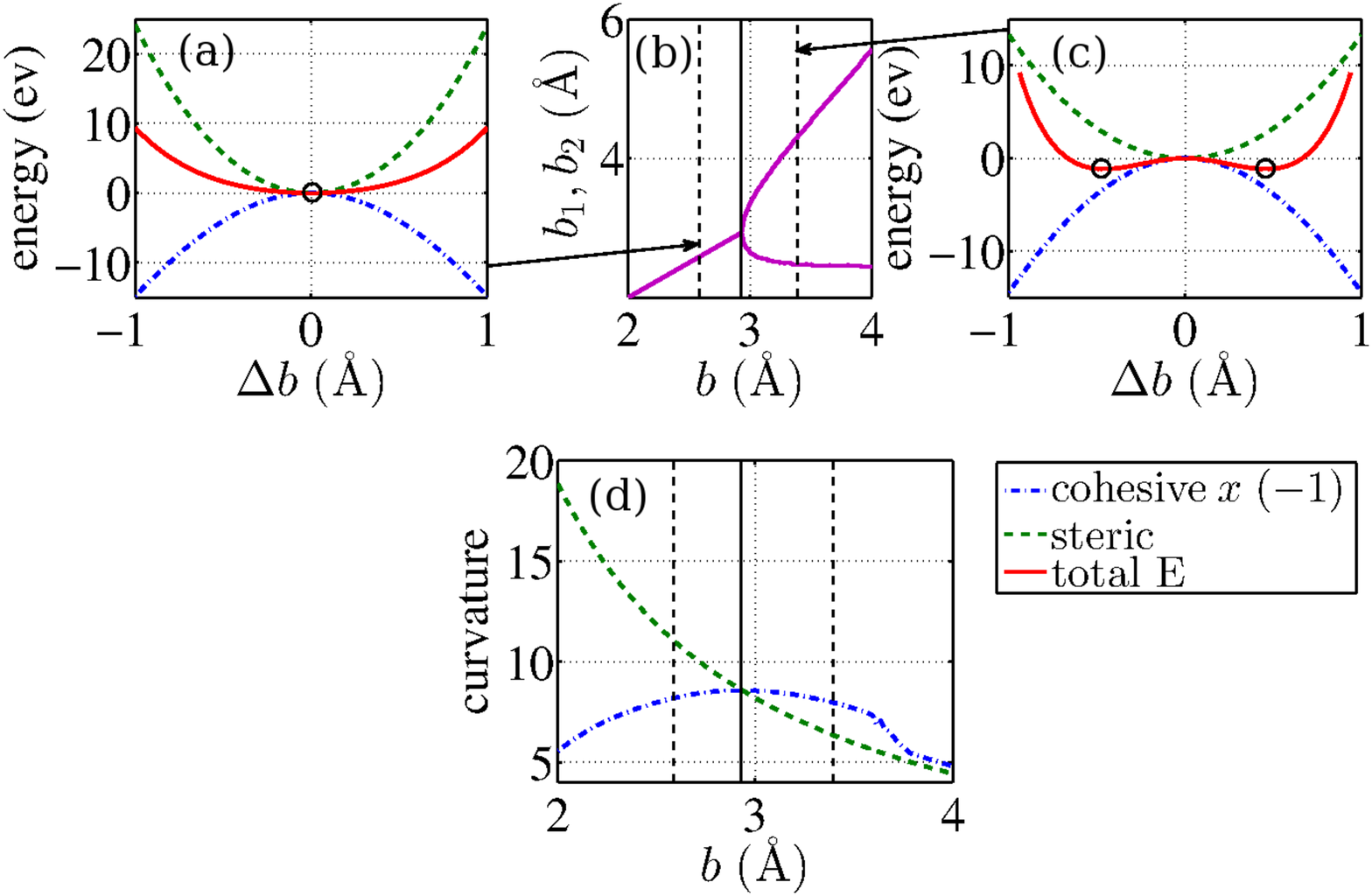}
  \caption{\label{LG} {\bf (a)} The dependence of the cohesive
    contribution (green), repulsive contribution (blue) and the total
    energy (red) of the 3/4 trimer AsH${_2}$-AsH-AsH${_2}$ on the
    location of the middle arsenic, at a high density favoring the
    symmetric state. The circle indicates the location of
    the stable minima. {\bf (b)} is the same as Fig.~\ref{b1b2}. {\bf (c)} Same as (a), but at a low density
    favoring symmetry breaking. There are two equivalent minima
    now. Panel {\bf (d)} displays the dependence of the absolute
    values of the curvatures of the repulsive and cohesive parts at
    $\Delta b = 0$, on the trimer length. Vertical dashed lines indicate $b$ values associated with energy contours (a) and (c).}
\end{figure}

In search for alternative explanation, we note that by construction,
MOPAC presents the full energy of the molecule as a sum of a cohesive
and repulsive part: The cohesive part is evaluated by solving the HF
problem using the valence electrons, as already mentioned. The steric
part is parametrized to model the repulsion between the ionic
cores. (We have verified that not much promotion of $3d$ electrons
takes place at the densities in question, see the Supplementary
Material.)  In Fig.~\ref{LG}, we plot the cohesive and repulsive
contributions, along with the total energy, in the symmetric and
symmetry-broken regime. The cohesive part is increasingly stabilized
for larger $\Delta b$. This is expected since the energy of the
shorter bond depends sensitively on the bond length while the cohesive
energy of the longer bond depends on the displacement already
relatively weakly.  The repulsive part is convex down and is minimized
at $\Delta b = 0$. At sufficiently high densities, the curvature of
the repulsive term exceeds that of the cohesive part, thus stabilizing
the symmetric, $b_1 = b_2$ state.  The opposite take place in a
sufficiently long trimer, thus leading to a bistable potential
corresponding to the emergence of two equivalent symmetry broken
states.

\begin{figure}[t]
  \centering
  \includegraphics[width = .8 \figurewidth]{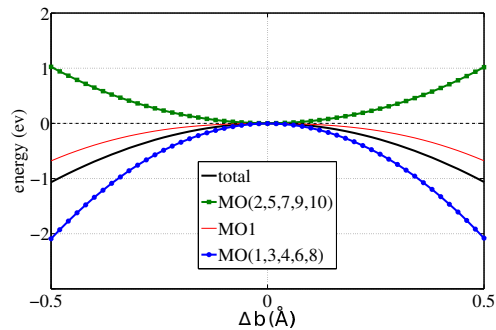}
  \caption{\label{MO_avg} The solid black line shows the dependence of
    the MOPAC-inferred cohesive energy as a function of the location
    of the middle arsenic for the 3/4 trimer
    AsH${_2}$-AsH-AsH${_2}$. The sum of the MOPAC terms from
    Fig.~\ref{Walsh} that stabilize and destabilize the symmetric
    state shown in green and blue respectively. The individual
    contribution of the lowest term is shown in red.}
\end{figure}

\begin{figure}[t]
  \centering
  \includegraphics[width = .9 \figurewidth]{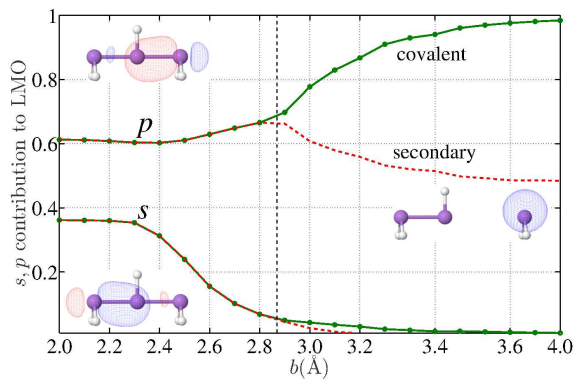}
  \caption{\label{43LMO} The $s$ and $p$ content of the localized
    molecular orbitals (LMOs), as functions of the trimer length per
    bond for the 3/4 trimer AsH${_2}$-AsH-AsH${_2}$.  The localized
    molecular orbitals (LMOs) on the opposite sides of the symmetry
    breaking are exemplified in the insets.}
\end{figure}

\begin{figure}[t]
  \centering
  \includegraphics[width = 0.9 \figurewidth]{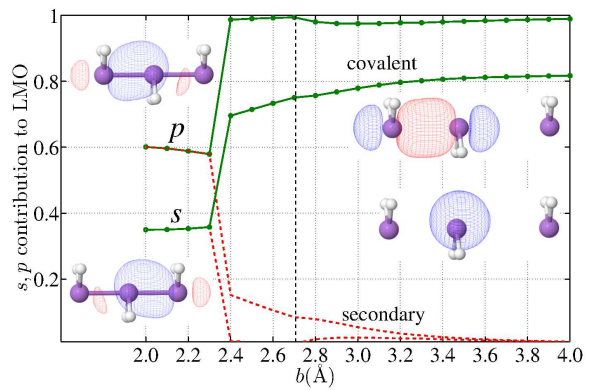}
  \caption{\label{33LMO} Same as Fig.~\ref{43LMO}, but for the 3/3
    trimer AsH${_2}$-AsH$_2$-AsH${_2}$.  The vertical dashed line
    indicates the length at which the {\em spatial} symmetry is
    broken. Note that the localized molecular orbitals exhibit a
    symmetry breaking while the molecule itself is still symmetric.}
\end{figure}

Fig.~\ref{MO_avg} provides a summary of the MOPAC-produced electronic
terms from Fig.~\ref{Walsh}; it displays the total cohesive energy,
the partial contributions of the terms that stabilize and destabilize
the symmetric state, and, separately, the lowest-energy term from
Fig.~\ref{Walsh}. The latter term clearly contributes most to the
destabilization. We note that of the four terms favoring symmetry
breaking, three terms, including the bottom one, stem from
$sp$-mixing. Conversely, only one of the $sp$-mixed orbitals
stabilizes the symmetric state. This suggests the symmetry breaking is
driven to a large extent by $sp$-mixing, consistent with solid state
precedents.~\cite{SeoHoffmann1999} The latter observation may seem in
conflict with the expectation that the amount of $sp$-mixing should
decrease with inter-atomic separation.  This expectation is directly
confirmed by the data in Figs.~\ref{43LMO} and \ref{33LMO}, where we
show the contribution of the $s$ and $p$ atomic orbitals to the
bond. We determine the latter contribution by a localization
procedure,~\cite{RevModPhys.32.296, RevModPhys.35.457,
  doi:10.1063/1.1677859,NBO6} which is discussed in great detail in
Section~\ref{localization}. We observe that upon dilation of the
4-electron trimer, $sp$-mixing largely peters out by the time the
spatial symmetry is broken. This process is even more dramatic for the
electron-poor trimer $($AsH$_2)_3$, for which the bonding orbitals
undergo a symmetry-breaking transition even as the trimer is still
spatially symmetric. This, inherently electronic transition has to do
with a transfer of electrons from the $pp\sigma$-bond to the lone pair
on the central arsenic and will be discussed in Section
\ref{localization}.

How does one reconcile the significance of $sp$-mixing for symmetry
breaking, as apparent from Figs.~\ref{Walsh} and \ref{MO_avg}, with
its decrease at low densities, where the actual symmetry breaking
occurs, Fig.~\ref{43LMO}? To resolve this apparent contradiction we
note that although the destabilization of the symmetric state due to
$sp$-mixing does decrease with interatomic separation for sufficiently
long trimers, the stabilization due to steric repulsion diminishes
even faster, see Fig.~\ref{LG}(d). This emphasizes a relatively subtle
feature of the symmetry breaking: The cohesive and steric interactions
both evolve similarly with density, at least at sufficiently low
values of the latter. As the inter-nuclear distance decreases,
electrons move toward the inter-atomic space because of the cumulative
effects of Coulomb attraction to the involved nuclei; this stabilizes
the cohesive component. At the same time, the steric repulsion {\em
  also} increases with density. The resulting bond enthalpy is
therefore a modestly-sized quantity resulting from a delicate balance
between two opposing, large quantities. We will observe a similar, but
richer pattern in Section~\ref{solids}, in the context of
density-driven coordination changes in solids.

\begin{figure}[t]
  \centering
  \includegraphics[width = .9 \figurewidth]{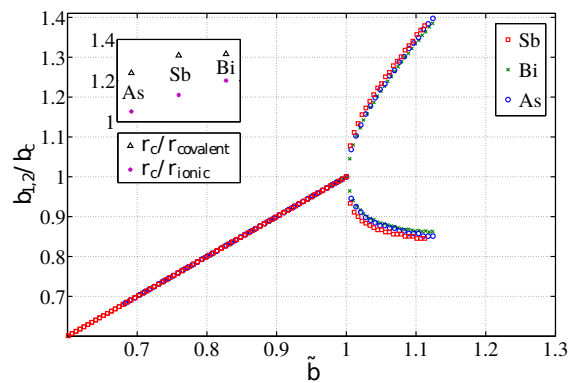}
  \caption{\label{AsSbBi} The dependences of the bond length for 3/4
    trimers XH$_2$-XH-XH$_2$ (X = As, Sb, Bi), normalized by the
    respective critical length. The inset displays the ratio of the
    critical length in terms of the corresponding covalent and ionic
    radii for each element. The following values, in \AA, for the
    covalent (ionic) radii for As, Sb, and Bi are adopted: 1.18
    (1.39), 1.36 (1.59),~\cite{WBP72} 1.52~\cite{CW80}
    (1.7~\cite{WBP72}). }
\end{figure} 

We next investigate whether our model trimeric units exhibit a
universality of the type in Fig.~\ref{LH}, using the atom size as the
control variable. Fig.~\ref{AsSbBi} shows the ``bifurcation'' plots
for three elements from group 15---As, Sb, and Bi---rescaled by the
corresponding critical length. In the inset, we show the ratio of the
latter length to the corresponding ionic and covalent radii for
individual substances. We observe a fair deal of universality upon
rescaling. At the same time, the critical length of the trimer, which
is the characteristic length scale in the problem, is not strictly
tied to common measures of the atomic size. In a systematic trend, the
departure increases with the atomic mass.  These conclusions are
consistent with above findings on $sp$-mixing driving the symmetry
breaking and the general notion that the amount of $sp$-mixing tends
to decrease as one goes down the group in the periodic
table.~\cite{PapoianHoffmann2000} In addition, the above trend is
consonant with Clementi et al.'s~\cite{doi:10.1063/1.1712084}
observation that the separation between the maximum charge-density
radius of the outermost $d$ shell and valence $s$ and $p$ orbitals
anti-correlates with the atomic number. In the present context, this
implies that the effective size of the ionic core is greater for
heavier atoms, relative to the extent of the frontier atomic
orbitals. The resulting enhancement in steric repulsion thus serves to
stabilize the symmetric configuration. Note that already rescaling
both $b$ and $\Delta b$ with the critical length $b_c$ largely
suffices in bringing all of the bifurcation graphs to a universal
form; we will see shortly this simplification does not apply
generally.

Similarly to the preceding discussion, we next study the effects of
varying the electron content of the three center bond on the symmetry
breaking. Specifically, we vary the number of passivating hydrogens on
the AsH$_2$-AsH$_n$-AsH$_2$ trimer, as mentioned in the beginning of
the Section. Similarly to Fig.~\ref{AsSbBi}, the bifurcation graphs
follow a universal shape. In contrast with that situation, both
variables $b$ and $\Delta b$ need to be rescaled, see the
Supplementary Material. The critical lengths depend on the population
of the $pp\sigma$ bond: $b_c = 2.71$, $2.87$, and $2.63$ for three,
four, and five electron bond, respectively.  This is consistent with
the view of the middle orbital of the $pp\sigma$ bond,
Fig.~\ref{Lewis}, as mildly bonding.~\cite{PapoianHoffmann2000} We
reiterate that the four-electron case satisfies the Lewis octet rule
in the symmetric state.  To avoid confusion we note that the electron
count cannot be generally regarded as an independent control parameter
but is determined self-consistently at given values of density and
electronegativity variation, as is already clear from
Figs.~\ref{43LMO} and \ref{33LMO}.

\begin{figure}[t]
  \centering
  \includegraphics[width = 0.9 \figurewidth]{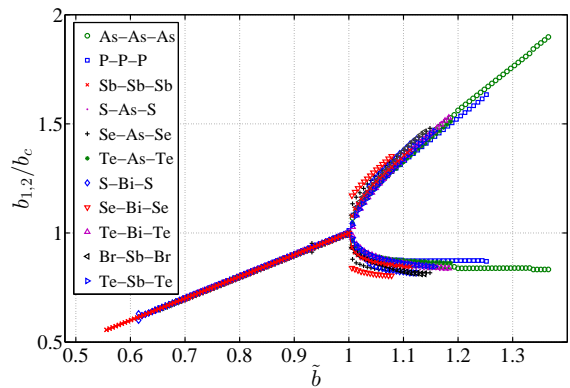}
  \caption{\label{43_universal} A number of bifurcation diagrams for
    pnictogen and chalcogen containing 3/4 trimers shown in reduced
    coordinates, as in Fig.~\ref{AsSbBi}. To maintain the 4 electron
    count, a chalcogen is passivated with one less hydrogen than a
    pnictogen.}
\end{figure}

In discussing effects of electronegativity variation, we first recall
that ionicity tends to suppress dimerization in extended
one-dimensional systems.~\cite{PhysRevLett.49.1455,
  PhysRevLett.45.926} A similar effect can be seen at the
extended-H\"uckel level for a hypothetical linear H${_3^-}$ molecule,
Chapter 6.4 of Ref.~\onlinecite{ABW}. We have checked that modifying
the electronegativity variation along the trimer does not
significantly affect the universality of the symmetry breaking, see
Fig.~\ref{43_universal} and Supplementary Material. There, we also
demonstrate that the leading effect of introducing additional
electronegativity variation is to shorten the bonds, that is, the
critical length for a mixed trimer X-Y-X or Y-X-Y is usually less than
the average of the critical lengths for the homoatomic trimers X$_3$
and Y$_3$. The effect is modestly stronger when the more
electronegative element is placed at the terminal positions,
consistent with the earlier notion that the three-center bond is
already partially ionic because the terminal atoms contribute more
electronic density than the central atom to the individual two-center
bonds.

The compilation of symmetry breakings for an extensive set of trimers,
in Fig.~\ref{43_universal}, juxtaposes effects of varying the atom
size, electronegativity, and valence.  We again observe that every
time continuous symmetry breaking takes place, a law of corresponding
states holds already when one rescales all lengthscales by the
critical length. If one were to rescale the bond-length difference
$(b_1 - b_2)$, the universality would be even more pronounced. The
rather extensive compilation in Fig.~\ref{43_universal} shows that
some of the trimers actually exhibit a {\em discontinuous} symmetry
breaking, c.f. the bottom of Fig.~\ref{LH}. The energy surface for
this more complicated situation now exhibits up to three minima, at
$b=\text{const}$, see the Supplementary Material. In contrast with
Fig.~\ref{LH}, the density range in which the symmetric and asymmetric
molecule coexist is very narrow. Most likely such a discontinuity in a
small molecule would be washed away by fluctuations, similarly to the
lowering of the critical point in Fig.~\ref{b1b2}. On the other hand,
in the solid state the discontinuity may will be significantly
stabilized by the crystal field.

% and a juxtaposition of the Landrum and Hoffmann data for Br-Sb-Br
% materials on the analogous data produced by MOPAC for the Br-SbH-Br
% trimer.

\section{Bond assignment as a result of electronic symmetry breaking}

\label{localization}

As discussed above, one may associate the spatial symmetry breaking in
the 3/4 trimer with regard to the bond {\em strength} between
neighboring arsenics with the spatial symmetry breaking $b_2 \ne b_1$,
at least for sufficiently large values of $|b_2 - b_1|$. On the other
hand, the relatively electron-poor 3/3 trimer exhibits an apparent
change in bonding already in the symmetric $b_1 = b_2$ configuration.
To elucidate this type of electronic transition we consider two
specific hydrogen-passivated trimers AsH$_2$-AsH-AsH$_2$ and
AsH$_2$-AsH$_2$-AsH$_2$ while preventing the breaking of the $b_1
\leftrightarrow b_2$ symmetry altogether. We shall impose an
artificial constraint $b_1 = b_2$ for {\em all} values of the trimer
length. The hydrogens are constratined so that the molecule has two
symmetry planes, one containing the 3-center bond and the other---call
it plane $R$---perpendicular to the bond and containing the middle
arsenic. Each MO thus must be either even or odd with respect to the
reflection in plane $R$.

% In contrast with what is suggested by the simple Lewis dot diagram
% in Fig.~\ref{Lewis}, the actual populations of the lone pairs and
% the $pp\sigma$ bond, if any, depend on the As-As bond length, as we
% will see shortly.

To quantify bonding in these symmetric trimers we use the localized
molecular orbital (LMO) formalism. A pedagogical overview of the
latter can be found in the Supplementary Material. Here, we only provide
definitions and brief descriptions for several pertinent
quantities. Originally inspired by symmetry adapted orbitals in
relatively symmetric molecules, the localized molecular orbitals can
be assigned for arbitrary geometries.  One starts out with the
molecular orbitals (MO) $\psi_i$
\begin{equation} \label{psiExp} \psi_i = \sum_A \sum_{\lambda_A}
  \widetilde{C}_{\lambda_A i} \phi_{\lambda_A},
\end{equation}
where $\phi_{\lambda_A}$ stands for an atomic orbital on atom $A$.
Given a set of {\em occupied} MOs, one transforms to an alternative
set of orthonormal orbitals $\chi_j$ 
\begin{equation} \label{localized_orbitals} \chi_j = \sum_A
  \sum_{\lambda_A} C_{\lambda_A j} \phi_{\lambda_A}
\end{equation}
such that a certain quantity reflecting the self-repulsion within
individual new orbitals is maximized.\cite{RevModPhys.32.296,
  RevModPhys.35.457, doi:10.1063/1.1677859, PerkinsStewart1980} The
quantities $\chi_{j}$, often called localized molecular orbitals
(LMO), represent an attempt by an interpreter to partition (the
already bound) electrons among orthonormal, maximally localized
orbitals, each of which thus binds together the smallest number of
atoms.  The number of atoms inolved in the corresponding bond, or the
bond center number $n_j$ is computed according to:
\begin{equation} \label{center_number} n_j\equiv \sum_A \left(
    \sum_{\lambda_A} C^2_{\lambda_A j} \right)^2.
\end{equation}
The contribution the $j$-th LMO to a bond, as opposed to a lone pair
for instance, is often called the ``bonding contribution''
\begin{equation} \label{BcontDef} C^\text{(bond)}_{jj} \equiv \la
  \chi_j | \widehat P | \chi_j \ra = 2 \sum_{\lambda \mu} C_{\lambda
    j} C_{\mu j} P_{\lambda \mu},
\end{equation}
where terms pertaining to the same atoms are excluded from the
summation and $P_{\lambda \sigma}$ is the density matrix element
\begin{equation} \label{densMat} P_{\lambda \mu} \equiv 2 \sum_i
  \widetilde{C}_{\lambda i} \widetilde{C}_{\mu i}.
\end{equation}
The summation in Eq.~(\ref{densMat}) is exclusively over occupied MOs.

It will be useful to complement the LMO formalism, which is quite
anthropocentric, with a more basic descriptor, viz., the number of
electrons shared by atoms A and B, $B_{AB}$, often called the Wiberg
index:\cite{Armstrong73}
\begin{equation} \label{Wiberg_index} B_{AB} \equiv \sum_{\lambda_A
    \!, \, \lambda_B} P^2_{\lambda_A \lambda_B}.
\end{equation}

\begin{figure}[t]
  \centering
  \includegraphics[width = \figurewidth]{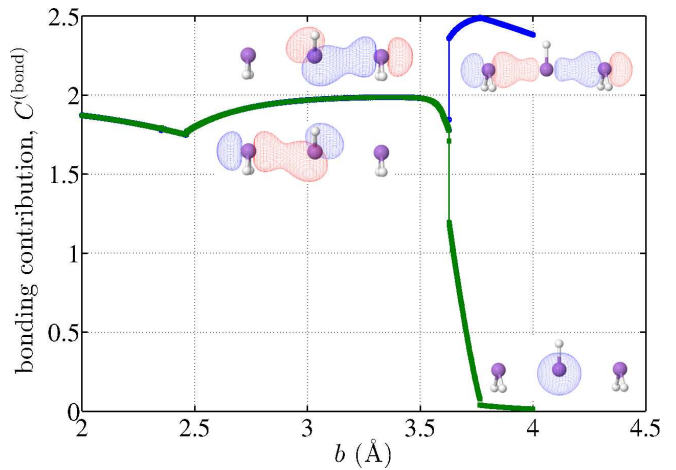}
  \caption{\label{43BcontExp} The bonding contribution of the LMOs as
    functions of the As-As bond length $b = b_1 = b_2$, for the 3/4
    case.}
\end{figure}

\begin{figure}[t]
  \centering
  \includegraphics[width
  =\figurewidth]{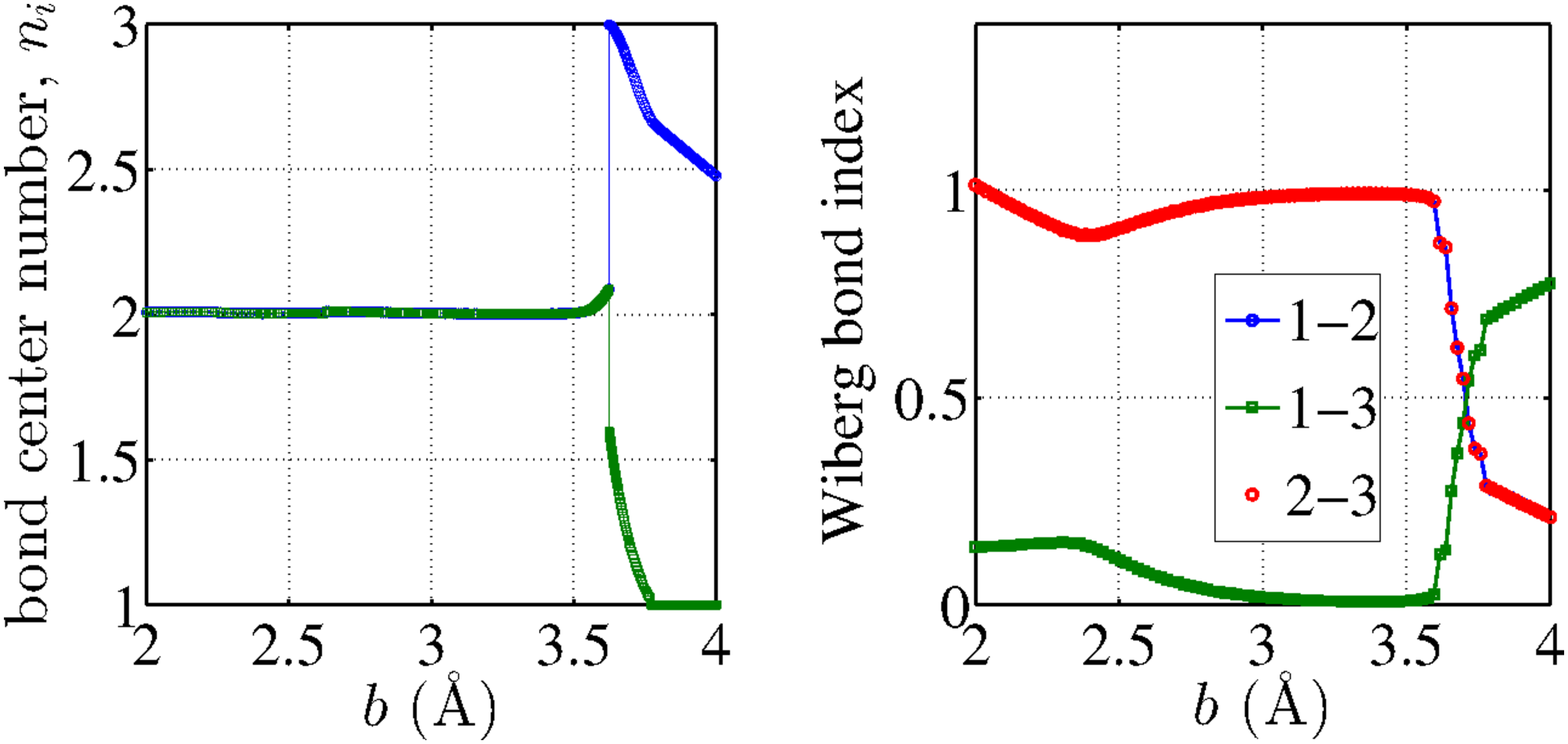}
  \caption{\label{BO} Displayed as functions of the As-As bond length
    $b = b_1 = b_2$: {\bf (a)}, the bond center number for the LMOs
    from Fig.~\ref{43BcontExp}; {\bf (b)}, the Wiberg bond index. Note
    the latter is computed using the density matrix and does not rely
    on the localization procedure.}
\end{figure}

First off, we identify the localized molecular orbitals (LMOs) that
are associated with the As-As-As subsystem, as opposed to the As-H
bonds. There are two such LMOs.  Their bond contributions,
Eq.~(\ref{BcontDef}), are shown in Fig.~\ref{43BcontExp} as functions
of the trimer-length.  In that same figure, we display the
corresponding localized molecular orbitals themselves. The
corresponding bond-center number, Eq.~(\ref{center_number}), and
Wiberg bond index, Eq.~(\ref{Wiberg_index}), are shown in
Figs.~\ref{BO}(a) and (b) respectively.

According to Figs.~\ref{43BcontExp}-\ref{BO}, the bond characteristics
of the electron rich 3/4 trimer remain steady within a remarkably
broad range of the inter-arsenic distance and are essentially the same
as those in the ground state.  The bond order eventually changes, at
sufficiently low densities, and does so in an abrupt fashion,
especially the bonding contribution.  Figs.~\ref{43BcontExp} and
\ref{BO}(a) indicate that the transition is truly discontinuous: Its
precise location exhibits a hysteretic behavior and depends on the
precise protocol such as the grid size for the quantity $b$ or the
tolerance of the self-consistent Hartree-Fock (SCHF) procedure; the
resulting ambiguity in $b$ is not large, however, a hundredth of an
angstrom or so.  The transition in the bond type is apparently
accompanied by a {\em symmetry change} in the LMO. On the high density
side, the LMOs are mirror images of each other and amount to the same
bond contribution. Neither of these LMOs contributes to an irreducible
representation of the molecule's point group, only a linear
combination does.  On the low density side, in contrast, both LMOs are
even functions with regard to the reflection in plane $R$. One LMO
stretches over the three centers, consistent with its bond-center
number and the bond contribution, while the other LMO is essentially a
lone pair. The nearest-neighbor As-As bonds on the high-density side
of the transition are unquestionably weak. The bond length at the
transition---$b=3.628$~\AA~for the specific realization in
Fig.~\ref{43BcontExp}---is about 50\% longer than its equilibrium
value and, in fact, is more appropriate for a closed-shell, secondary
interaction.~\cite{ZLMicro1} Still, the interaction is not
closed-shell since the formally antibonding orbital in the $pp\sigma$
bond is vacant, see graphical three-orbital representation in
Supplementary Material.  In any event, we have confirmed the symmetry
breaking transition using a higher end approximation and alternative
methods of conceptualizing molecular bonding, viz. the Natural Bond
analysis~\cite{g09, NBO6} and the QTAIM
theory,~\cite{doi:10.1021/jp983362q,MolinaMolina2001,BaderAIM,MacchiCDA}
see the Supplementary Material. There, we also show that the molecular
orbitals do eventually localize on the respective centers for
sufficiently long trimers.

\begin{figure}[t]
  \centering
  \includegraphics[width =
  \figurewidth]{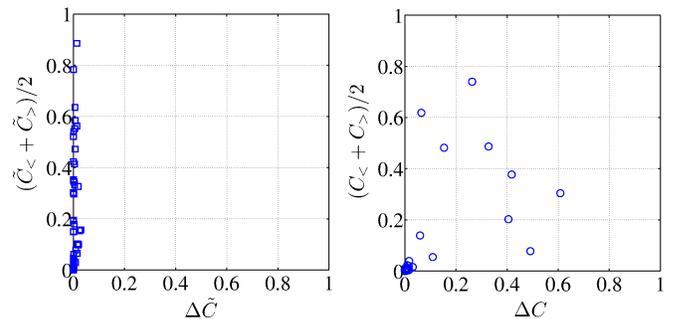}
  \caption{ \label{43scatter} Scatter plots of the changes in the
    expansion coefficients of the MOs, {\bf (a)}, and LMOs, {\bf (b)},
    upon the symmetry breaking transition in Fig.~\ref{43BcontExp},
    vs. their average values.}
\end{figure}

We next inquire whether the apparently abrupt change in the localized
molecular orbitals is caused by an abrupt change in the {\em
  canonical} molecular orbitals, if any. Already a small molecule such
as the AsH$_2$-AsH-AsH$_2$ has many molecular orbitals. Specifically,
at the MOPAC level, each arsenic has nine orbitals per atom ($4s$,
$p$, and $d$) and each hydrogen has one, that make substantial
contributions to the MOs.  Individually plotting a large number of
molecular terms, as functions of density, is hardly
illuminating. Instead, we make two scatter plots,
Fig.~\ref{43scatter}(a) and (b), in which we show the {\em change} in
the expansion coefficients $C$ and $\widetilde{C}$ from
Eqs.~(\ref{psiExp}) and (\ref{localized_orbitals}) across the
transition vs. their arithmetic average. The two sampling points are
at $b =$ 3.61315 \AA~ and 3.61335 \AA, respectively; this specific
realization of the LMO-switching transition takes place for a slightly
shorter trimer than in Fig.~\ref{43BcontExp}. According to the scatter
plots, the changes in the MO expansion coefficients $\widetilde{C}$
are gradual; these changes are commensurate with the magnitude of the
change in the bond length across the transition. In contrast, the
coefficients $C$ of the LMO expansion change discontinuously. (A
specific pair of LMOs were chosen for presentation in
Fig.~\ref{43scatter}(b); others show the same trend.) One can thus
rule out the possibility that the rapid change in the bond assignment
is caused by a term-crossing, consistent with the results of direct
inspection of the molecular terms, see the Supplmentary Material. Note
that the Wiberg bond index from Fig.~\ref{BO}(b) changes noticeably,
but continuously near the transition.

The appearance of a discontinuous transition for the LMOs is not
unexpected considering that the localization procedure, discussed in
the Supplementary Material, is a non-linear problem. For the sake of
concreteness, we will designate those transitions as
``LMO-transitions,'' to distinguish them from any transitions that
happen already at the Hartree-Fock level, if any. We will see below
that LMO-transitions occurring in the absence of an underlying
transition at the HF level are an exception rather than the rule for
larger systems.
 
The qualitative change in the LMO in Fig.~\ref{43BcontExp} is
consistent with the apparent destabilization of the MO corresponding
to the $pp\sigma$ interaction, on the one hand, and stabilization of
the MO corresponding to the lone pair on the middle arsenic, on the
other hand. (The respective molecular terms are shown in
%Figs.~\ref{34MOWalsh} and \ref{MOexplanation} in the Supplementary
Figs.~S15 and S16 in the Supplementary Material.)

\begin{figure}[H]
  \centering
  \includegraphics[width = \figurewidth]{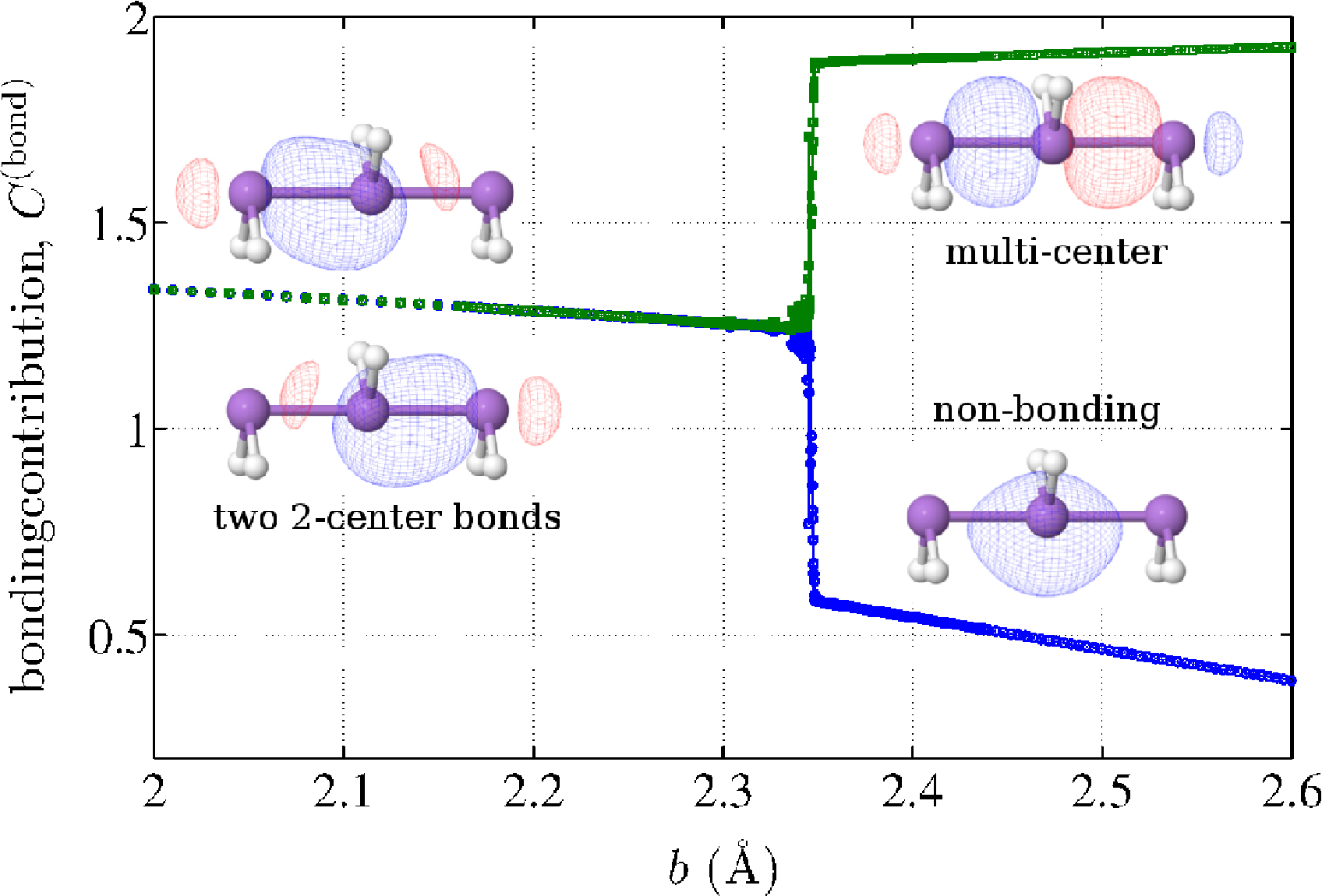}
  \caption{\label{33BO} The bonding contribution of the LMOs as
    functions of the As-As bond length $b = b_1 = b_2$, for the 3/3
    molecule.}
\end{figure}

The above findings for the 3/4 trimer are consistent with our results
for the 3/3 case AsH$_2$-AsH$_2$-AsH$_2$, which is of interest in its
own right, see Fig.~\ref{33BO} and further discussion in the
Supplementary Material.  Because the $pp\sigma$ bond in the 3/3 trimer
is one electron short, the molecule is only {\em metastable} in the
symmetric geometry. Nevertheless, the molecule near its metastable
minimum exhibits the bond lengths and LMO characteristics similar to
those of the 3/4 trimer near its ground state. We thus observe that
bringing the nuclei closer together induces sufficient amount of
electron transfer out of the lone pair on the central arsenic so as to
fill the $pp\sigma$ bond. This is witnessed by the formation of two
two-center LMOs.

\begin{figure}[t]
  \centering
  \includegraphics[scale=0.30]{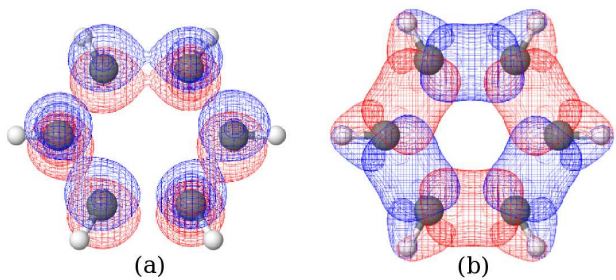}
  \caption{\label{benzeneLMO} The localized molecular orbitals for the
    ground state of benzene.}
\end{figure}

In both 3/4 and 3/3 case, we observe that a stable (or metastable)
three center bond is signaled by the presence of two {\em two}-center
LMOs. One may formulate this notion in the form of the following,
tentative rule: {\em In a stable (or metastable) molecule, the bonding
  LMOs should be two-center and cover each nearest neighbor bond.}
Conversely, when the molecule is not fully covered by {\em two}-center
LMOs, the molecule is subject to a structural instability. One may
benefit from further illustration of these notions by considering the
familiar example of the benzene molecule, see
Fig.~\ref{benzeneLMO}. Here we observe that of the 18 available
valence electrons, 12 fill the $\sigma$ bonded network, leading to the
formation of six two-center LMOs. The remaining six electrons amount
to three non-two-center LMOs, which by itself would imply an
instability. The instability is second order Jahn-Teller, and a
finite-size analog of the Peierls instability.~\cite{Bersuker2006,
  ABW} The benzene molecule can be thought of as a superposition of
the bonding situations on the opposite sides of the transition shown
in Fig.~\ref{43BcontExp}.  The amount of instability stemming from the
non-bonding LMOs is however insufficient to break the spatial symmetry
favored by the bonding LMOs, consistent with more sophisticated
analyses;~\cite{ABW} benzene thus retains the six-fold, not three-fold
symmetry in its ground state.

One may ask how extended an LMO could be.  The most extended LMO we
have succeeded in generating spans five centers, see the Supplementary
Material. Consistent with the coverage rule above, such multi-center
LMOs would appear at densities where the molecule is unstable.

% We finish the Section by briefly discussing the low-density side of
% the transition in the trimer. Formally, the delocalization of the
% LMO---so as to span nearly three centers---can be traced to the
% relatively small amount of charge contained within the interatomic
% space ???

%\section{Bonding in Extended Systems as a Symmetry Breaking}
%\label{localization}

The above ideas can be profitably applied to extended
systems. Arguably the simplest example of such an extended system is a
one-dimensional chain of equivalent orbitals, at half-filling. Already
at the H\"uckel level, this model exhibits a rich behavior, if the
bonds are allowed to deform, subject to a restoring force from the
lattice. Using the creation (annihilation) operator $c_{n,
  s}^\ddagger$ ($c_{n, s}$) for an electron on site $n$ with spin $s$,
the energy function can be written as:
\begin{eqnarray} \cH = \sum_n \sum_{s=\pm1/2} & \left[t(x_{n},
    x_{n+1}) \, (c^\dagger_{n,s} c_{n+1,s} + c^\dagger_{n+1,s}
    c_{n,s}) \right. \nonumber \\ +& \left. (-1)^n \epsilon \,
    c_{n,s}^\dagger c_{n,s} \right] + \cH_\text{lattice} (\{x_n\}),
  \label{SSH_E}
\end{eqnarray}
where $t(x_{n}, x_{n+1})$ is the hopping matrix element between sites
$n$ and $n+1$, $(-1)^n \epsilon$ on-site energy, and
$\cH_\text{lattice}$ accounts for the elastic response of the lattice
and the kinetic energy of the nuclei. In the lowest order expansion in
bond deformation, $t(x_{n}, x_{n+1}) = t^{(0)} - \alpha (x_{n+1} -
x_{n} - a)$, $\cH_\text{lattice} = \sum_{n} k (x_{n+1} - x_n - a)^2/2
+ M\dot{x_n}^2/2$. In the absence of electronegativity variation,
$\epsilon = 0$, Eq.~(\ref{SSH_E}) gives the venerable
Su-Schrieffer-Heeger Hamiltonian for
trans-polyacetylene.~\cite{RevModPhys.60.781} A non-zero $\epsilon \ne
0$ was introduced by Rice and Mele to study heteropolar
polymers.~\cite{PhysRevLett.49.1455, PhysRevLett.45.926}

At half-filling and in the absence of electronegativity variation,
$\epsilon=0$, the system (\ref{SSH_E}) held at uniform spacing between
nearest sites is a metal. It is, however, Peierls-unstable with
respect to dimerization,~\cite{Peierls, RevModPhys.60.781, ABW}
$x_{n+1} - x_{n} = \pm (-1)^n \text{const}$, upon which it becomes an
insulator. The gap is approximately proportional to the differential
$\Delta t \equiv |t(x_{n}, x_{n+1}) - t(x_{n}, x_{n-1})|$ in the
hopping matrix element between the stronger and weaker bond; the gap
is caused by scattering of the electrons near the wavelength
corresponding to the unit cell of the distorted lattice, which now
contains two lattice sites.  In contrast, when the electronegativity
variance is sufficiently large, the dimerization is suppressed while
the material can be thought of as an ionic insulator; the gap is
largely due to the work needed to transfer an electron from the more
electronegative atom to its less electronegative counterpart, as is
the case in sodium chloride for instance. As a rule of thumb,
Peierls-dimerization will be suppressed when the electronegativity
variation is comparable to $\Delta t$ the system would exhibit when
$\epsilon = 0$.~\cite{PhysRevLett.49.1455, PhysRevLett.45.926} For a
chain of passivated arsenics, this is numerically close to 1-2
eV. Whether the chain from Eq.~(\ref{SSH_E}) becomes a Peierls or
ionic insulator, it develops a charge density wave
(CDW).~\cite{CanadellWhangbo1991} The presence of dimerization implies
that the CDW has an off-site component, often called bond-order wave,
while non-vanishing electronegativity variation implies the CDW has an
on-site component.  Thus at sufficiently low mass density and
electronegativity, the one-dimensional chain from Eq.~(\ref{SSH_E})
will be a Peierls insulator.
% Interaction effects may also lead to the emergence of the insulating
% state.\cite{SS53, NT55, CH78} \verb+check this+

To set the stage for the discussion of bonding in extended
one-dimensional systems we note that the 3/4 trimer is unique in that
it is the only case, in which the $n$-center, $(n+1)$ electron
bond---corresponding to half-filling---is actually stable.  This can
be understood by using the two-center LMO coverage rule formulated
above. By that rule, an $n$-center, $(n+1)$ electron bond is stable if
there are enough electrons to fill $(n-1)$ LMOs, i.e., $n+1 = 2
(n-1)$. (This is assuming there is no electron transfer from lone
pairs or there are no additional sub-bands that could provide
stability as in the benzene example above.) The equation is solved
only by $n=3$, and so the multi-center bond becomes electron deficient
for any molecule longer than $3$ centers. In the $n \to \infty$ limit,
one obtains one electron per bond; a $($AsH$_2)_n$ chain will
dimerize, if let go,~\cite{ZLMicro1} owing to the aforementioned
Peierls instability.~\cite{Peierls, RevModPhys.60.781, ABW, ZLMicro1,
  ZLMicro2}

\begin{figure}[t]
  \centering
  \includegraphics[width =
  \figurewidth]{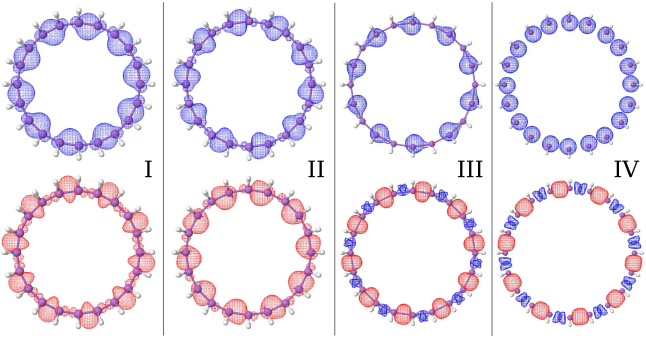}
  \caption{ \label{ring20highDens} The localized molecular orbitals on
    the (AsH$_{2}$)$_{20}$ ring molecule shown for clarity as two sets
    of equivalent orbitals. The two colors correspond with the
    opposite signs. The ranges for each of the four regimes are shown
    in Fig.~\ref{CDW}.  The LMOs are shown for specific values of the
    bond length $b$: (I) $b=1.81$~\AA, (II) $b=2.06$~\AA, (III)
    $b=2.58$~\AA, and (IV) $b=3.09$~\AA. The specific value
    $b=2.58$~\AA~for regime III was chosen because at yet greater
    values of $b$, the ring becomes unstable with respect to
    dimerization. }
\end{figure}

We now consider a chain of passivated arsenic atoms with uniform
spacing between nearest neighbors. To prevent symmetry-lowering due to
open ends we consider a closed ring, viz., $($AsH$_2)_{20}$. The ring
length 20 is sufficiently large so that the effects of $sp$-mixing due
to the curvature of the $pp\sigma$ network are modest.  The
passivating hydrogens are fixed so that the molecule has $D_{20h}$
symmetry. Within the studied range of densities, we find four distinct
bonding regimes which we label by roman numerals; representative LMOs
are shown in Fig.~\ref{ring20highDens}.

\begin{figure}[t]
  \centering
  \includegraphics[width =
  \figurewidth]{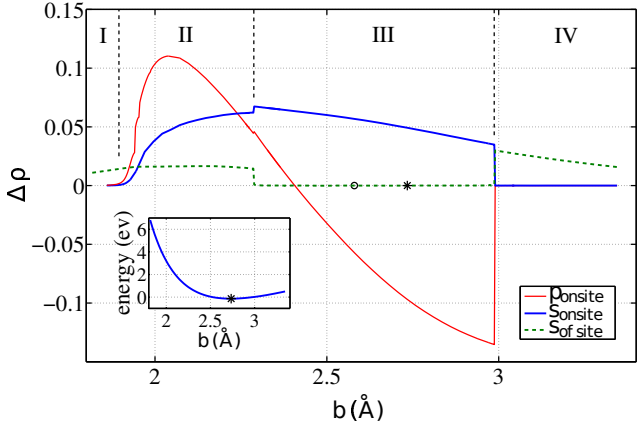}
  \caption{\label{CDW} Dependence of the strength of the
    charge-density wave in the AsH$_{2}$)$_{20}$ ring on the bond
    length $b$. The $s$- and $p$-contributions to the on-site
    component are shown as solid lines, while the $s$-component of the
    off-site, bond order wave with the dashed line. Minimum energy
    configuration is marked with the asterisk; the bond length
    associated with the onset of dimerization, $b=2.58$~\AA, with the
    circle. }
\end{figure}

Unlike in the trimer case, the LMO transitions for the 20-member ring
are entirely due to symmetry breaking already at the level of the
canonical molecular orbitals, as we demonstrate in Fig.~\ref{CDW}.
Visualizing changes in the molecular orbitals is difficult because of
their large number.  Here we take advantage of the fact that in the
full density range in question, the electrons form a charge-density
wave (CDW) commensurate with the periodicity of the chain. Indeed,
consistent with the expectation that a 1D metal at half-filling is
Peierls-unstable toward dimerization, the charge distribution exhibits
a periodic pattern whose periodicity is twice that exhibited by the
chain. Because the chain itself is fixed at uniform spacing between
nearest neighbors, the symmetry breaking is entirely due to
electron-electron repulsion.  The symmetry lowering manifests itself
via the formation of either on-site or off-site CDW, or both.  To
quantify the strength of the on-site charge density wave, we compute
the variation in the diagonal entry of the density matrix:
\begin{equation} \Delta \rho^\text{(on)} = \frac{1}{2} \sum
  (P_{\lambda_{2n+1}, \lambda_{2n+1}} - P_{\lambda_{2n},
    \lambda_{2n}}),
\end{equation}
where the summation is over a subset of orbitals of interest on any
even-numbered and odd-numbered center, $2n$ and $(2n+1)$,
respectively.  The $s$ and $p$ contributions to the CDW strength are
shown with the red and blue solid lines in Fig.~\ref{CDW}. To quantify
the off-site, bond-order wave, we limit ourselves to the $s$-orbitals,
to avoid ambiguity stemming from arbitrary mutual signs of the
respective contributions of the $p$ orbitals to the off-diagonal
elements of the density matrix. Thus we compute the variation in the
inter-atomic charge density of the $s$-electrons according to
\begin{equation} \Delta \rho^\text{(off)} = \frac{1}{2} (P_{2n+1,
    2n}^{(s)} - P_{2n, 2n-1}^{(s)}),
\end{equation}
where the superscript $(s)$ signifies that only the $s$-orbital
contribution to the density matrix is used. The quantity $\Delta
\rho^\text{(off)}$ is shown in Fig.~\ref{CDW} with the dashed line.

The ground state of the system---subject to the aforementioned
geometric constraints---falls in regime III, where only the on-site
CDW is present. Of the three transitions between states with distinct
CDW states, one is continuous (I~$\leftrightarrow$~II) and the rest
are discontinuous. Consistent with the preceding Section, the
localized-molecular orbitals represent a sensitive indicator of charge
redistribution: The CDW changes gradually between regimes I and II;
the symmetry of the off-site CDW does not change, while the on-site
component begins to gradually develop a pattern with a lower, ten-fold
symmetry on the r.h.s. of the transition. In contrast, the LMOs show a
pronounced symmetry breaking.  On the lower density, larger $b$ side
of the transition, the LMOs become chiral. To avoid any possible
confusion we point out that the {\em canonical} MOs cannot and do not
experience such a drastic symmetry breaking, given the geometric
constraints. And so for every chiral MO, if any, there is a
corresponding MO at the same energy but opposite handedness. Note that
there appear to be a few, continuous ``microtransitions'' within
regime II, however these do not modify the shapes of the LMOs
qualitatively.

The presence of the charge density wave, Fig.~\ref{CDW}, automatically
implies that the Hartree-Fock ground state of the molecule is doubly
degenerate, since rotation of the molecule by 360/20=18 degree results
in a distinct yet equivalent electronic configuration. The presence of
such a degeneracy is a consequence of our using a single-determinant
wave-function. A better approximation for the actual ground state
would be a superposition of those two distinct solutions. In physical
terms, the presence of the degeneracy is a sign of the Peierls
instability, as already mentioned. If equal spacing between nearest
neighbors were not enforced, the molecule would readily convert into a
ring made of ten equivalent dimers; there are two equivalent ways to
dimerize.  Incidentally, we note that unlike in the trimer case study
in Section~\ref{trimer}, convergence of the self-consistent
Hartree-Fock (SCHF) procedure for the 20-member ring was less
robust. Decreasing the tolerance from the default value of
$10^{-4}$~kcal/mol to a smaller value of $10^{-6}$~kcal/mol was
helpful in finding lower energy solutions. Still, we cannot be certain
that the solutions shown in Figs.~\ref{ring20highDens} and \ref{CDW}
correspond to the lowest available Hartree-Fock energy.

As in the trimer case, we observe that at sufficiently high densities,
the LMOs are two-center and cover every bond, see
Fig.~\ref{ring20highDens}. Likewise, the coverage becomes less
complete for lower densities. Most significantly, one set of the LMOs
becomes largely lone pairs by the time the ring becomes unstable
toward dimerization, thus supporting the two-center coverage rule
formulated above. For the reader's reference, we provide in
Fig.~S24 of the Supplemental Material, the values for the bonding contribution, Wiberg bond
number, and the center number for the four configurations in
Fig.~\ref{ring20highDens}. We anticipate that to develop an automated
tool to quantify the localization of bonding electrons, one may have
to use a properly weighted combination of several bond
characteristics.  Two more examples of LMOs are provided in the
Supplementary Material.

% The relatively extended system in the form of a 20-member ring helps
% bring home the notion that distinct values of bond order correspond
% with different patterns of charge density waves. The CDW can
% establish even if the nuclei are constrained to be in a very
% symmetric configuration. Already the electron-electron interaction
% suffices for forming such a CDW. Transitions between distinct CDW
% patterns were anticipated by Kohn for macroscopic solids with nuclei
% fixed in relatively symmetric
% configurations;~\cite{PhysRevLett.19.789, PhysRevLett.19.439} he
% specifically pointed out that such transitions signify a structural
% instability and likely lead to displacive transitions.  The present
% results provide a specific realization for the microscopic picture
% advanced by that author.

\section{Hierarchy of Chemical Interactions: The Solid State}
\label{solids}

\begin{figure}[t]
  \centering
  \includegraphics[width = .9 \figurewidth]{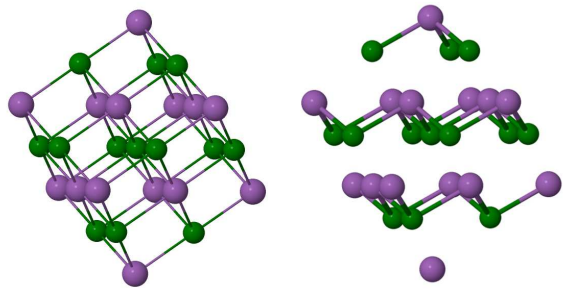}
  \caption{\label{bulkAs} The structure of rhombohedral arsenic at
    normal conditions (r.h.s.) can be thought of as two superimposed
    face-centered cubic lattices, shown in purple and green, shifted
    relative to each other along the [111] direction. At sufficiently
    high pressures, the magnitude of the shift vanishes thus yielding
    the simple-cubic lattice (l.h.s.).}
\end{figure}

\begin{figure}[t]
  \centering
  \includegraphics[width = \figurewidth]{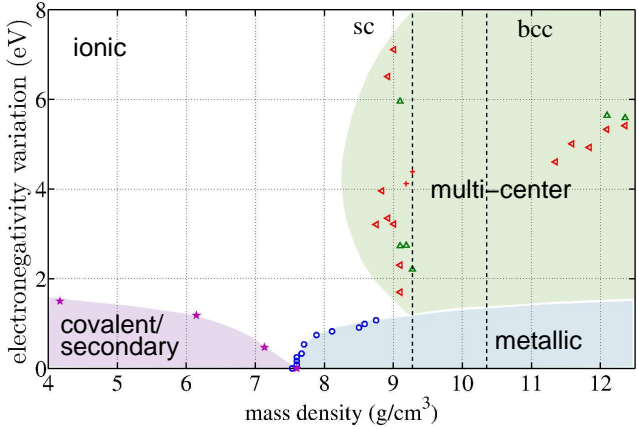}
  \caption{\label{phase_diag} Phase diagram of chemical
    interactions. The abscissa corresponds with the mass density,
    assuming each atom has the same mass as arsenic. The ordinate
    gives the absolute value of the electronegativity difference
    $2\epsilon$ between pseudo-arsenic and pseudo-antimony atoms; the
    two artificial elements form either the simple-cubic (sc) or
    body-centered cubic (bcc) structure, by construction; the dashed
    lines delineate the coexistence of the two phases. The left-bottom
    corner is occupied by a relatively low-symmetry structure, such as
    rhombohedral arsenic, in which the bonding is spatially
    non-uniform. The $\epsilon=0$ end of its boundary matches that for
    the transition between rhombohedral and simple-cubic arsenic. See
    text for explanation of the symbols.}
\end{figure}

To extend the preceding notions to 3D materials we first note that the
propensity of systems exhibiting substantial electronegativity
variation to be insulators is not specific to a particular
coordination pattern and is generic to any number of spatial
dimensions; it is characteristic of many oxides and halides, for
instance. In contrast, the Peierls instability is specific to
one-dimensional systems, except in some idealized situations, such as
when $sp$ mixing and $pp\pi$ interactions can be
neglected,~\cite{burdett5764, burdett5774, BurdettLeePeierls} see also
Refs.~\cite{PhysRevB.76.052301, SeoHoffmann1999, ZLMicro1} Here we
take a phenomenological approach and simply use the experimental fact
that rhombohedral arsenic becomes simple-cubic given sufficient
pressure,~\cite{PhysRevB.77.024109, silas:174101} see
Fig.~\ref{bulkAs}. Alongside, the material undergoes an
insulator-to-metal transition. (The self-consistent emergence of
uniform, metallic bonding at high densities is consonant with the
seminal ideas of Wigner~\cite{TF9383400678} and
Mott,~\cite{0370-1298-62-7-303, Mott1990} of course.)  Regardless of
the precise mechanism of this electronic and mechanical instability,
the formal reason why materials exhibiting bond-order CDW are
insulating is a lack of electronic function overlap and, hence, lack
of bonding within a significant subset of inter-atomic spaces. The
latter subset houses the troughs of the off-site charge density wave,
analogously to the Peierls-dimerized chain.  Thus we identify the low
mass density, low $\epsilon$ corner on our phase diagram of chemical
interactions, Fig.~\ref{phase_diag}, with an insulating phase
characterized by a bond-order charge-density wave, in which both
covalent and secondary interactions are generally present.

\begin{figure}[t]
  \centering
  \includegraphics[width = \figurewidth]{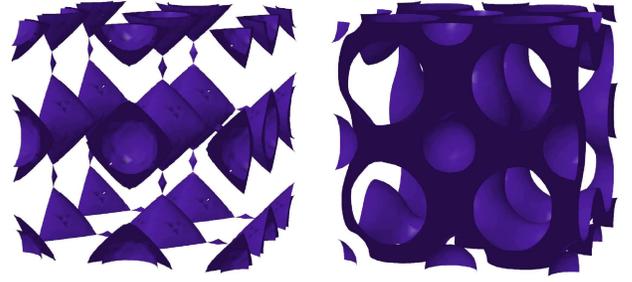}
  \caption{{\em Left}: The ELF=1/2 surface at the onset of
    metallicity.  {\em Right}:~The ELF=1/2 surface in the metallic
    regime. In both panels, the metallic and ionic regions are shown
    as filled and empty region, respectively. The cavities correspond
    to the ionic cores.\label{cubic_ELF}}
\end{figure}

To estimate the location of the boundary, we assess the uniformity of
bonding by computing the so called electron localization function
(ELF).\cite{Silvi1994} The ELF is determined by the curvature of the
spherically averaged conditional pair probability for the electrons.
By construction, the ELF varies between zero and
one,\cite{doi:10.1021/jp992784c} the two extremes corresponding to
perfect localization and a uniform electron gas respectively. Surfaces
of constant ELF=1/2 may be thought of as separating metallic and
insulating regions.\cite{doi:10.1021/jp992784c} The metallic regime
sets in when the valence electrons percolate the space, see
Fig.~\ref{cubic_ELF}; conversely, there cannot be uniform bonding in
the absence of the percolation. We have obtained the electronic
wavefunctions for simple-cubic arsenic using the plane wave
pseudo-potential method (PSPW) method in NWChem~\cite{Valiev20101477}
and PBE96 exchange-correlation potential.  A 64 atom repeat unit was
used.  We find that the percolation sets in very close to the
experimentally known value~\cite{PhysRevB.77.024109, silas:174101} of
the mass density at which elemental arsenic undergoes the
rhombohedral-to-simple cubic transition. This is consistent with the
expectation that the lack of connectivity in the valence electronic
wavefunction will lead to a structural instability and, ultimately,
regions of weak, secondary bonding. This is in full agreement with the
results of Section~\ref{localization}.

To estimate the mass density when such percolation takes place in the
presence of electronegativity variation, we employ rock salt
structures for the di-pnictogen compounds SbAs, BiAs, and BiP.  The
compounds were chosen to cover a broad range of electronegativity
variation; note that AsP, BiAs and SbAs solids have been reported
experimentally.\cite{doi:10.1021/cm960606f} Since the phase diagram is constructed with
two arsenics in mind, we must readjust our results to account for the
difference in the atomic size and mass of the diatomic compounds from
elemental arsenic. In view of our results for mixed-pnictogen trimers
from the preceding Section, we used the ionic radii as the rescaling
factor for the length.  The so obtained points for the emergence of
structural instability are shown with stars in
Fig.~\ref{phase_diag}. The points happen to fall onto a smooth line.
Since to the left of the line, the density is too low to provide for
bonding, the line gives an approximate bound on the mass density below
which uniform bonding could not take place.  Thus on the low density,
low-$\epsilon$ side, the structure is characterized by a co-existence
of (two-center) covalent and secondary bonding, as in the structure of
rhombohedral arsenic, Fig.~\ref{bulkAs}, or As$_2$S$_3$,
Fig.~\ref{coex}(b).  At the boundary and beyond, the bonding is
expected to be more uniform. For instance, on the low $\epsilon$ end,
it could be simple cubic, as in elemental arsenic or phosphorus. On
the low density end, given large electronegativity variation, and for
1:1 stoichiometry, one could have the rock salt or CsCl structure,
depending on the ion size ratio. Structures with lower
point-symmetries can be envisioned, depending on the stoichiometry.

To extend the phase diagram in Fig.~\ref{phase_diag} to broader ranges
in density and electronegativity variation, we will employ the
following, unabashedly artificial device, as afforded by MOPAC:
Imagine the rock salt structure, in which the anionic sites are
occupied by arsenic, and cationic sites by antimony. Now increase the
on-site energies for arsenic by $\epsilon$ upward so that the new
value is $E'_\text{As} = E_\text{As} + \epsilon$ for each orbital. At
the same time, move all the antimony orbitals in energy so that they
are $\epsilon$ below the unperturbed {\em arsenic} orbitals:
$E'_\text{Sb} = E_\text{As} - \epsilon$. Thus we impose an
electronegativity variation entirely analogously to how it is done in
the model Hamiltonian (\ref{SSH_E}).
% The electronegativity is, of course, not a continuous variable, nor
% can it be varied experimentally---or using a mainstream quantum
% chemistry package---without changing other parameters such as the
% atomic size and the number of valence electrons.
Next, adjust the orbital exponents on the pseudo-antimony atoms so
that the spatial dependences of Sb-As and Sb-Sb inter-orbital matrix
elements fit maximally those of the corresponding As-As matrix
elements.  The resulting fit is good, but not perfect (see the
Supplementary Material), because of the distinct angular dependences
of the (valence) orbitals on As and Sb, which correspond to the
principal quantum number $n=4$ and $n=5$ respectively.  Lastly, make
the Sb-Sb, Sb-As, and As-As ionic core repulsions identical while
re-parametrizing them so that the density at which our hypothetical
rock salt structure exhibits two transitions: the
rhombohedral-to-simple cubic (rh-to-sc) and sc-to-bcc---both match
their experimental values for elemental arsenic at $\epsilon=0$. The
need for this re-parametrization is easy to understand since MOPAC's
default parameters are optimized for conventional pressures but do not
necessarily account for the deformation of the ionic core at high
pressures; further detail can be found in the Supporting Material. We
reiterate that even at $\epsilon = 0$, our pseudo arsenic and antimony
are not strictly equivalent, because of the aforementioned difference
in the angular dependence of the valence wavefunctions.

Now, to move vertically on the phase diagram one varies the quantity
$\epsilon$; the quantity $2\epsilon$ is thus the externally imposed
portion of the electronegativity differential between the cationic and
anionic sites. The latter is the ordinate in Fig.~\ref{phase_diag} by
construction.  The horizontal axis corresponds with the mass density;
the particle mass is set equal to that of arsenic for concreteness.

From here on, we limit ourselves to the area {\em outside} the
covalent-secondary sector, the latter shown in lilac. One should
expect a variety of structural transitions involving coordination
changes as one moves about on the diagram. We will limit ourselves to
just one such transition, viz., 'simple cubic'-to-'body centered
cubic' (sc-to-bcc), which is known to occur in elemental
arsenic.\cite{PhysRevB.77.024109} The latter transition is convenient
for modeling pressure-induced coordination changes in that all bonds
in the nearest coordination shell are equivalent in both structures,
while the coordination itself changes meaningfully during the
transition, viz., between 6 and 8. Away from the transition, we fix
the structure to be rock-salt and cesium-chloride on the low and high
density sides of the phase boundary, respectively; we vary only the
density, not structure within individual phases. The
(zero-temperature) transition is detected in the standard fashion, by
computing the formation enthalpies of the two phases, as functions of
volume, and finding the common-tangent to the two curves,~\cite{L_AP}
see the Supplementary Material. This procedure amounts to mutually
equating the pressures and chemical potentials of the two phases at
zero temperature, aside from any error due to possible differences in
the zero-point vibrational energies of the two structures. Because the
transition is discontinuous in volume, there are actually a pair of
phase boundaries, see Fig.~\ref{phase_diag}, the two enclosing the
phase-coexistence region; the latter is known to exhibit
incommensurate host-guest structures.~\cite{PhysRevB.77.024109} We
observe that the transition density is not sensitive to the value of
the electronegativity variation. This lack of sensitivity likely
indicates a limitation of our pseudo-pnictogen construct. Indeed, an
ionic crystal with stoichiometry 1:1 and comparable sizes of the
cation and anion is expected to have the CsCl structure at ordinary
densities. The rock salt structure would require negative pressure or
sufficient mismatch in the ion sizes.

Nonwithstanding its limitations, our artificial procedure allows one
to monitor, even if qualitatively, charge redistribution in response
to changes in density, electronegativity variation, and coordination.
To quantify the spatial distribution of the frontier electrons, we
directly compute the (average) excess charge $\Delta \rho$ on the more
electronegative element relative to the less electronegative
element. The amount of charge on an atom is determined by adding the
diagonal entries of the density matrix pertaining to the $s$, $p$, and
$d$ orbital on that atom. The charge differential $\Delta \rho$ gives
the strength of the on-site charge density wave.  Near the horizontal
axis on the diagram, $\epsilon=0$, $\Delta \rho$ is small, which we
associate with metallic bonding in the absence of a bond-order
wave. Conversely, at large values of $\epsilon$, the quantity $\Delta
\rho$ is large, implying ionic bonding. Because the quantity
$\epsilon$ couples linearly to the electron density, as in
Eq.~(\ref{SSH_E}), the following statement holds:
\begin{equation} \Delta \rho = - \frac{\prtl E}{\prtl \epsilon},
\end{equation}
where $E$ is the exact energy of the system per atom. (This can be
shown explicitly using Eq.~(11.16) of Ref.~\cite{LLquantum}.)
% (Note the bonding is relatively isotropic in both cases.)

\begin{figure}[t]
  \centering
  \includegraphics[width = \figurewidth]{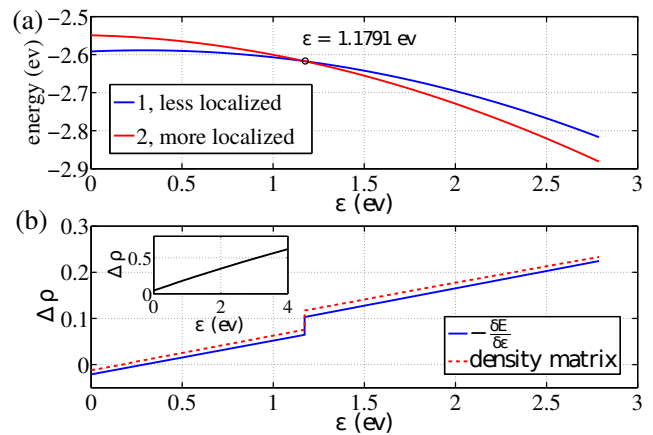}
  \caption{\label{Eeps} {\bf (a)} A specific instance of term-crossing
    for the ground state of the pseudo-pnictogen compound from
    Fig.~\ref{phase_diag}. {\bf (b)} The corresponding dependence of
    the electron density differential $\Delta \rho$ as determined
    directly using MOPAC-computed density matrix and by numerically
    differentiating the $E(\epsilon)$ curves from panel (a). The
    differentiation was performed using the best parabolic fit to the
    $E(\epsilon)$ curves. Although numerically similar, the two curves
    are not identical because MOPAC does not solve the Schr\"odinger
    equation exactly.}
\end{figure}

We next prepare our pseudo-pnictogen sample at distinct values of the
density and electronegativity variation, by performing $\epsilon$
sweeps from $2 \epsilon = 0$ to $8$~eV, while keeping the density
fixed.  The sample size is 64 atoms for the sc structure, 54 atoms for
the bcc structure; the finite size effects are treated according to a
standard procedure.~\cite{PerkinsStewart1980} We observe that below a
certain threshold density, the energy and wavefunction of the system
depend smoothly on the value of epsilon.  Above that threshold
density, the dependence becomes more complicated. The sample can now
transition between distinct energy terms, as in Fig.~\ref{Eeps}(a).
Each term crossing is accompanied by a discrete change in charge
distribution, see Fig.~\ref{Eeps}(b). Because of such crossings, the
wavefunction and energy show a hysteretic behavior during $\epsilon$
sweeps (see the Supplementary Material). The locations of the
corresponding term-crossings can be detected; some are indicated on
the phase diagram in Fig.~\ref{phase_diag} with the triangles.

These transitions were anticipated by Kohn a while
ago~\cite{PhysRevLett.19.789, PhysRevLett.19.439} in the context of
the metal-insulator transition and are generally expected for solids
exhibiting conduction and valence bands with distinct anisotropies.
In the picture advanced by Kohn, the canonical metal and insulating
states are separated by a chain of consecutive transitions that
mutually stabilize distinct sets of electrons and holes.  Here we
observe a noteworthy situation that term crossings (at the ground
state energy) begin to occur {\em only} above a certain value of mass
density.
% Below that threshold density, there is no sharp distinction between
% states characterized by relatively low and high local polarization,
% the latter reflected in the value of the charge differential $\Delta
% \rho$.
Two states corresponding to two different energy terms may be regarded
as distinct {\em phases} corresponding to distinct degrees of electron
localization. This is brought home by introducing the Legendre
transform of the energy, $E^{\text{(inh)}} (\Delta \rho)$, which can
be treated naturally as a function of $\Delta \rho$:
\begin{equation} \label{Helm} E^{\text{(inh)}}(\Delta \rho) =
  E[\epsilon(\Delta \rho)] + \epsilon(\Delta \rho) \, \Delta \rho,
\end{equation}
where
\begin{equation} \label{epsrho} \epsilon(\Delta \rho) = \frac{\prtl
    E^{\text{(inh)}}(\Delta \rho)}{\prtl \Delta \rho}.
\end{equation}
The label ``inh'' signifies that the thermodynamic potential
$E^{\text{(inh)}}(\Delta \rho)$ is inherently a functional of the
electronic density distribution.  The existence of such a functional
is guaranteed by the Hohenberg-Kohn theorem.~\cite{PhysRev.136.B864}
In such a density-based approach,~\cite{L_AP} one can unambiguously
determine the field necessary to achieve a specific strength of the
CDW if desired, via Eq.~(\ref{epsrho}). The thermodynamic potentials
$E^{\text{(inh)}}(\Delta \rho)$ and $E(\epsilon)$ are analogous to the
Helmholtz and grand-canonical free energy, respectively.~\cite{CL_LG}

\begin{figure}[t]
  \centering
  \includegraphics[width = \figurewidth]{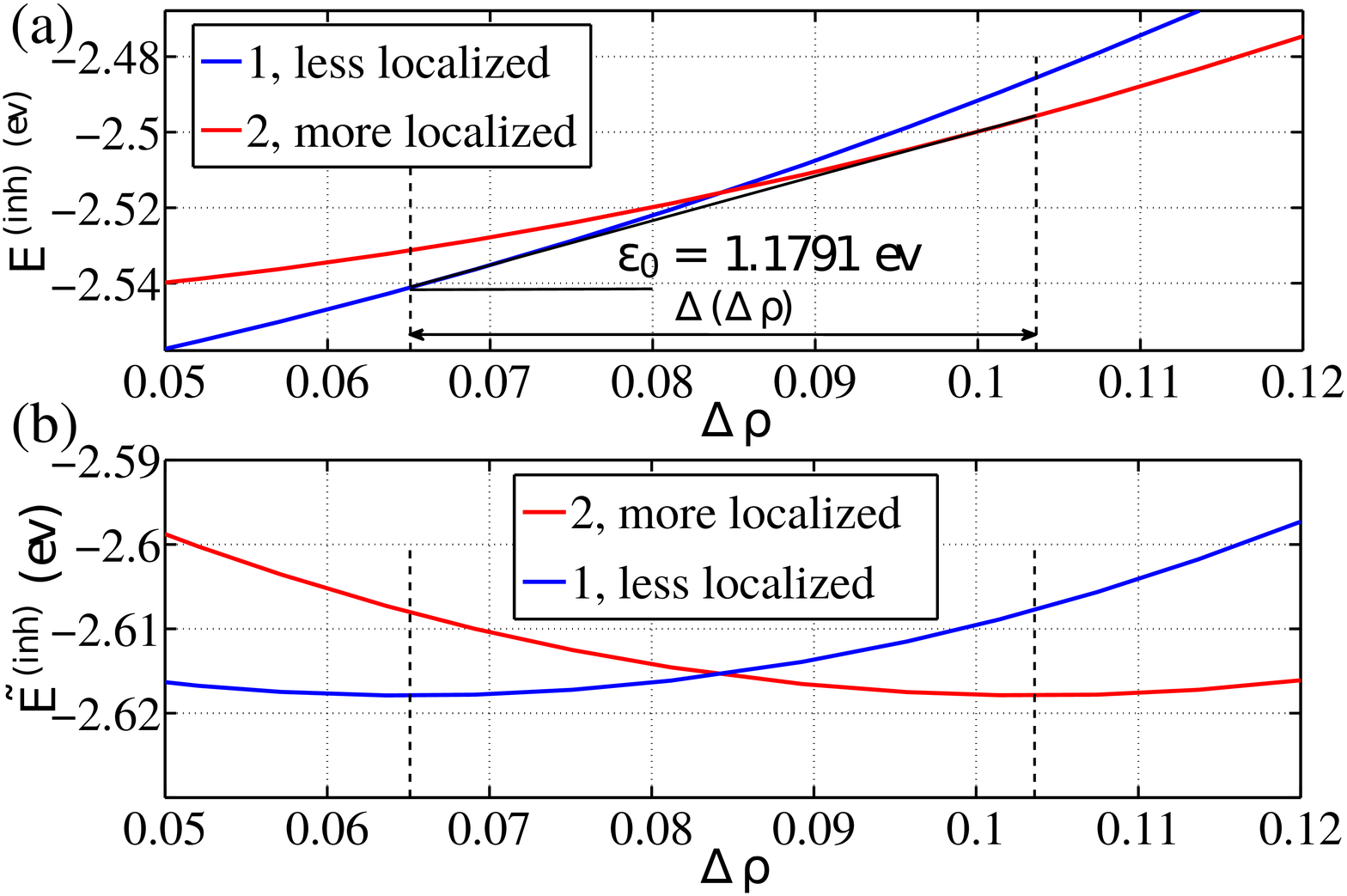}
  \caption{\label{Erho} {\bf (a)} The inherent electronic energy from
    Eq.~(\ref{Helm}) as a function of the electron density
    differential $\Delta \rho$, corresponding to Fig.~\ref{Eeps}. {\bf
      (b)} The modified energy $\widetilde{E}^\text{(int)}(\Delta
    \rho) \equiv E^\text{(int)}(\Delta \rho) - \epsilon_0 \, \Delta
    \rho$ whose minima correspond to the equilibrium values of $\Delta
    \rho$ when the externally imposed electronegativity variation is
    equal to $\epsilon_0$. The locations of the minima are indicated
    with the dashed lines.}
\end{figure}

Fig.~\ref{Erho}(a) displays the term crossing from Fig.~\ref{Eeps} in
terms of the inherent energy $E^\text{(inh)}(\Delta \rho)$, at mass
density 9.01 g/cm$^3$. The latter figure explicitly illustrates a
discontinuous transition between two states, where the quantity
$\epsilon_0$ gives the slope of the common tangent to the two terms.
Particularly illuminating is the graph of the quantity
$\widetilde{E}^\text{(inh)}(\Delta \rho) \equiv E^\text{(inh)}(\Delta
\rho) - \epsilon_0 \, \Delta \rho$, where $\epsilon_0$ is a constant,
not a function of $\Delta \rho$; see Fig.~\ref{Erho}(b).  When the
conditions for equilibrium for any pair of states are met, the two
minima of $\widetilde{E}^\text{(inh)}(\Delta \rho)$ corresponding to
the states have the same depth.  Note that the term crossing in
Fig.~\ref{Erho} is only possible because there are more than one
orbital per site. Otherwise, the variable $\Delta \rho$ would specify
the charge distribution uniquely, which would then imply that to the
same density distribution, there correspond more than one energy
functional, a physical impossibility.~\cite{PhysRev.136.B864} We have
checked that the individual contents of the $s$, $p$, and $d$ orbitals
corresponding to the two terms in Fig.~\ref{Erho} do exhibit
discontinuities at the term crossing.

The l.h.s. and r.h.s.  minima in Fig.~\ref{Erho}(b) correspond to
states with a lesser and greater degree of electron localization. The
presence of a barrier separating those states can be traced to a very
familiar phenomenon, viz., the lack of mutual miscibility of oil and
water. Indeed, states characterized by distinct degrees of electron
localization will also exhibit distinct polarizabilities, hence the
analogy. The hysteretic region on the phase diagram formally
corresponds to a {\em coexistence} of two phases. The coexistence
region exhibits an electronic pattern that interpolates between those
typical of metallic and ionic bonding, the two formally corresponding
to maximally extended and localized bonding orbitals, respectively.
Thus we identify the ionic/metallic coexistence region with the
intermediate case of the {\em multi-center} bond, see also the
discussion in Section~\ref{discussion}.

We emphasize that although we have treated local electronegativity as
a continuous variable, in practice the choice of $\epsilon$ is limited
to that afforded by specific chemical elements. The presence of
discontinuous transitions of the kind shown in Fig.~\ref{Erho}
indicates an additional complication for the materials chemist: Even
if realizable chemically in principle, a specific value of
electronegativity variation may not be achievable in an actual
compound, if the resulting charge density wave with a spatially
uniform strength is unstable toward phase separation. In the best case
scenario, the compound will be a collection of stripes whose CDW
strength alternates between $\Delta \rho_1$ and $\Delta \rho_2$. In
the worst case, the compound will not form altogether. Conversely, no
such complications arise at sufficiently low mass densities.

The present calculations thus indicate that depending on the density,
a transition between states with distinct magnitudes of the CDW can be
generally either continuous or discontinuous.  This means, in
particular, that the metal-insulator transition can be either
continuous or discontinuous. (Note we are not considering effects of
disorder.~\cite{PhysRevLett.42.673, Mott1993}) The possibility of a
continuous localization transition in electronic systems is in
contrast with classical liquids made of rigid particles, which always
solidify discontinuously.~\cite{LandauPT1, Brazovskii1975, dens_F1,
  L_AP} These contrasting behaviors may come about for the following
reason: Rigid particles create a bounding potential for each other in
a self-consistent fashion; there is no other source of particle
confinement. In contrast, electrons in solids are subject to a field
due to largely stationary nuclei. This field already lowers the full
translational symmetry that would have been present in a weakly
interacting electron gas, the precise extent of symmetry-breaking
determining whether electric conductance vanishes.  Incidentally, note
that the transition in Figs.~\ref{Eeps} and \ref{Erho} is only weakly
discontinuous; its analog in an actual material may not be easily
distinguished from a continuous transition or a soft crossover.

As a consistency check, we have tested the stability of our
pseudo-pnictogen rock salt structure with respect to deformation
toward the rhombohedral structure, Fig.~\ref{bulkAs}, see details in
the Supplementary Material. This instability line, shown with circles
in Fig.~\ref{phase_diag}, is seen to lie under the ionic-metallic
coexistence region.  On the one hand, this implies the propensity for
electron localization (or lack thereof) in symmetric structures is
deeply connected to structural instability, as was
anticipated by Kohn.~\cite{PhysRevLett.19.789, PhysRevLett.19.439} On
the other hand, the instability implies that in a proper treatment,
the CDW pattern and lattice must be determined self-consistently.

In parallel with Section~\ref{trimer}, we now address the interplay of
cohesive and repulsive interactions during density-driven transitions
accompanied by coordination changes. The strengths of the two
interactions often move in the same direction: Both the electron
density in the inter-atomic space and the repulsion between the ionic
cores tend to increase with density. A qualitative, Le Chatelier-like
criterion can be stated for whether a change in coordination is driven
primarily by steric repulsion or cohesive interactions. First note
that the interatomic distance $r$ can either decrease or increase upon
a discontinuous volume change. For instance, it is easy to convince
oneself that the following simple formula interpolates the relation
between the specific volume $v$ and coordination number $n$ for the
diamond, simple cubic, and bcc lattice: $r \approx [0.16 v n]^{1/3}$,
($n \le 8$). Now, consider a pressure-driven transition with $\Delta v
< 0$ and $\Delta n > 0$.  If the distance $r$ {\em increased}, then
the increase in the coordination was to counteract the effects of
steric repulsion; hence the latter can be cited as the dominant factor
in the coordination increase, while the reverse transition can be
thought of as driven by cohesive interactions.  For instance, for a
pressure-driven sc-to-bcc transition, $n_\text{bcc}/n_\text{sc} =
4/3$. For such a transition to be driven by cohesive forces, the
volume decrease would have to be 25\%; we are not aware of such large
density changes for phase transitions in condensed phases. In
contrast, for the bcc-to-fcc transition, $r_\text{fcc}/r_\text{bcc}
\approx (1.09 \, v_\text{fcc}/v_\text{bcc} )^{1/3}$, and so already a
10\% change in volume would imply cohesive interactions are
important. An elegant discussion of pertinent electronic processes can
be found in Ref.~\cite{doi:10.1021/ja0114557}

The above ideas help to resolve a seeming contradiction (see also a
discussion in Ref.~\cite{ANIE:ANIE200602485}): On the one hand, there
is an empirical correspondence rule stating that lighter elements
under high pressures tend to form structures typical of elements down
the group at normal pressures.~\cite{ISI:000077643800009,
  ANIE:ANIE200602485} At the same time, heavier elements tend to
exhibit less $sp$-mixing and higher
coordination,~\cite{PapoianHoffmann2000, doi:10.1021/ja0032576} the
di-pnictogen tri-chalcogenides representing a good
example,~\cite{ZLMicro1} as in Fig.~\ref{coex}. On the other hand, we
have seen that the amount of $sp$-mixing should increase with
density. This seeming contradiction may be resolved in the following
way: The $sp$-mixing should indeed increase with density for a fixed
or mildly deforming structure. Eventually, however, a density driven
structural transition causes a coordination increase that leads to a
longer bond and lesser amount of $sp$-mixing. It is this kind of
transition that underlies the {\em general} trend that the amount of
$sp$-mixing should decrease with mass density.

The increased coordination on the one hand, and the decreased amount
of bond directionality (caused by $sp$-mixing) on the other hand, are
hallmarks of the metallic interactions. Thus we identify the high
density, $\epsilon \to 0$ limit with metallic interactions and
delocalized electrons. We reiterate that the low and high density
limit, for sufficiently low $\epsilon$, should correspond to
insulating and metallic behaviors on general
grounds.~\cite{TF9383400678, 0370-1298-62-7-303, Mott1990} Now, for
any value of density, one may always choose a large enough
electronegativity difference to force the valence electrons to
localize on the more electronegative atoms. Thus we identify the
$\epsilon \to \infty$ limit with the ionic bond and conclude that the
coexistence region widens, in terms of $\epsilon$, with density. The
above reasoning does not straightforwardly apply when some bonds
become shorter and others longer with density changes, as in the
covalent-secondary sector in Fig.~\ref{LH}. At least in the case of
rhombohedral arsenic, it is still possible to argue, see the
Supplementary Material, that the amount of $sp$-mixing decreases with
density.

\begin{figure}[t]
  \centering
  \includegraphics[width = .7
  \figurewidth]{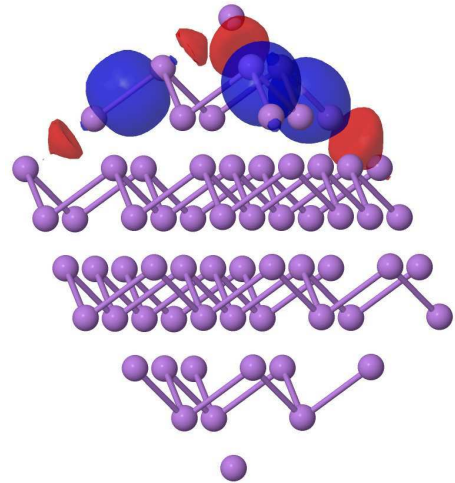}
  \caption{\label{rhAs} Examples of the LMOs obtained for a specific
    HF solution for elemental arsenic arranged in the simple-cubic
    structure, indicated by the spheres. The HF procedure was seeded
    by the solution for rhombohedral arsenic, see text for
    explanation. Each link denotes where a two-center LMO is present;
    only three LMOs are actually shown, to avoid crowding. The link
    pattern clearly following the bond pattern of rhombohedral
    arsenic.}
\end{figure}

\begin{figure}[t]
  \centering
  \includegraphics[width = .7
  \figurewidth]{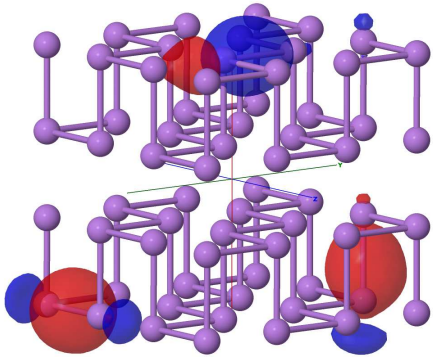}
  \caption{\label{bP} Same as Fig.~\ref{rhAs}, but with the HF
    solution obtained by using the black phosphorus structure as the
    seed.}
\end{figure}

We finish by displaying some of the CDW patterns that emerge in the
solid state at the single Slater-deteminant level as a result of
Kohn-like transitions.  Here we use MOPAC and the LMO formalism to
describe bonding in an extended sample of simple cubic arsenic, see
Figs.~\ref{rhAs} and \ref{bP}. The molecular orbitals are built for a
modestly-sized repeat unit, which we chose to be a $4 \times 4 \times
4$ cube; the crystal field of an infinite solid is inferred by MOPAC
according to a procedure explained in Ref.~\cite{JS556} Already in
this caricature solid, we find that the search for the true lowest
energy HF-solution often becomes impractical.  The result of the
calculation depends on the initial guess for the wavefunction. We
limit ourselves to two specific initial guesses by first creating
lower-symmetry structures, namely, rhombohedral arsenic and
black-phosphorus. Then, we gradually move the nuclei toward their
positions in the simple-cubic structure, while recomputing the
wavefunction at each step using the result from the preceding geometry
as the initial guess for the SCHF procedure. We find that the CDW that
appears to be the likely true ground state for the symmetry-broken
nuclear configuration persists even in the symmetric, simple-cubic
nuclear configuration. The resulting LMOs are shown in Fig.~\ref{rhAs}
and \ref{bP} for the rhombohedral arsenic and black phosphorus seed,
respectively. The mass density employed in these calculations is
5.89~g/cm$^3$, which is just above the experimental value of
5.75~g/cm$^3$ for rhombohedral arsenic at normal conditions and well
below the density of simple-cubic arsenic, viz., 7.48~g/cm$^3$.  We
observe that the two-center LMOs cover only a half of the
nearest-neighbor spaces, consistent with the experimentally known fact
that simple-cubic arsenic is unstable toward a displacive transition
at ambient pressure.~\cite{silas:174101, PhysRevB.77.024109} We have
also found that in the absence of deliberate ``seeding,'' the
``classical'' HF solution spontaneously acquires the symmetry of
rhombohedral arsenic, at sufficiently low densities. This suggests
that at least in some cases, the solution of the Hartree-Fock problem
for an unstable, symmetric atomic arrangement may may predict the
geometry of the actual, symmetry-lowered structure

\section{Discussion}

\label{discussion}

The present results demonstrate the inherent relation and, at the same
time, distinction between fundamental chemical forces. Both the type
and strength of the chemical bond are established as a result of phase
transitions resulting from charge redistribution.  The canonical
chemical interactions are shown to underly distinct thermodynamic
phases and thus can be viewed as distinct sectors on a phase diagram.
We have argued that in the most minimal description, such a phase
diagram is in the space formed by two variables, particle density and
local electronegativity variation. The conventional, two-center
covalent bond and closed-shell interactions can be thought of as
mutually-complementary, intrinsic counterparts that occupy the very
same sector on the phase diagram of chemical interactions.  The
intrinsic connection between the covalent bond and closed shell
interactions is that the two are symmetry broken versions of the
multi-center bond. In turn, the multi-center bond can be thought of as
a coexistence, or hybrid, between the metallic and ionic bond.

The metallic and ionic bond thus emerge as the two most fundamental
interactions that become entirely distinct in the high density
limit. The distinction can be understood qualitatively as poor mutual
miscibility of delocalized and localized electrons. The former can be
thought of as an electron fluid that can flow despite the partial
breaking of the translational symmetry due to the nuclei.  In
insulators, the translational symmetry breaking is complete.  Phase
separation in electronic systems and associated striped phases has
been reported by Schmalian and Wolynes,~\cite{SWmayo} who have
suggested such separation is pertinent to metal ammonia solutions and
high $T_c$ superconductors alike.  The present model analysis shows
the distinction between the metallic and ionic interactions decreases
as the density is lowered, until the two become two opposite limits of
a {\em continuum} of interactions. At lower yet densities, the number
of electrons in the inter-atomic space becomes insufficient to support
the multicenter bond; the latter thus breaks into the covalent bond
and the closed-shell interactions, provided the electronegativity
variation is not too large. The resulting charge-density wave can be
thought of as a bond-order wave.

That changes in bond order in {\em small molecules} should be
analogous to phase transitions can be viewed as a limiting case of the
results obtained for bulk samples. This notion is consistent with the
finite-size-scaling philosophy of bonding and dissociation due to
Kais, Herschbach, and others.~\cite{KaisSerraACP,
  doi:10.1063/1.1637581} Here we observe that changes in bond order
are associated with transitions in the strength, but also {\em
  symmetry} of the charge density wave.  Discrete changes in bond
order come about naturally in venerable theories of bonding, such as
the molecular orbital theory or the early ideas of G.~N.~Lewis. Bond
{\em breaking} during molecular dissociation is often viewed as a
term-crossing event that occurs provided the ground and an excited
state molecular term switch their identities at separations beyond a
certain threshold value.~\cite{LLquantum} The ``sharpness'' of the
transition transition depends on the strength of coupling between the
two terms at the crossing point.  Yet already for modestly sized
molecules, let alone solids, the spacing between molecular terms
becomes very small making it difficult to identify term
crossings. Instead, we take the charge density wave (CDW)
perspective,~\cite{gruner2009density, CanadellWhangbo1991} which
itself is rooted in the venerable density functional
theory~\cite{PhysRev.136.B864, PhysRev.140.A1133} The Coulomb
attraction to the nuclei and the electron-electron repulsion favor
electron localization; the delocalization is driven by minimization of
the kinetic energy. The interplay of these opposing forces gives rise
to a variety of CDW patterns; each pattern is associated with a
specific bond-order assignment.  Transitions between respective
patterns turn out to be well defined. Localized molecular orbitals
(LMOs) represent a particularly convenient way to monitor and
visualize how the bonding electrons are redistributed during those
transitions.  Unlike the canonical MOs, the LMOs are not required to
comprise irreducible representations of the molecule's symmetry group
and are more prone to symmetry changes.

The similarity in behavior between distinct HF-derived CDW states in
small molecules, on the one hand, and thermodynamic phases, on the
other hand, is not surprising on formal grounds. The Hartree-Fock
derived electronic energy $E$, for a finite basis-set, is a minimum of
a quartic polynomial made of expansion coefficients such as the
$\widetilde{C}$s from Eq.~(\ref{psiExp}); the fourth-order terms come
from the electron-electron interactions. Clearly, such a polynomial
will exhibit a myriad solutions already for modestly sized
molecules. Distinct solutions are often separated by barriers, hence
the discontinuous transitions discussed above. {\em Continuous}
transitions are somewhat less common but will still take place when
the Jacobian $\prtl^2 E/\prtl \widetilde{C}_i \prtl \widetilde{C}_j$
computed at the minimum in question vanishes. This is quite similar to
how one may describe macroscopic phase transitions at a meanfield
level using the Landau-Ginzburg classical density functional
theory.~\cite{Goldenfeld} The CDW perspective suggests that the
connection between the present results and macroscopic phase
transitions is not only formal but stems from the intrinsic propensity
of mutually repulsive particles to localize or delocalize in a
cooperative fashion.~\cite{L_AP}

The interatomic spacing is a natural lengthscale in bonded systems; it
is, in turn, largely determined by the Bohr radius. Structures
characterized by {\em supra}-atomic lengths can be often thought of as
made of small ``building blocks'' that are perturbed only weakly when
the solid is assembled; this is the case with many oxides and halides,
for instance. Complex inorganic solids made of individually stable
subunits can have unit cells that contain as many as tens of thousands
of atoms.~\cite{Weber:sn5082, Dshemuchadse:sn5123} Alternative types
of ordering are possible---even if on a modest length scale---where
the type of bonding within the building blocks will depend on whether
the blocks are standalone or bonded relatively intimately, as in
Fig.~\ref{coex}(a). This indicates that there is an emergent length
scale in condensed phases that is not directly tied to the atomic
scale. Kohn designated such ordered structures ``superlattices,''
which could arise, in principle, during a series of transitions
separating the canonical metallic and insulating
states.~\cite{PhysRevLett.19.789} Consistent with those early ideas,
such supra atomic length scales emerge in the present picture as
widths of interfaces separating states with distinct CDWs---or
distinct chemical interactions---that coexist spatially. The simplest
formalism to describe such a coexistence is via the Landau-Ginzburg
free energy functional:~\cite{CahnHilliard, RowlinsonWidom, Bray,
  CL_LG}
\begin{equation} \label{FLG} F = \int dV \left[ \frac{\kappa}{2}
    (\nabla \phi)^2 + {\cal V}(\phi) \right],
\end{equation}
where $\kappa > 0$ is a phenomenological coefficient reflecting the
free energy penalty for spatial inhomogeneity in the order parameter
$\phi$. The bulk free energy density ${\cal V}(\phi)$ has two or more
minima corresponding to coexisting phases, such as that in
Fig.~\ref{Erho}.

Sufficiently far from the critical point, the meanfield description
embodied in the functional (\ref{FLG}) becomes quantitative; the
interface width $l_\text{intf}$ scales roughly as
$(\kappa/g^\ddagger)^{1/2}$, where $g^\ddagger$ is the height of the
barrier in the bulk energy density ${\cal V}(\phi)$ that separates the
corresponding pair of phases. The interface tension coefficient is of
the order $(\kappa g^\ddagger)^{1/2}$. At the critical point, which
formally corresponds to the $g^\ddagger \to 0$ limit, the meanfield
approximation is only qualitative, however it still correctly predicts
that the interface width diverges, while its tension vanishes. The
correlation length also diverges at the critical point and so does the
susceptibility,~\cite{Goldenfeld} implying a structural instability,
such as that at the simple cubic-to-rhombohedral transition in
elemental arsenic.  It is thus likely that in actual materials, which
form subject to many perturbations, the length scale associated with
the coexistence will always remain finite. Consequently, the hybrid of
the metallic and ionic bond is the multi-center bond.  The latter is
directional, even if its progenitors are not. The directionality comes
about already because of the stiffness of the interface; the presence
of the interface breaks the isotropy of space. These ideas are
consonant with the classic explanation of bond directionality via
hybridization of atomic orbitals.

The interface width decreases with separation from the critical point;
once the width becomes less that the interparticle-spacing, there are
at least two alternative possibilities: At low densities (or for
lighter elements), the multi-center bond is replaced by a coexistence
of the two-center covalent bond and secondary interactions, as in
Fig.~\ref{coex}(b). The Bi$_2$Te$_3$ structure in Fig.~\ref{coex}(a)
corresponds to intermediate densities; here the multicenter bond is
still present and forms an interface between covalent/secondary
regions.  Exclusively covalent bonding is also possible for particular
electron counts, as in the diamond structure. (Still, the C-C bond
length in diamond is longer than that in graphene.) At higher
densities, metallic and ionic interactions become well-defined
individually, but could also spatially coexist as in layered compounds
where current is purveyed within select layers.

One thus arrives at the notions of the familiar two-center bond and
bond directionality---intrinsically atomistic concepts---starting from
a {\em coarse-grained} picture, such as that described by the
(classical) density functional in Eq.~(\ref{FLG}). Such a
coarse-grained view is far from new; it underlies the venerable
density-functional theory,~\cite{PhysRev.136.B864, PhysRev.140.A1133}
of course.  Here, we find that some results of {\em classical} density
functional treatments are transferable to electrons, such as poor
miscibility of localized and delocalized particles. The Heusler and
half-Heusler compounds, and incommensurate phases seem to exemplify
such poor miscibility, as mentioned ealier.  Furthermore, the idea of
coarse-graining is central to the renormalization group (RG) theory of
phase transitions. Coarse-graining transformations can be thought of
as movement in the space defined by coupling
constants.~\cite{Goldenfeld} Distinct phases come about as attractive
fixed points for such transformations. The present results suggest the
ionic, metallic, and covalent/secondary interactions are such
attractive fixed points, the latter essentially corresponding with the
corners of the venerable Arkel-Ketelaar
triangle.~\cite{doi:10.1021/ed012p53, vanArkel, Ketelaar, Burdett1995}

Laws of corresponding states arise when the system is close to a
critical point, where the RG flows of the coupling constants are not
sensitive to system-specific detail but only to the symmetry and range
of interactions because the correlation length now greatly exceeds
molecular lengthscales. Long wavelength, possibly disordered patterns
that may arise as a consequence of such instabilities are difficult to
predict using ab initio methods, because the size of the repeat unit,
if any, of the distorted structure is not a priori known. Examples of
such complex distortions of recent note are represented by hybrid
organic-lead halide perovskites~\cite{doi:10.1021/acs.jpclett.6b01648}
and oxide-ion conducting perovskite
derivatives.~\cite{doi:10.1021/jacs.6b10730} It is quite possible that
the multiplicity of mutual orientations of rigid molecular units in
those compounds, such as octahedra or tetrahedra, scales exponentially
with the sample size thus limiting the utility of brute force, ab
initio approaches. A hybrid strategy seems more profitable, in which
one uses ab initio methods to infer the parameters for a
coarse-grained description in the form of a free-energy functional,
such as those that have been used to study displacive transitions and
multiferroic behavior.~\cite{PhysRevLett.3.412, Dove1997} (Examples of
such inference for simple liquids and spin systems can be found in
Ref.~\cite{L_AP}) The resulting functional can be tested, in
principle, for the multiplicity of states.~\cite{MP_Wiley,
  SchmalianWolynes2000, BL_6Spin, PhysRevB.91.100202} Conversely,
textbook approximations to canonical bonding schemes should work close
to attractive fixed points. Near phase boundaries---not too close to
critical points, if any---one should expect ordering on supra-atomic
length scales.

Of course the present, coarse-grained view of the chemical interaction
cannot replace detailed quantum-chemical analyses for the stability of
known compounds.  Rather, we believe the present findings complement
those detailed analyses and, in particular, may provide an additional
tool for predicting novel compounds. For instance, the
density-functional perspective may help one to predict whether
specific ingredients will bind to form a stable or metastable solid,
as opposed to segregating into unwanted products.  Indeed, insofar as
localized electrons are tied to regions with significant
electronegativity variation---and vice versa for the delocalized
electrons---the mutual miscibility of the corresponding moieties is
tied to that of the respective electronic subsystems.  The latter
mutual miscibility seems a natural variable for machine-learning
algorithms that have been used for predicting the existence and
properties of compounds.~\cite{Dyer17052013, 2178621820060801,
  Villars200126} The present results also suggest that multiferroic
materials and any other solids prone to structural instabilities
should be relatively close to the critical points and phase boundaries
on the phase diagram in Fig.~\ref{phase_diag}. Note that while
achieving uniformly high densities in the bulk may require substantial
compression, locally-dense environment can be created using chemistry
even at normal conditions, for instance, by substituting elements by
their heavier neighbors down the group.

The present results on specifics of electronic localization already
suggest a concrete way to speed up search for stable geometries. 
% The diversity of bulk structures is reflected in the degeneracy of
% the Hartree-Fock ground state, as was the case with the relatively
% extended systems in Figs.~\ref{ring20highDens}-\ref{bP}.  While in
% the thermodynamic limit, one expects the degeneracy of distinct {\em
%   metastable} HF-solutions to scale exponentially with the system
% size;~\cite{SchmalianWolynes2000} the issue of the degeneracy of the
% HF {\em ground} state is subtler. One expects that such a degeneracy
% should be lifted by quantum fluctuations.  Formally, the degeneracy
% of the HF ground state we have encountered is a consequence of using
% a single Slater determinant as our variational wavefunction. The
% present analysis suggests
We have seen that already in the symmetric, simple cubic structure of
arsenic, the LMOs predict the actual distorted structure without the
need for costly geometric optimization. This suggests a way to
efficiently screen very large numbers of candidate compounds for
stability and possible isomers. In this procedure, a computer first
generates a variety of structures for a molecule or solid alike,
subject only to steric constraints. The candidate structures may be
relatively symmetric, thus greatly reducing the pool of trial
configurations.  In the next step, a computationally efficient,
semi-empirical approximation is used to obtain the molecular orbitals,
after which the localized molecular orbitals are generated. (Both
functionalities are already implemented in MOPAC, for instance.) If
the two-center LMOs happen to cover inter-atomic spaces within a
sufficiently large fraction of nearest neighbors, the candidate
compound and structure are rated as a high probability target for
further screening.  Indeed, since every nearest neighbor bond is
covered by a two-center LMO, one may expect the putative geometry to
be either stable or metastable.  If, on the other hand, the coverage
is not complete, the nuclear configuration is expected to distort so
as to follow the electronic instability. If, furthermore, there are
more than one equivalent yet distinct coverage patterns, then we
arrive at the venerable notion of a resonance.~\cite{PaulingBond,
  doi:10.1021/cr100228r} The resulting distorted geometries may or may
not be separated by a barrier surmountable by zero-point molecular
vibrations, the former and latter cases exemplified by ammonia and
trans-polyacetylene,~\cite{RevModPhys.60.781} respectively.  The
candidate compounds may be also generated subject to preset
constraints of interest in specific applications, such as the design
of a docking site for a given ligand of interest. These ideas should
be suitably modified for bonding involving $d$ and higher order
orbitals and for coordination numbers exceeding six; this is work in
progress.

{\bf Supplementary Material}

The Supplementary Material is organized by Section of the main text
and contains supporting grahics and discussion on: Potential energy
surfaces of the trimers; effects of zero-point vibrations,
electronegativity varation, and electron count on symmetry breaking;
and lack of promotion of core electrons
(Section~\ref{trimer}). Pedagogical review of the LMO formalism;
additional aspects of the LMO and stability analysis for trimers and
20-member ring; and NBO and QTAIM perspective on the dissociation of
the trimer (Section~\ref{localization}). Details of MOPAC
re-parametrization, and $sp$-mixing in bulk arsenic
(Section~\ref{solids}).

{\bf Acknowledgments}: We thank Peter G. Wolynes, Eric Bittner, Arnold
Guloy, Zhifeng Ren, and J.~J.~P.~Stewart for discussions and insight.
This work has been supported by the National Science Foundation Grants
CHE-0956127, CHE-1465125, and the Welch Foundation Grant E-1765.  The
authors acknowledge the use of the Maxwell/Opuntia Cluster and the
advanced support from the Center of Advanced Computing and Data
Systems at the University of Houston. The authors acknowledge the
Texas Advanced Computing Center (TACC) at The University of Texas at
Austin for providing HPC resources. URL: http://www.tacc.utexas.edu.

%\bibliographystyle{ieeetr} 
%\bibliographystyle{jpc}
%\bibliography{lowT,ms}
%\bibliography{lowT,ms}

$^\dagger$ Present address: UT Southwestern Medical School; Dallas, TX
75390-9003 

%\bibliography{/Users/vas/Documents/tex/ACP/lowT,ms}
%\bibliography{lowT,ms}

\bibliographystyle{unsrt}
\putbib[lowT,ms]
\end{bibunit}

\begin{bibunit}[unsrt]
  
\clearpage

\setcounter{section}{0}
\setcounter{equation}{0}
\setcounter{figure}{0}
\setcounter{table}{0}
\setcounter{page}{1}
\makeatletter
\renewcommand{\theequation}{S\arabic{equation}}
\renewcommand{\thefigure}{S\arabic{figure}}
\renewcommand{\thetable}{S\arabic{table}}
\renewcommand{\bibnumfmt}[1]{[S#1]}

\begin{widetext}
 
  \begin{center} {\bf \large {\em Supplementary Material}\/: The chemical bond as an emergent phenomenon} \medskip

    {\large Jon C. Golden$^{2}$, Vinh Ho$^{1, 3}$ and Vassiliy Lubchenko$^{1, 2}$}
    \medskip

    {\normalsize $^1$Department of Chemistry, University of Houston,
      Houston, TX 77204-5003

     $^2$Department of Physics, University of Houston, Houston, TX
      77204-5005

     $^3$Present address: UT Southwestern Medical School; Dallas, TX 75390-9003}

  \end{center}
  
\end{widetext}

\date{\today}

% \SectionsOn
% \SectionNumbersOn

% \begin{document}

%\maketitle

% \tableofcontents

\setcounter{equation}{0}
\setcounter{figure}{0}
\setcounter{table}{0}
\setcounter{page}{2}
\makeatletter
\renewcommand{\theequation}{S\arabic{equation}}
\renewcommand{\thefigure}{S\arabic{figure}}
\renewcommand{\bibnumfmt}[1]{[S#1]}
\renewcommand{\citenumfont}[1]{S#1}

\section{Potential energy of the
  $\text{AsH}_2-\text{AsH}-\text{AsH}_2$ trimer}

\begin{figure}[h]
  \centering
  \includegraphics[width = 0.9 \figurewidth]{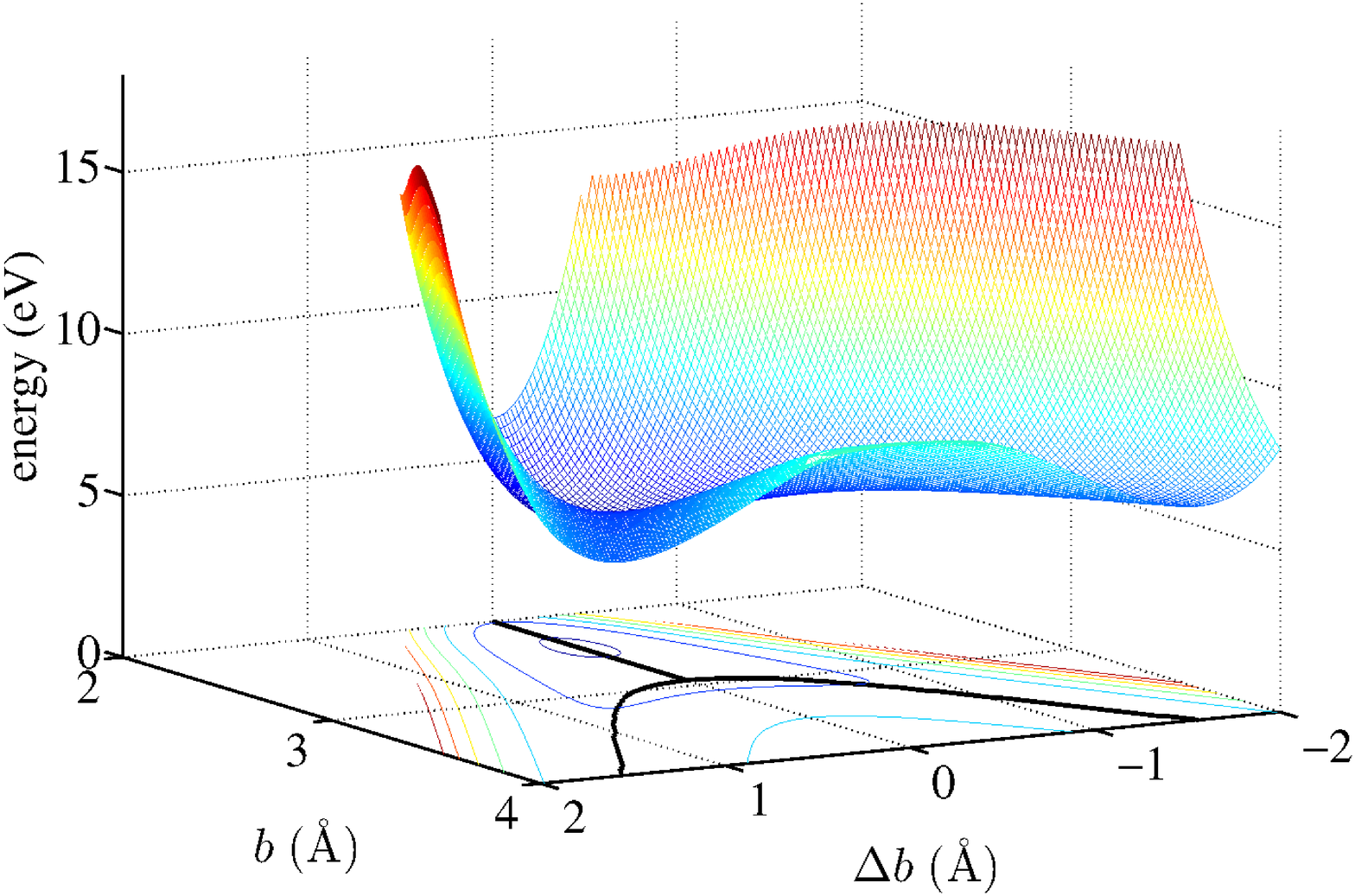}
  \caption{\label{PES} Potential energy surface for the linear 3/4
    trimer AsH$_2$-AsH$_1$-AsH$_2$ as a function the displacement
    $\Delta b \equiv (b_2 - b_1)/2$ of the central As atom off the
    trimer's midpoint and the chain length per bond $b \equiv (b_1 +
    b_2)/2$. The quantities $b_1$ and $b_1$ stand for the As-As bond
    lengths.  The minimum energy contour, shown in bold on the contour
    plot beneath the surface, is determined by minimizing the energy
    at a fixed value of $b$. The surface was obtained using the
    semi-empirical quantum-chemistry package MOPAC with PM6
    parametrization.\cite{JS12}}
\end{figure}

\section{Fluctuation-induced lowering of the symmetry breaking}

\begin{figure}[t]
  \centering
  \includegraphics[width =
  \figurewidth]{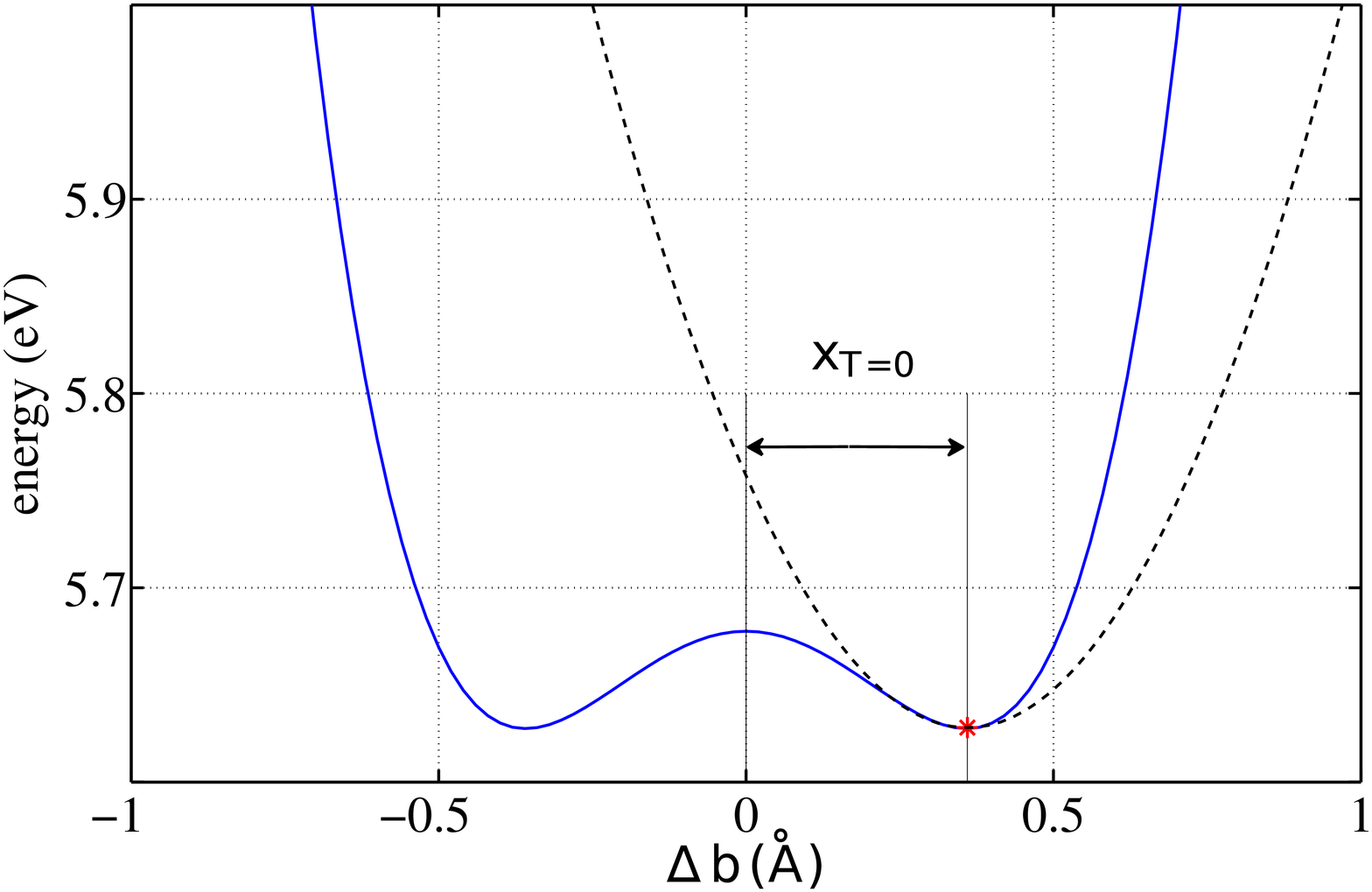}
  \caption{\label{fluctuation} The solid line is a slice of the
    potential energy function from Fig.~\ref{PES}. The location of the
    slice is chosen so that zero-point vibrations in the harmonic
    potential approximating a metastable minimum on the actual
    potential surface do not exceed, in magnitude, the half-width of
    the barrier separating the two minima on the symmetry broken
    energy surface. }
\end{figure}

The potential energy surface in Fig.~\ref{PES} was obtained using the
Born-Oppenheimer approximation; its minima determine the
zero-temperature location of the nuclei in the classical limit. A
slice of the energy surface in the symmetry-broken region, at a
constant value of the overall trimer length, is exemplified in
Fig.~\ref{fluctuation} by the solid line. Already zero-point vibration
will be sufficient to overcome the barrier separating the two minima
on the bistable energy profile, provided the barrier is sufficiently
low.  To determine qualitatively the trimer length where the barrier
can be still overcome by zero-point vibrations, we consider the
symmetric stretch for concreteness.  A harmonic oscillator with
frequency $\omega$ and mass $m$ exhibits zero-point vibrations of
magnitude~\cite{LLquantum}
\begin{equation} \label{x} x_{T=0} = \left(\frac{\hbar}{m \omega}
  \right)^{1/2}.
\end{equation}
For a quadratic potential whose force constant $\kappa$ matches the
curvature of a symmetry-broken minimum, $\omega = (\kappa/m)^{1/2}$,
where $m$ is the mass of a terminal atom.  Next, we find the value of
the trimer length at which the vibrational magnitude (\ref{x}) exactly
matches the half-width of the barrier separating the two minima, see
the graphical explanation in Fig.~\ref{fluctuation}. This length is
indicated with the asterisk on Fig.~4 of main text.

\section{Lack of promotion of $3d$ electrons}

\begin{figure}[tph]
  \centering
  \includegraphics[width = 0.9 \figurewidth]{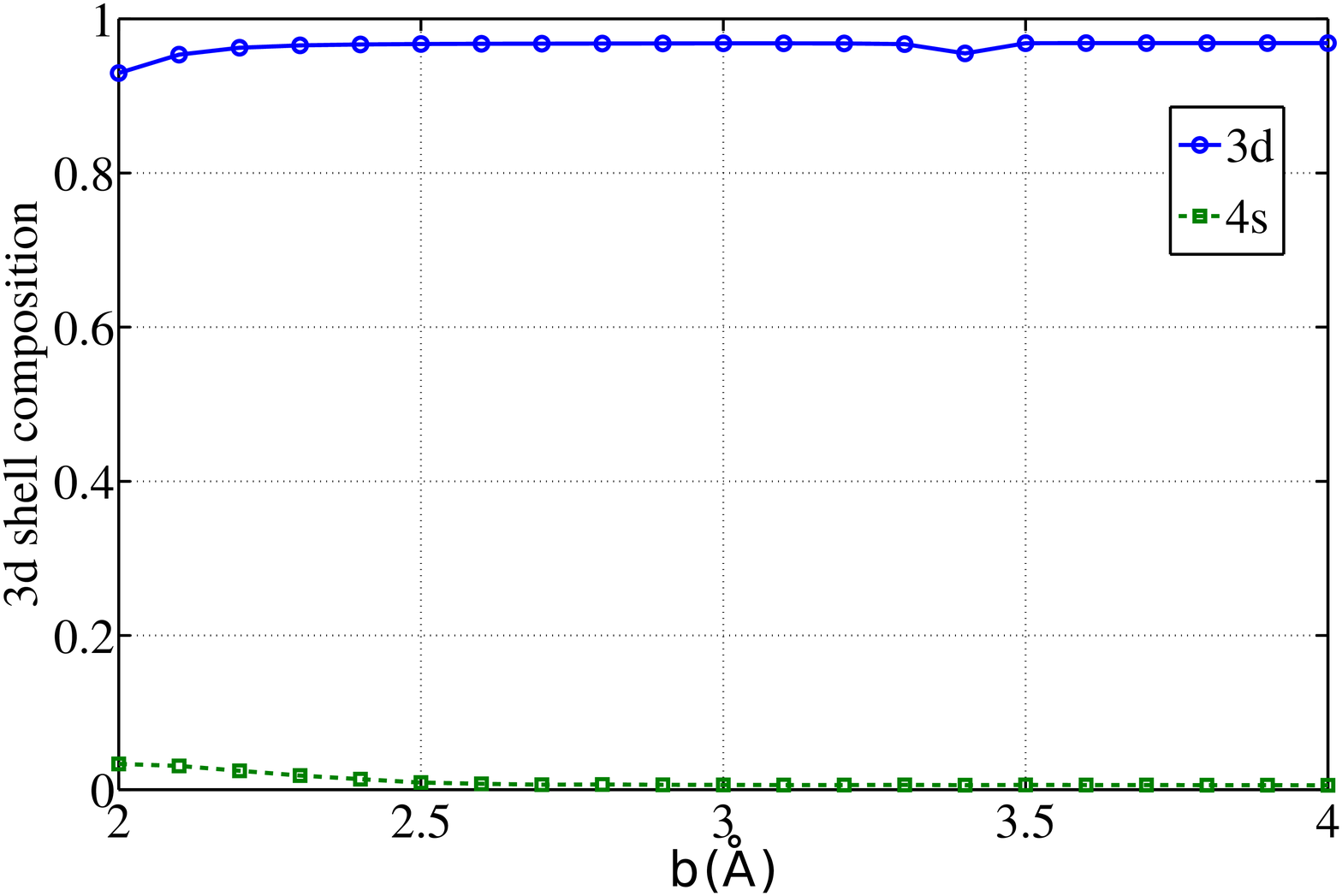}
  \caption{Fractional contribution of atomic $3d$ and $4s$ orbitals to
    the trimer's molecular orbital that can identified as primarily
    consisting of $3d$ atomic orbitals at low densities. The
    horizontal axis is the trimer length per bond. The trimer is the
    linear 3/4 molecule AsH$_2$-AsH$_1$-AsH$_2$. The calculations were
    performed with all-electron HF calculation as implemented in the
    package Orca.~\cite{PN137,Libint2} \label{AE_3d}}
\end{figure}

It is desirable to check how much the outermost filled $d$ shells are
modified following contraction of the trimer, to test whether
MOPAC's~\cite{JS12} implicit treatment of the core electrons
introduces substantial error.  To this end, we perform an all-electron
HF calculation using the package Orca.~\cite{PN137,Libint2} We
consider a trimer at low density and identify the molecular orbital,
call it X, that primarily consists of the $3d$ atomic orbitals. We
next compress the trimer and, alongside, monitor the contribution of
the atomic $3d$ atomic orbitals to the molecular orbital X. This
contribution is shown in Fig.~\ref{AE_3d} with circles. We observe
that the atomic $3d$ orbitals do not contribute substantially to the
bonding orbitals even at relatively high densities. For the sake of
comparison, we also show the contribution of the atomic $4s$ orbitals
to the very same molecular orbital, with the squares.

\section{Dependence of symmetry breaking on the $pp\sigma$ bond
  electron count}

\begin{figure}[h]
  \centering
  \includegraphics[width = 0.9
  \figurewidth]{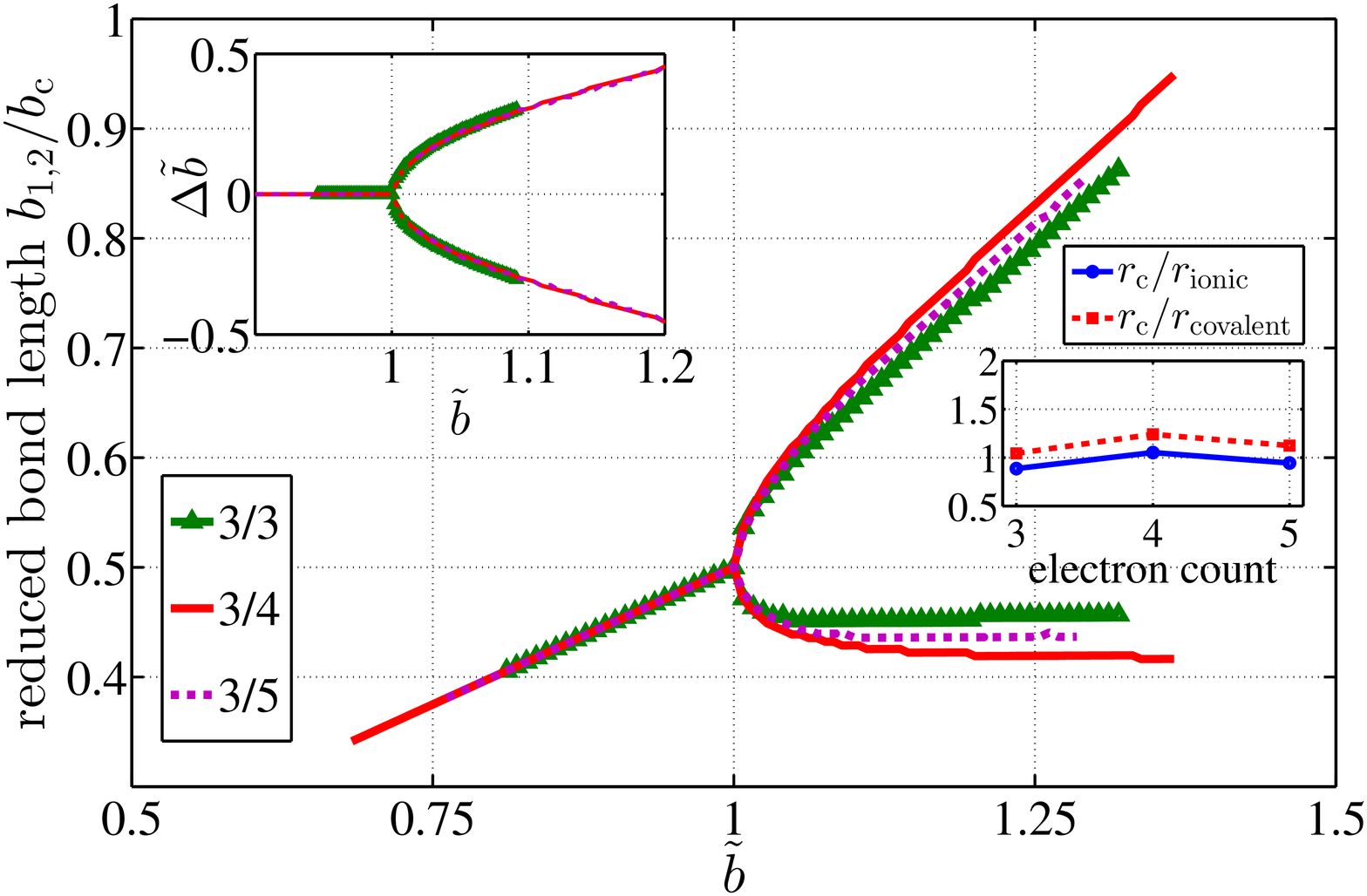}
  \caption{\label{electron_count_reduced} Bond lengths for three arsenic trimers,
    AsH$_2$-AsH$_n$-AsH$_2$ trimer, as functions of the overall trimer
    length per bond. The three cases, $n = 0, 1, 2$, correspond to a
    five, four, and three electron $pp\sigma$ bond bond,
    respectively. Both axes are scaled by the corresponding critical
    bond length $b_c$. Additional scaling is needed to bring the three
    curves to one universal
    dependence. }
\end{figure}

The symmetry-breaking graphs for three values of electron count in the
linear trimer AsH$_2$-AsH$_n$-AsH$_2$, $n = 0, 1, 2$, indicate that
the symmetric state becomes less stable when the electron count
differs from its ideal value of four, which yields perfect Lewis
octets for each arsenic. (This situation is realized for a
singly-hydrogenated central arsenic, $n=1$.) In addition, we observe
that in contrast with the situations shown in Fig.~4 and 10 of the
main text, the critical length is no longer the only characteristic
length in the problem. The three bifurcation graphs, shown in
Fig.~\ref{electron_count_reduced}, can still be rescaled so as to
follow one universal curve, inset of
Fig.~\ref{electron_count_reduced}, however doing so requires {\em two}
independent lengths, one needed to fix the lateral position of the
critical point, the other to fix the strength of symmetry breaking.

\newpage

\section{Effects of electronegativity variation}

\begin{figure}[h]
  \centering
  \includegraphics[width =
  \figurewidth]{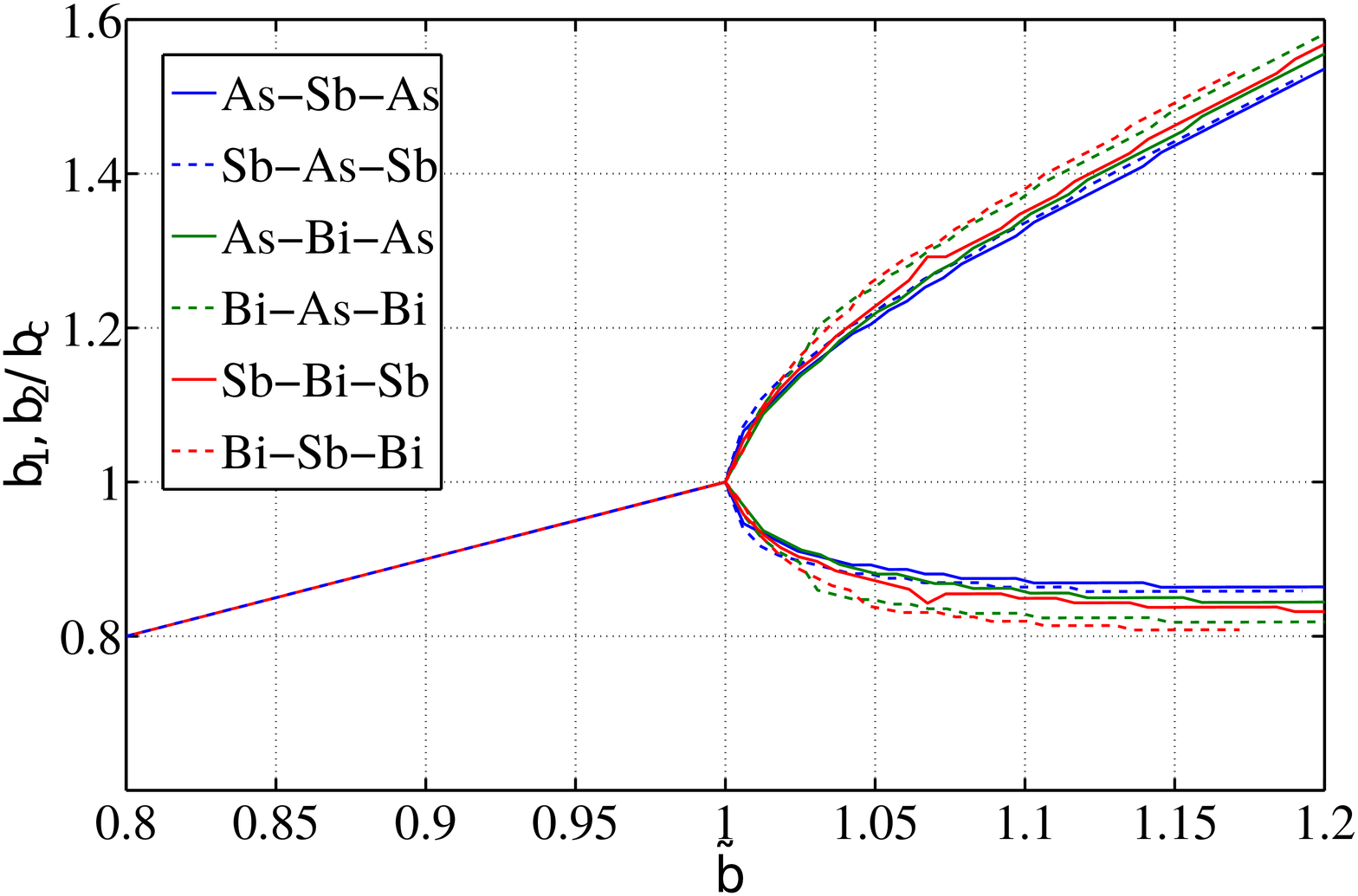}
  \caption{\label{hetero} Bond lengths for several heteronuclear 3/4
    trimers XH$_2$-YH$_1$-XH$_2$ and YH$_2$-XH$_1$-YH$_2$, as
    functions of the overall trimer length per bond. Both axes are
    rescaled by the corresponding critical radii, no additional
    rescaling of the $\Delta b$ coordinate was performed.}
\end{figure}

\begin{figure}[h]
  \centering
  \includegraphics[width =
  \figurewidth]{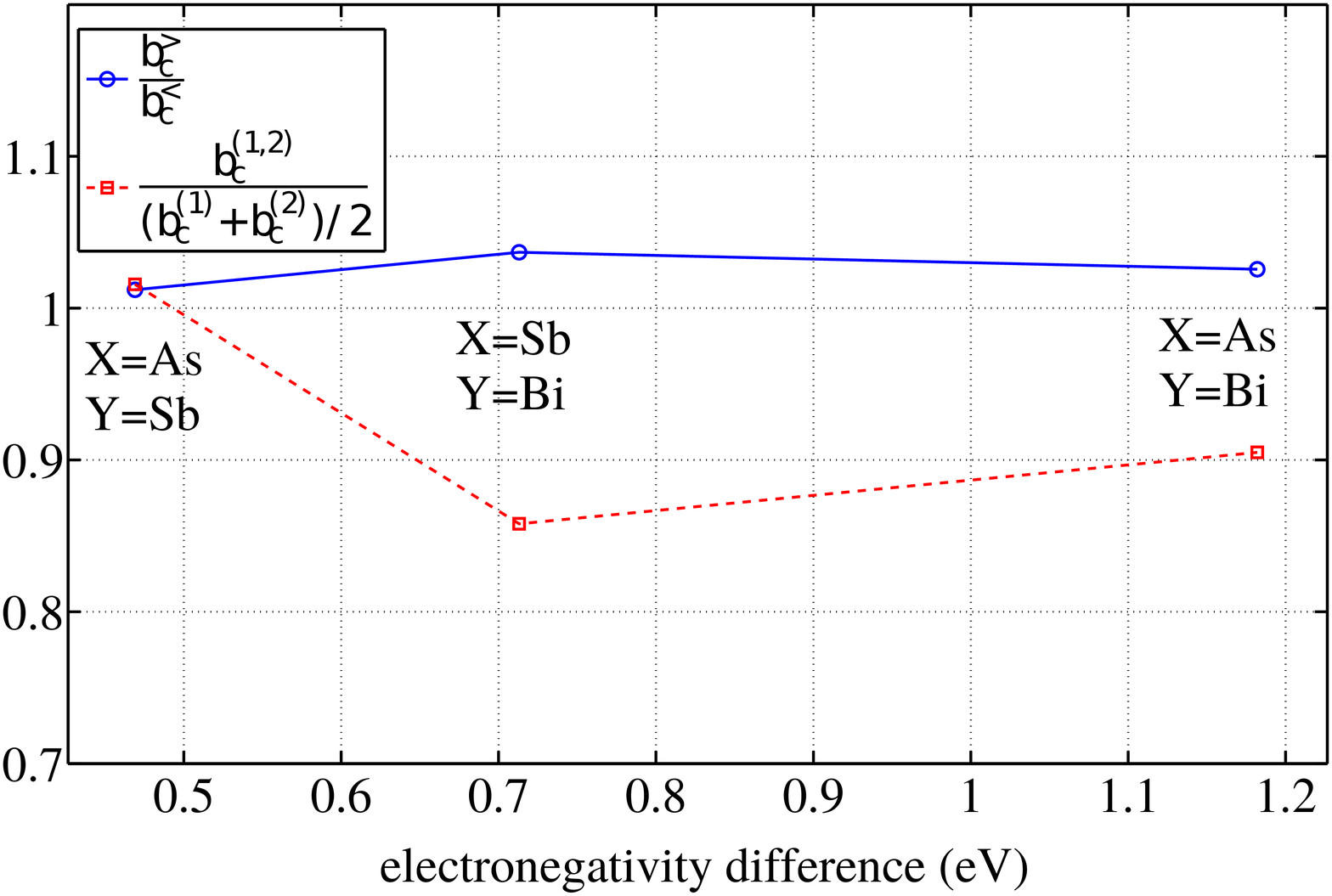}
  \caption{\label{rc_hetero} Critical lengths for the heteronuclear
    3/4 trimers XH$_2$-YH$_1$-XH$_2$ and YH$_2$-XH$_1$-YH$_2$ from
    Fig.~\ref{hetero}.   $b_c^>$ and $b_c^<$ correspond to the situation
    where the more electronegative element (X) is placed in the central
    and terminal position, respectively. $b_c^1$ and $b_c^1$ denote
    the critical lengths for the homonuclear trimer
    XH$_2$-XH$_1$-XH$_2$ or YH$_2$-YH$_1$-YH$_2$.}
\end{figure}

%Several molecular examples are considered in
%Figure~\ref{epsilon_variation} representing a range electronegativity
%difference between the central and terminal atoms spanning zero to
%nearly 2 ev. To offset the effects of atomic number and ion size on
%the actual transition location, we have normalized the difference
%between the spatial and electronic critical lengths by the spatial
%critical length $b_c^{\mathrm{(spatial)}}$ pertinent to each molecule.
%\begin{figure}[h]
%  \centering
%  \includegraphics[width =
%  \figurewidth]{43_Trimer_electronegativity_variation_A.eps}
%  \caption{Dimensionless separation length $\delta$ between the
%    electronic and spatial transitions, normalized by the critical
%    bond length $b_c^{\mathrm{(spatial)}}$ to suppress to the atomic
%    number and size influences, shown as a function of
%    electronegativity variation $\varepsilon$ between the terminal and
%    central atoms.\label{epsilon_variation}}
%\end{figure}

\newpage

\section{Example of the Landau-Ginzburg free energy near a 1st order
  transition}

\begin{figure}[h]
  \centering
  \includegraphics[width =
  \figurewidth]{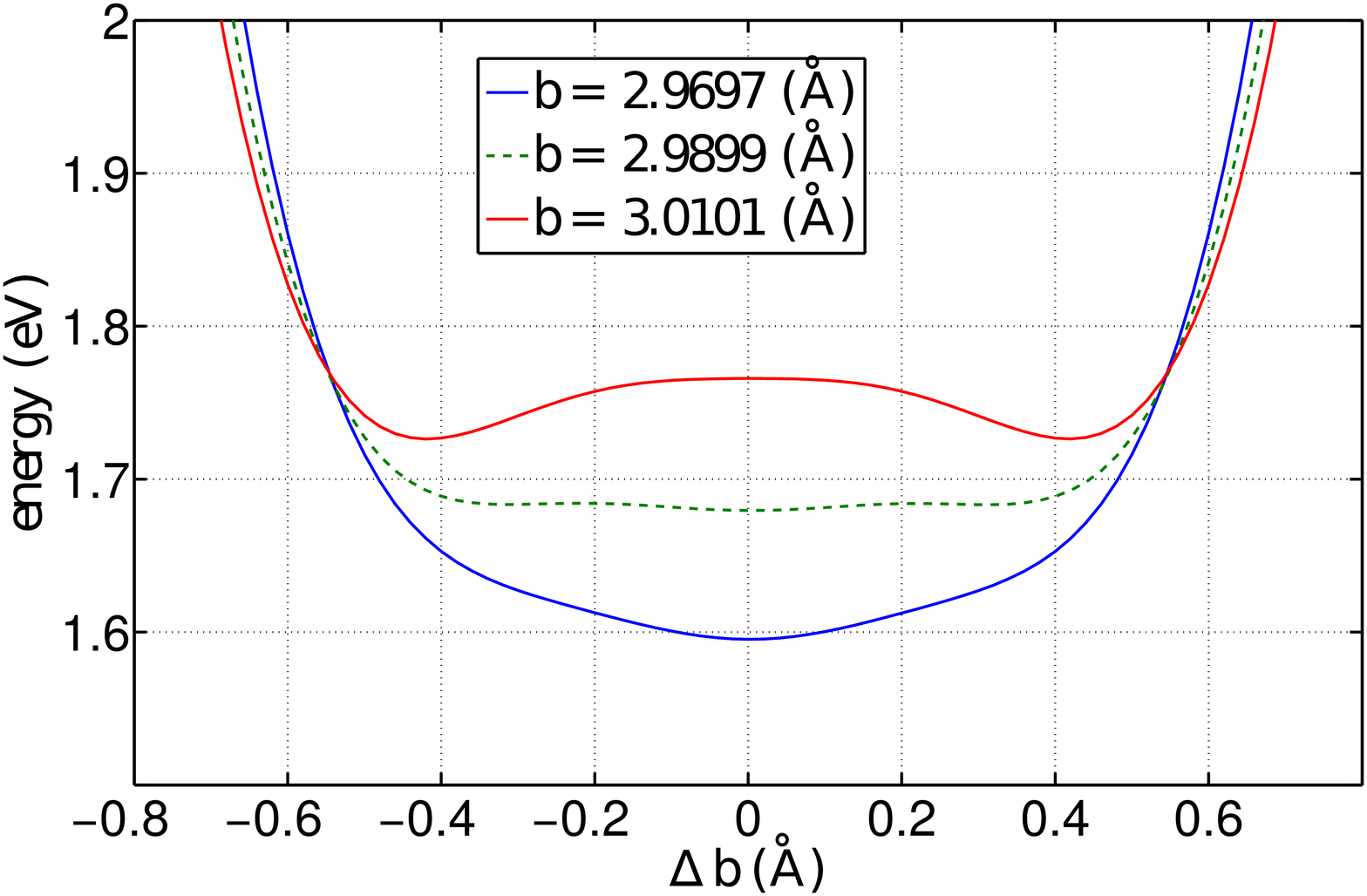}
  \caption{Slices of the full energy surface at $b=\text{const}$, for
    three values of the latter, for the 3/4 heterotrimer TeH-Sb-TeH
    near the symmetry breaking transition. The three distinct minima
    are present in one slice but are very shallow; the corresponding
    barriers would be easily erased by already zero-point vibrations.}
\end{figure}

\section{Re-parametrization of antimony in MOPAC}

\begin{figure}[h]
  \centering
  \includegraphics[width = 0.84 \figurewidth]{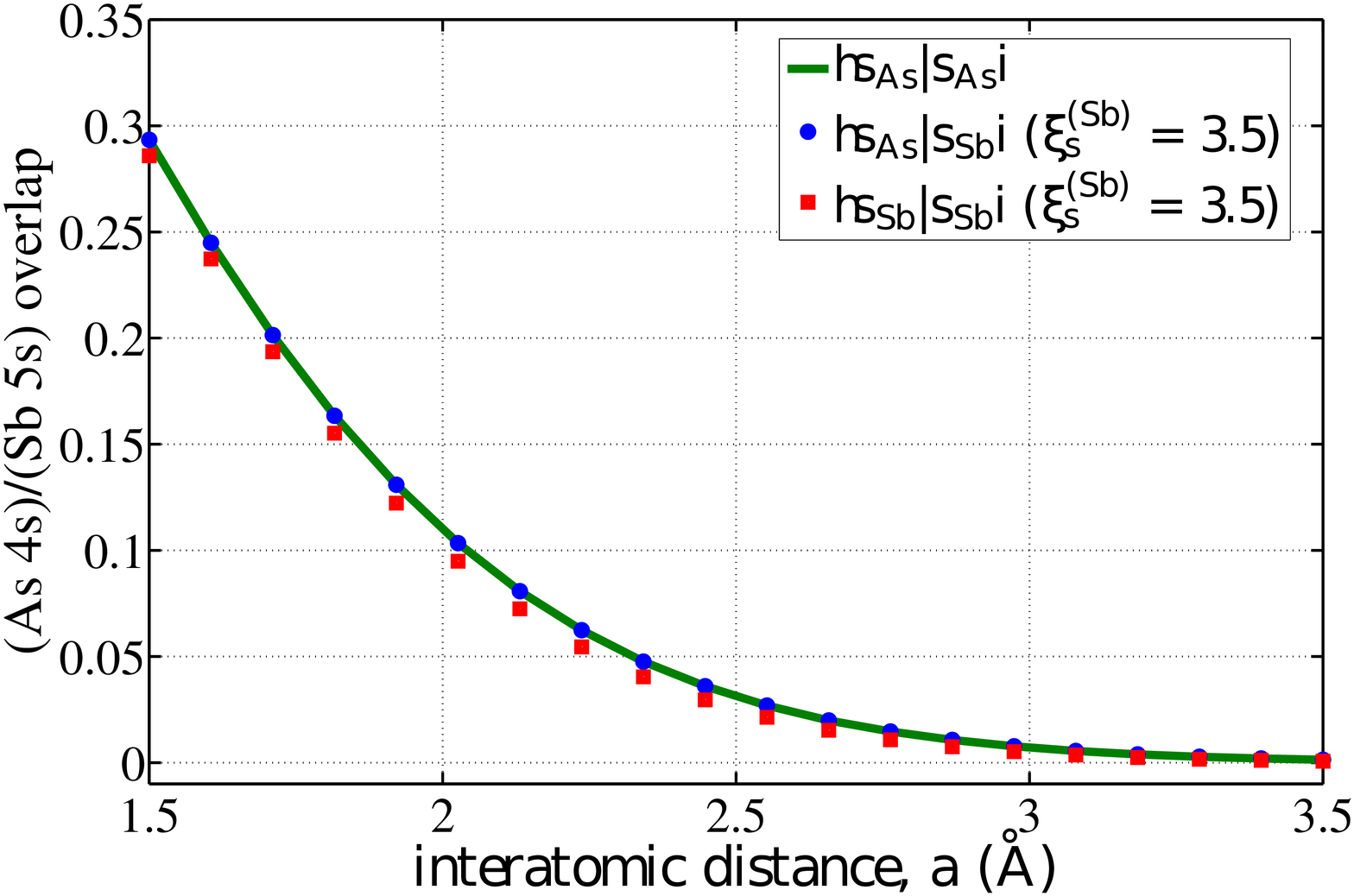}
  \caption{\label{shrinking1} The $s$ wavefunction overlaps as 
    functions of separation for As-As, As-Sb and Sb-Sb bonds.}
\end{figure}

\begin{figure}[h]
  \centering
  \includegraphics[width = 0.84 \figurewidth]{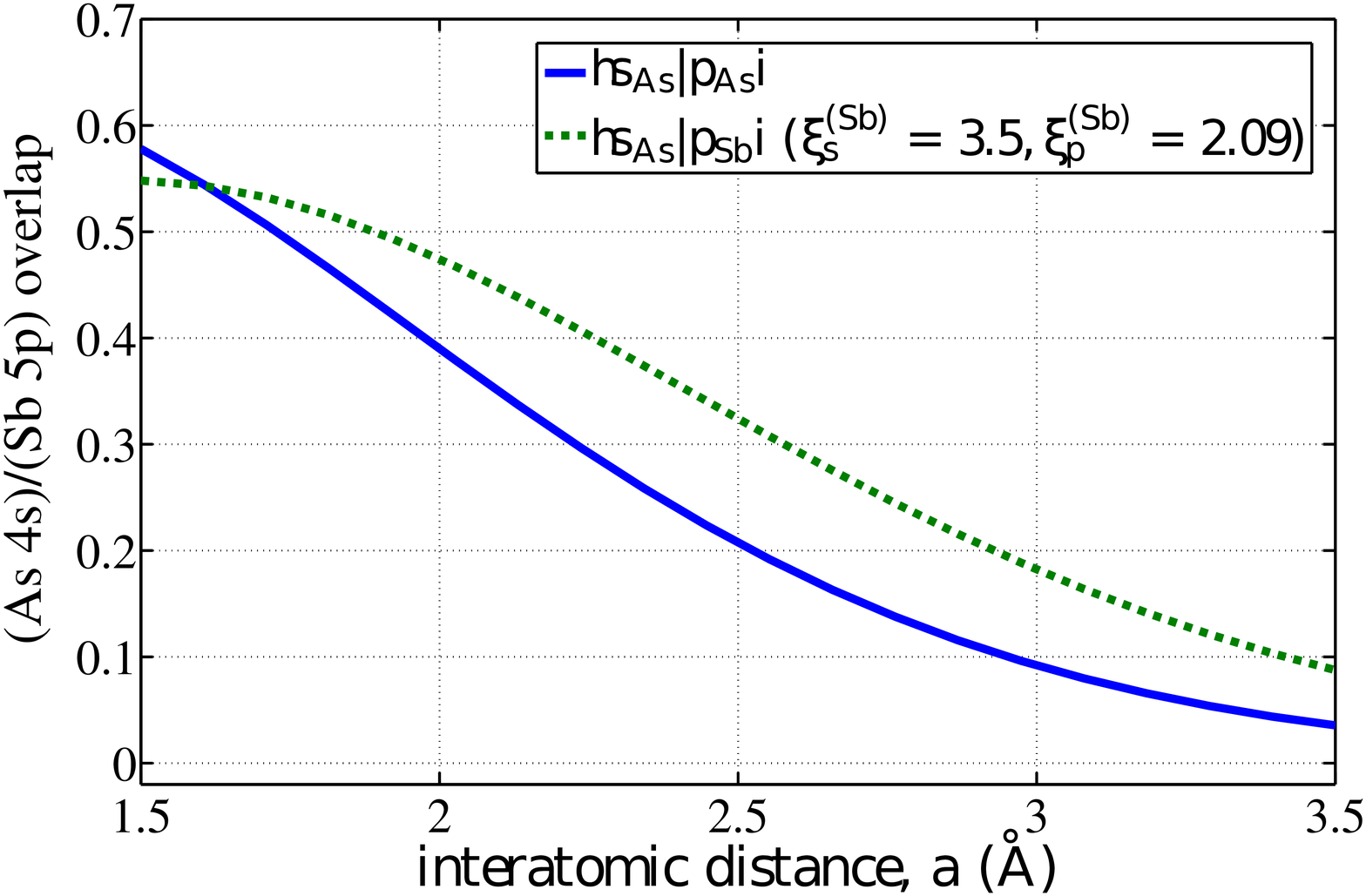}
  \caption{\label{shrinking2} The $sp$ wavefunction overlaps as
    functions of separation for As-As and As-Sb bonds.}
\end{figure}

\begin{figure}[h]
  \centering
  \includegraphics[width = 0.84 \figurewidth]{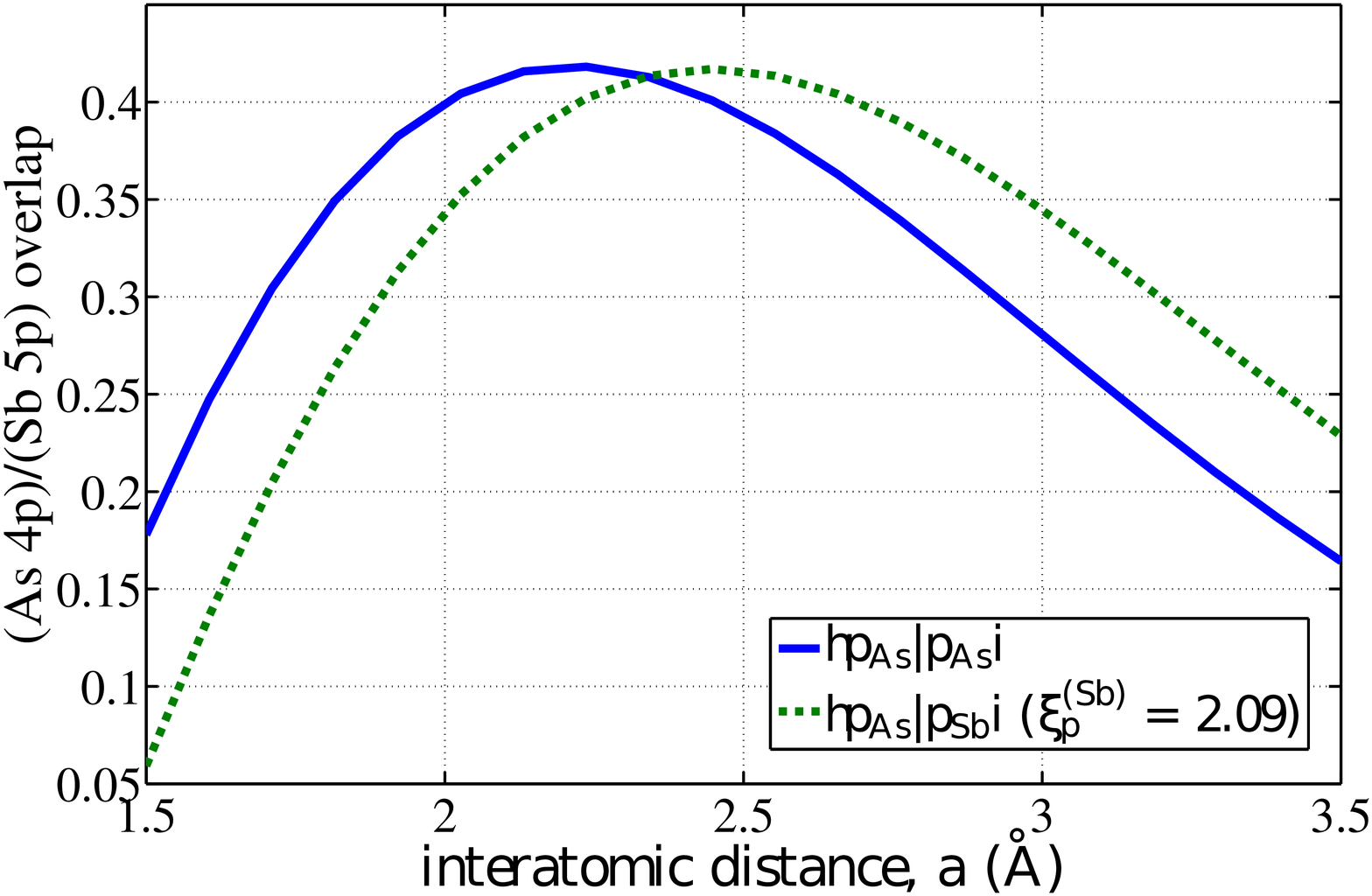}
  \caption{\label{shrinking3} The $pp$ wavefunction overlaps as 
    functions of separation for As-As and As-Sb bonds.}
\end{figure}

\begin{figure}[h]
  \centering
  \includegraphics[width = 0.84 \figurewidth]{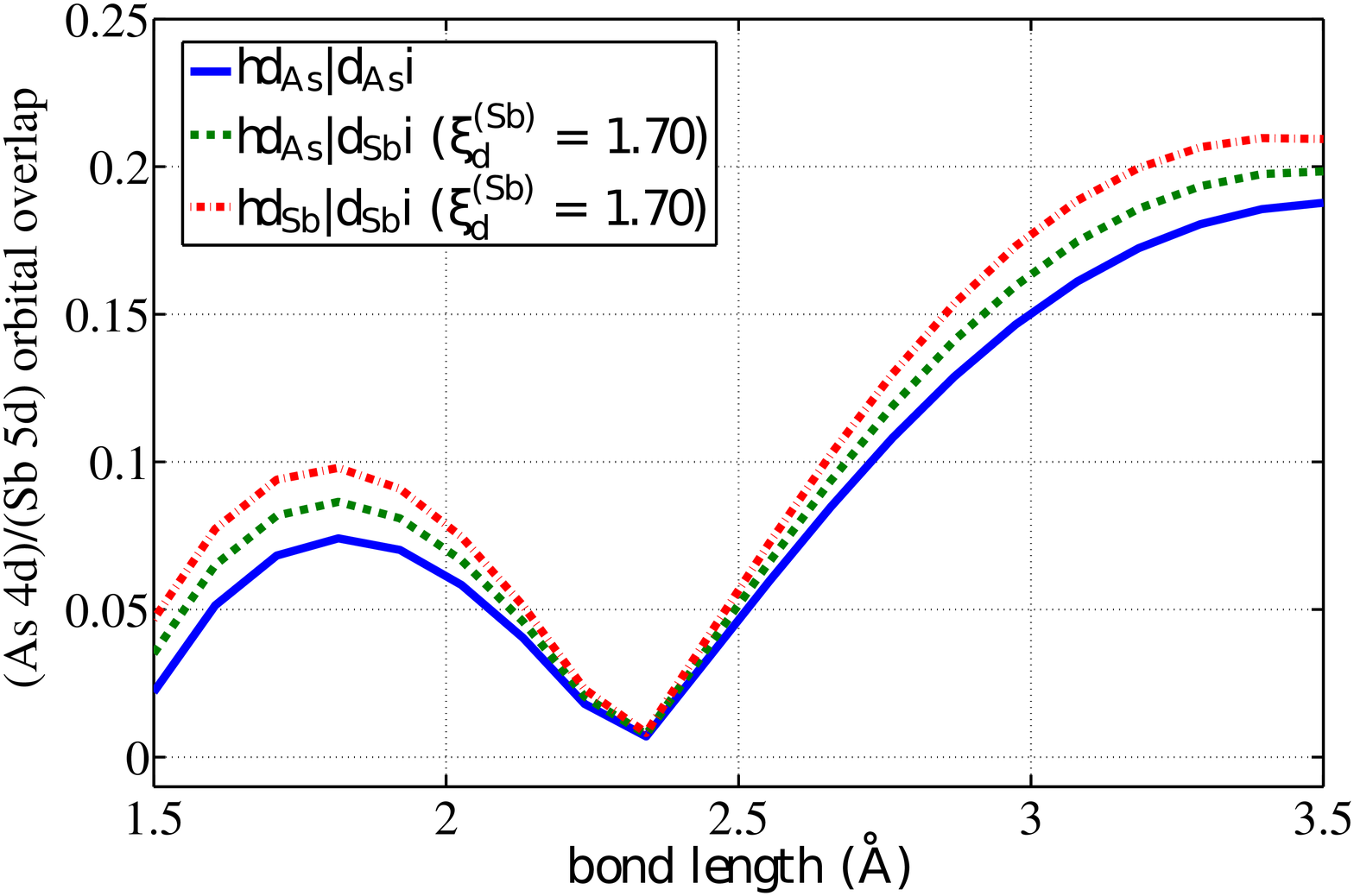}
  \caption{\label{shrinking4} The $dd$ wavefunction overlaps as
    functions of separation for As-As, As-Sb and Sb-Sb bonds.}
\end{figure}

MOPAC evaluates the matrix elements of the Hamiltonian using a
approximation, in which their coordinate dependence is fully
determined by coordinate dependence of the overlaps of the
corresponding (atomic) basis orbitals.~\cite{WolfsbergHelmholz}
Figs.~\ref{shrinking1}-\ref{shrinking4} show the wavefunction overlaps
involving the pseudo-antimony with itself and with arsenic, and, for
comparison, with MOPAC's built-in As-As overlaps. MOPAC uses
Slater-type orbitals. The lengthscale $\xi$ refers to the orbital
exponent that determines the inverse rate of the exponential decay of
the wavefunction overlap.

\newpage

\section{Re-parametrization of MOPAC PM6 basis set under pressure}

MOPAC employs one, Slater-type wave function per atomic orbital with a
fixed extent. The latter is optimized to match geometries and
formation enthalpies at normal conditions. Such a single-wavefunction
parametrization may not be quantitative at high pressures, however. To
test for this potential complication, we have solved for the HF
wavefunctions of the 3/4, homonuclear arsenic trimer using a Quantum
Chemistry package that approximates atomic wavefunctions with a linear
combination of five Gaussian type functions (GTF); the widths of the
individual Gaussians span a substantial range. We then monitor the
coefficients at individual GTF's as functions of density.

\begin{figure}[h]
  \centering
  \includegraphics[width = \figurewidth]{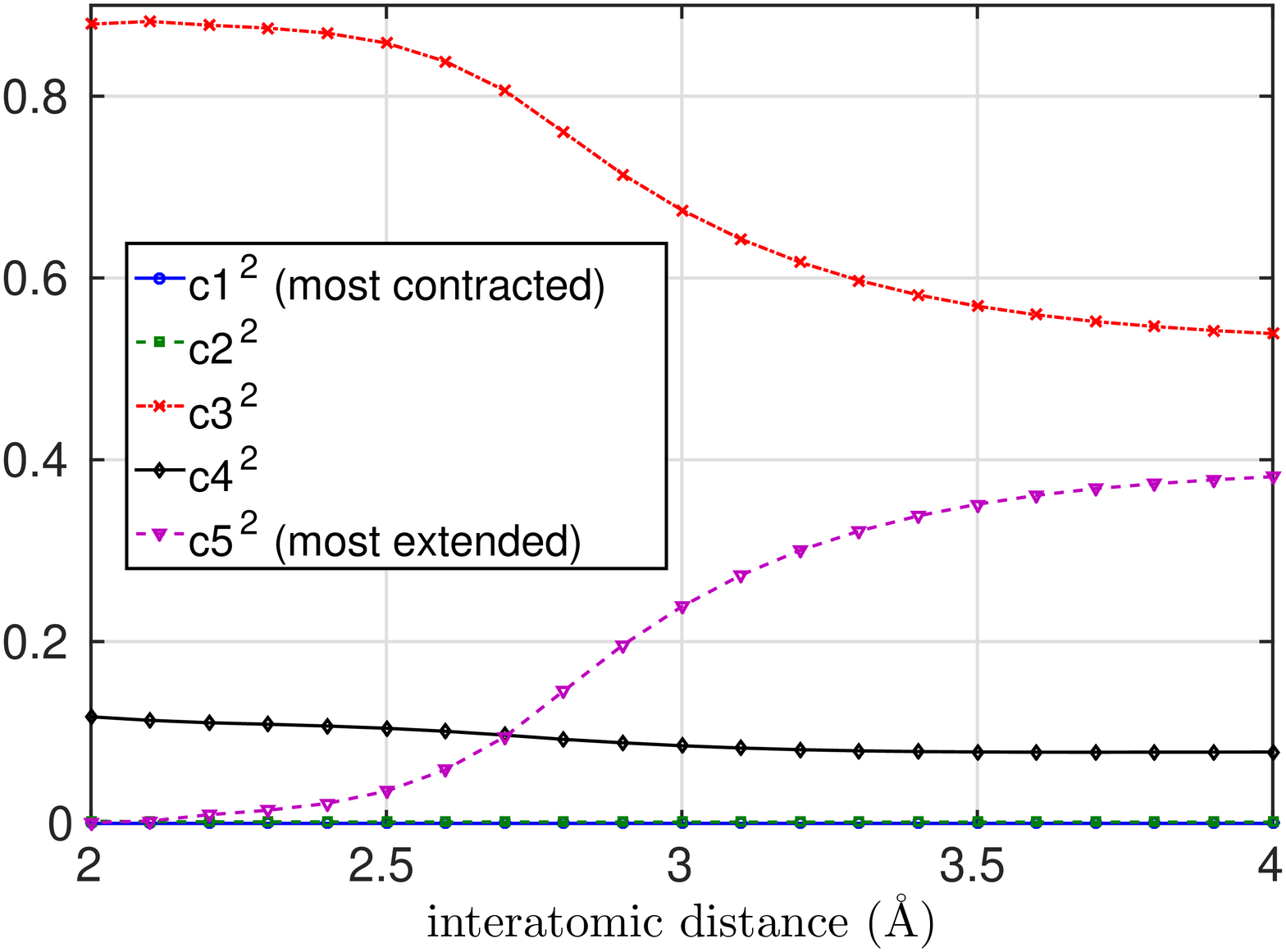}
  \caption{\label{shrinking} Composition of the $p$ atomic orbital in
    $pp\sigma$ bonding MO in arsenic trimer as a function of
    interatomic separation, from ORCA aug-cc-pvtz all electron HF
    calculations.}
\end{figure}

Fig.~\ref{shrinking} shows the makeup of a $p_{z}$ atomic orbital in
the $pp\sigma$ bond as a function of the As-As bond length. The
vertical axis shows the square of coefficients $c_{1}$-$c_{5}$, 1
representing the least extended and 5 representing the most extended
Gaussian function. As the ion cores approach each other, the
wavefunction becomes less diffuse; $c_{3}$ grows while $c_{5}$
shrinks.

In addition, MOPAC's default parametrization of the hard core
repulsion between the ionic cores may not be fully adequate in the
broad density range of interest. (Within the MOPAC PM6 method, the
Pauli repulsion between closed shells is approximated with Voityuk's
diatomic expression\cite{VR104}.)

\begin{figure}[h]
  \centering
  \includegraphics[width = \figurewidth]{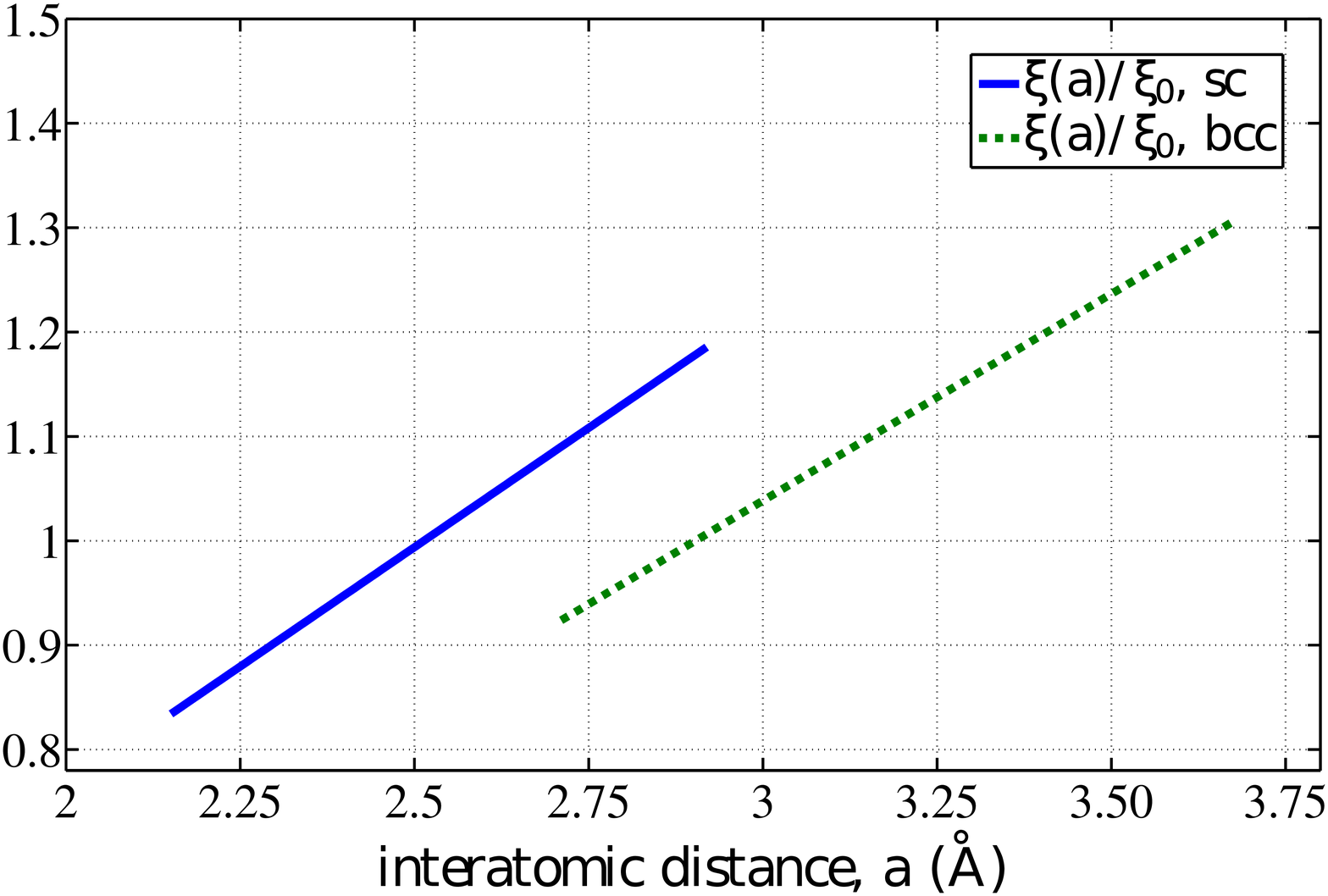}
  \caption{Orbital exponent scaling function behavior for simple cubic
    and body centered cubic lattices, illustrated using the ratio of
    the scaled exponent $\xi(a)$ to the default PM6 value
    $\xi_{0}$\label{fz}}

\end{figure}

\begin{figure}[h]
  \centering
  \includegraphics[width = \figurewidth]{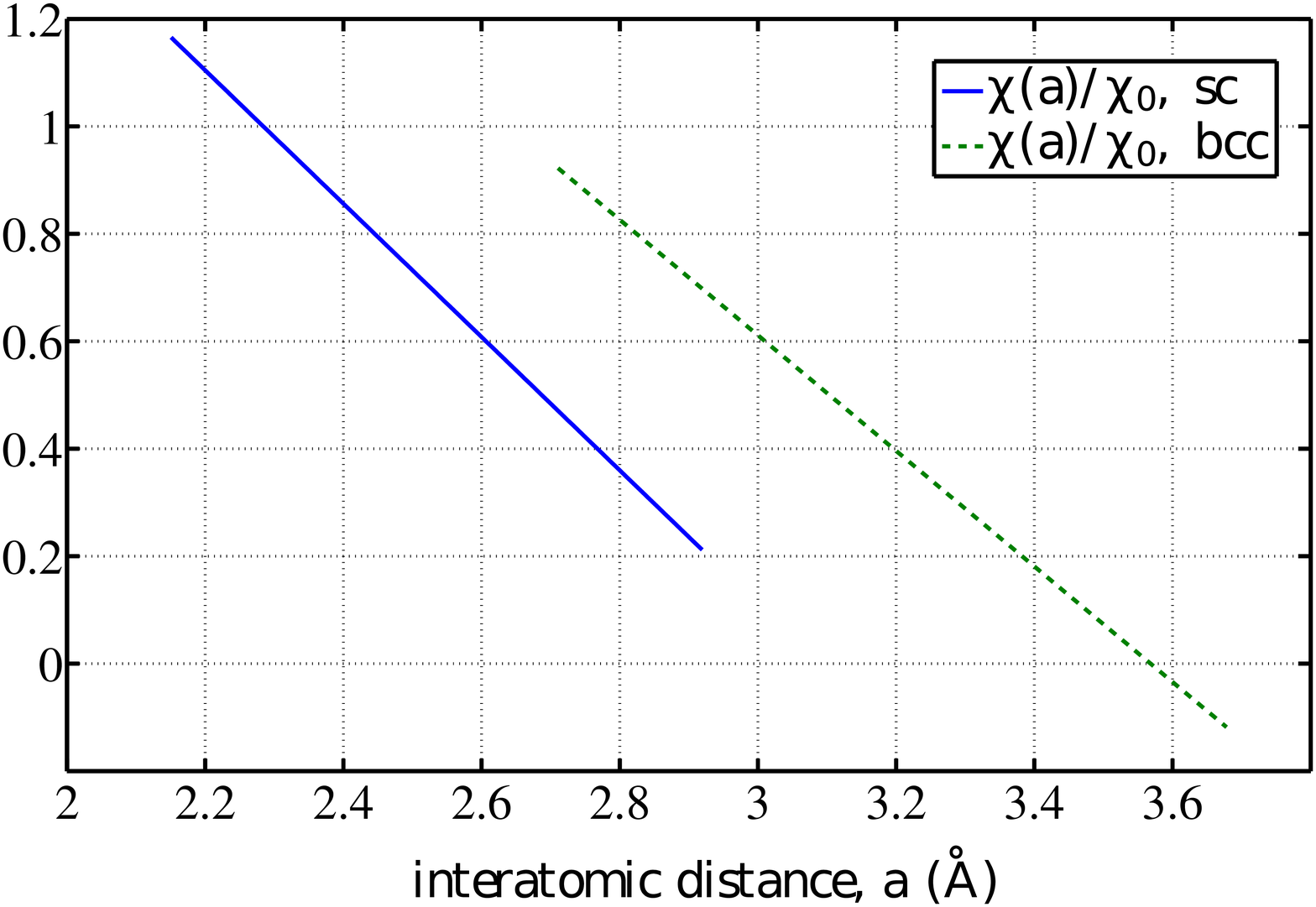}
  \caption{Linear core-core repulsion pseudo-potential scaling
    functions for simple cubic and body centered cubic
    lattices.\label{fx}}
\end{figure}

\begin{figure}[h]
  \centering
  \includegraphics[width = \figurewidth]{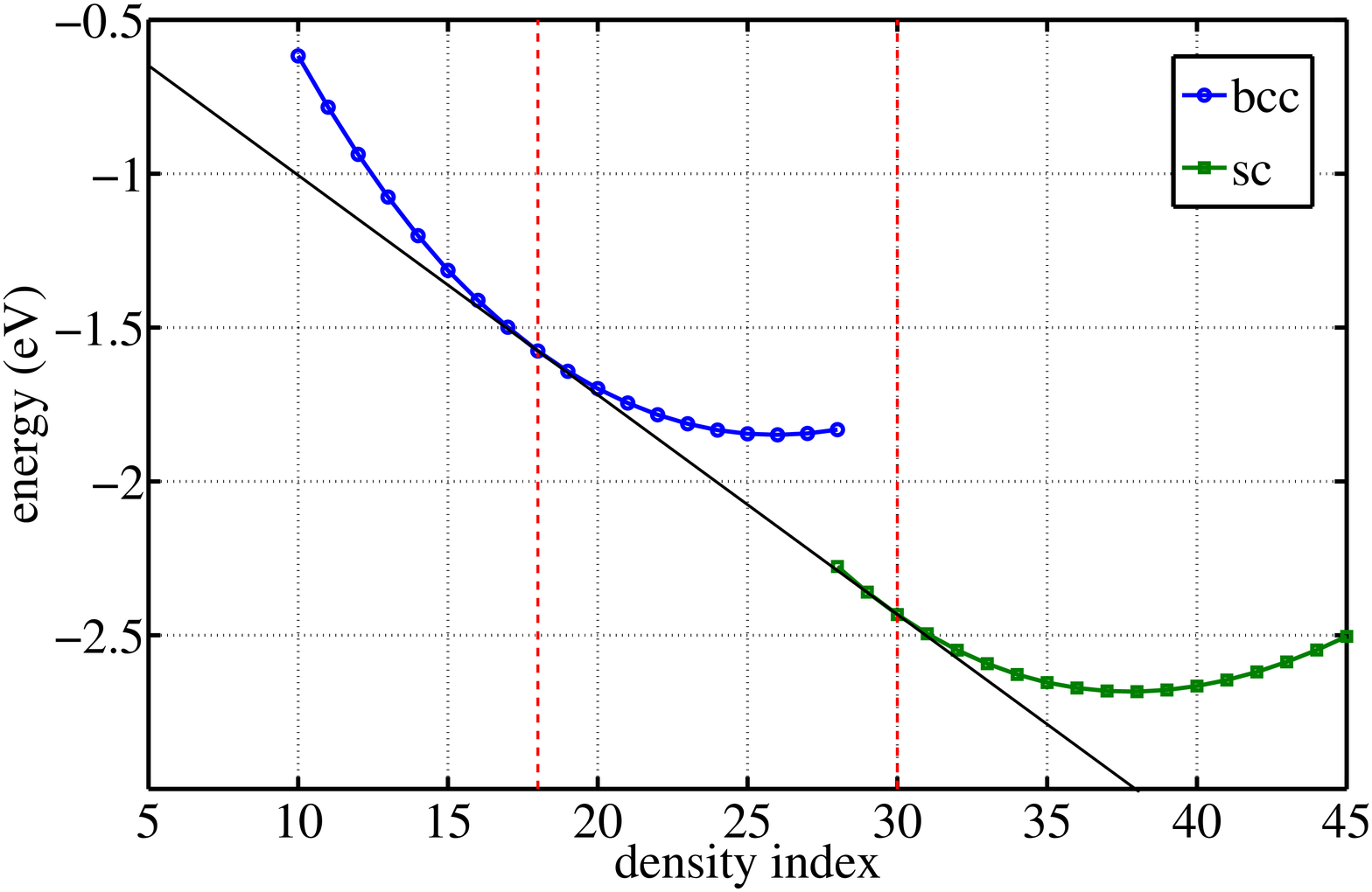}
  \caption{Common tangent construction of the first order simple cubic
    to bcc phase transition for the pseudo-arsenic material at
    $\epsilon$ = 0. The density index shown on the horizontal axis is
    an internal variable used in the PM6 calculations to set a common
    density in both the simple cubic and bcc lattices, the values of
    $18$ and $30$ marked with vertical dashed lines correspond to the
    experimentally observed boundaries of the first order
    transition.\label{pt}}
\end{figure}

The above notions are consistent with our findings that MOPAC's
built-in parametrization does not correctly locate the transition
densities for the sc-to-bcc transition in elemental arsenic
(sc='simple cubic', bcc='body-centered cubic'). As a stopgap solution,
we have introduced additional parametrization to both the orbital
exponent $\xi$ of the wavefunction overlap and the parameter $\chi$
which is the overall scaling factor for the core-core repulsion. The
resulting parametrization of $\xi$ and $\chi$ as functions of the
lattice constant $a$ is shown in Figs.~\ref{fz} and \ref{fx},
respectively.  Using the common tangent method, Fig.~\ref{pt}, we
confirm a first order simple cubic to body centered cubic transition
at 97~GPa and the experimental volume fraction $V_T/V_0$ reported in
high pressure xray crystallography studies. The equation of state for
bcc-As can be found in Ref.~\cite{PhysRevB.51.597} Note there is a
coexistence region between the simple cubic and bcc phases where the
structure is thought to be incommensurate.~\cite{PhysRevB.77.024109}
In the absence of information on the pressure dependence of the
density within the inter-transition region, we use a simple linear
form of the scaling functions in that region. By the same token, the
above re-parametrization is consistent with a continuous transition
between rhombohedral and simple cubic at 25 Gpa, while the
experimental $V_T/V_0$ is also recovered.~\cite{PhysRevB.41.5535}

\newpage
 
\newpage

\section{Hysteretic behavior of the energy as a function of the
  electronegativity variation}

\begin{figure}[h]
  \centering
  \includegraphics[width =
  \figurewidth]{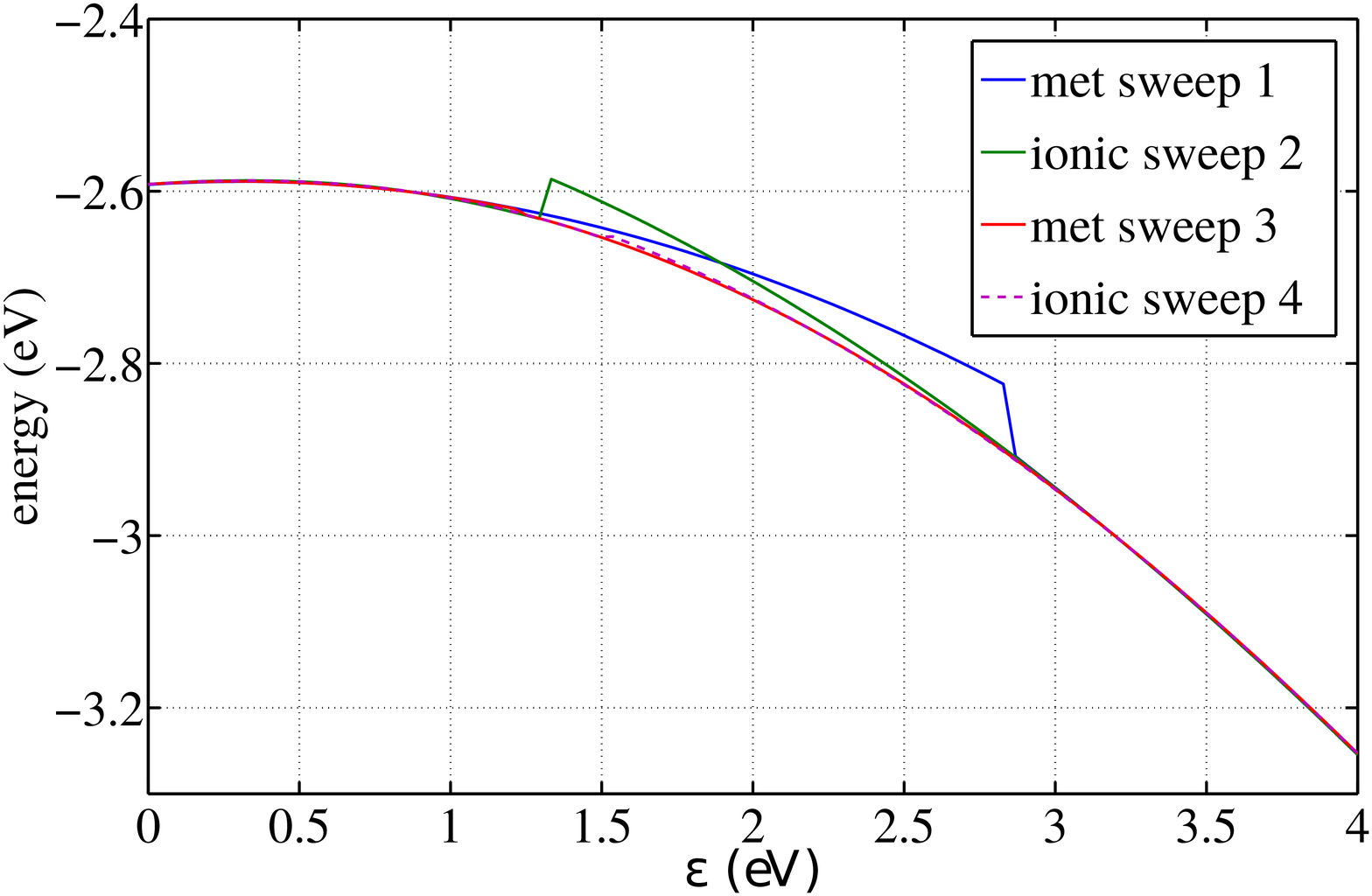}
  \caption{\label{busy_plot} A sample of the ``sweep'' protocol
    demonstrating an overshoot of a term crossing. Labels ``met'' and
    ``ionic'' correspond with sweeps that begin at the $\epsilon = 0$
    and $\epsilon = 4$~eV ends of the $\epsilon$ range.}
\end{figure}

We have found that the self-consistent Hartree Fock procedure does not
always converge to the variational bound on the true ground state that
is actually accessible to MOPAC. Given this uncertainty, we have
performed many $\epsilon$ ``sweeps,'' whereby the electronegativity
differential $2 \epsilon$ is cycled between values of 0 and 8 eV.  The
result of this protocol is exemplified in Fig.~\ref{busy_plot}. The
cycling was terminated when two consecutive sweeps back and forth did
not produce a jump to a lower yet term.

\newpage

\section{Stability of the simple cubic structure to rhombohedral
  distortion}

\begin{figure}[h]
  \centering
  \includegraphics[width = \figurewidth]{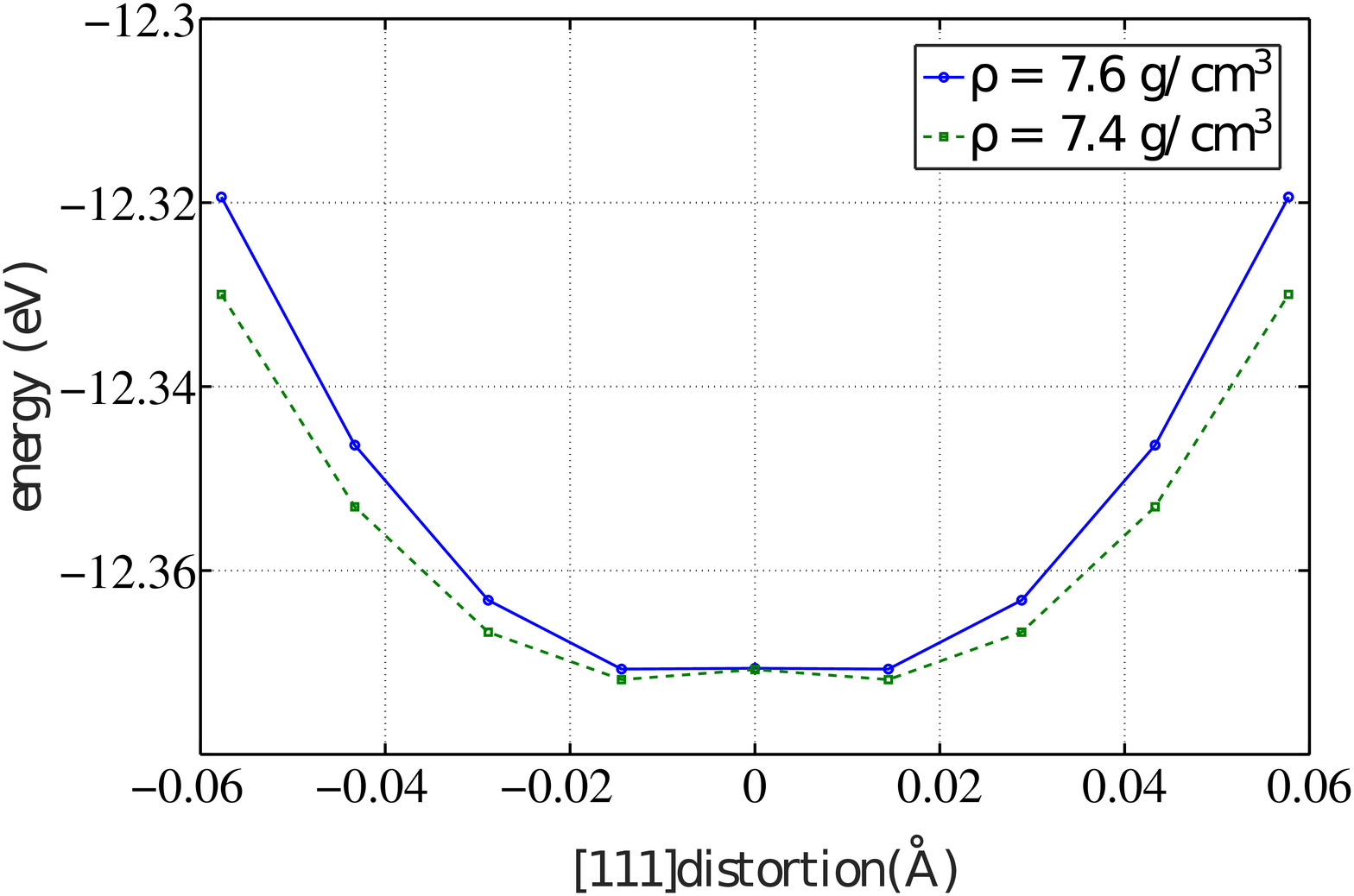}
  \caption{Onset of susceptibility to small distortion towards
    rhombohedral structure, as probed by relative shift of the
    cationic and anionic sublattices comprising the rock-salt
    structure of the pseudo di-pnictogen
    compound.\label{cubic_i51_k1}}
\end{figure}

Fig.~\ref{cubic_i51_k1} shows the energy of a 64 atom pseudo-arsenic
solid as a function of the magnitude of a distortion toward the
rhombohedral lattice, as in Fig.~12 of the main text. Hereby, we shift
the anionic and cationic sublattice relative to each other along the
[111] direction. At the higher of the two densities shown in
Fig.~\ref{cubic_i51_k1}, the presence of a central minimum indicates
that the distortion is destabilizing. This is approximately the
density where the rhombohedral-to-simple cubic transition is observed
experimentally.~\cite{PhysRevB.41.5535}

\section{Decrease in $sp$-mixing in rhombohedral arsenic with density}

To examine $sp$-mixing in rhombohedral arsenic we have used the
projected crystal orbital Hamilton population (pCOHP) analysis tool
Lobster~\cite{doi:10.1021/j100135a014,doi:10.1021/jp202489s,JCC:JCC23424} to extract chemical bonding
information from the results of plane-wave calculations.  pCOHP is a
close relative of the crystal orbital overlap population (COOP)
analysis, originally known as overlap population density of states
(OPDOS).

\begin{figure}[h]
  \centering
  \includegraphics[width = \figurewidth]{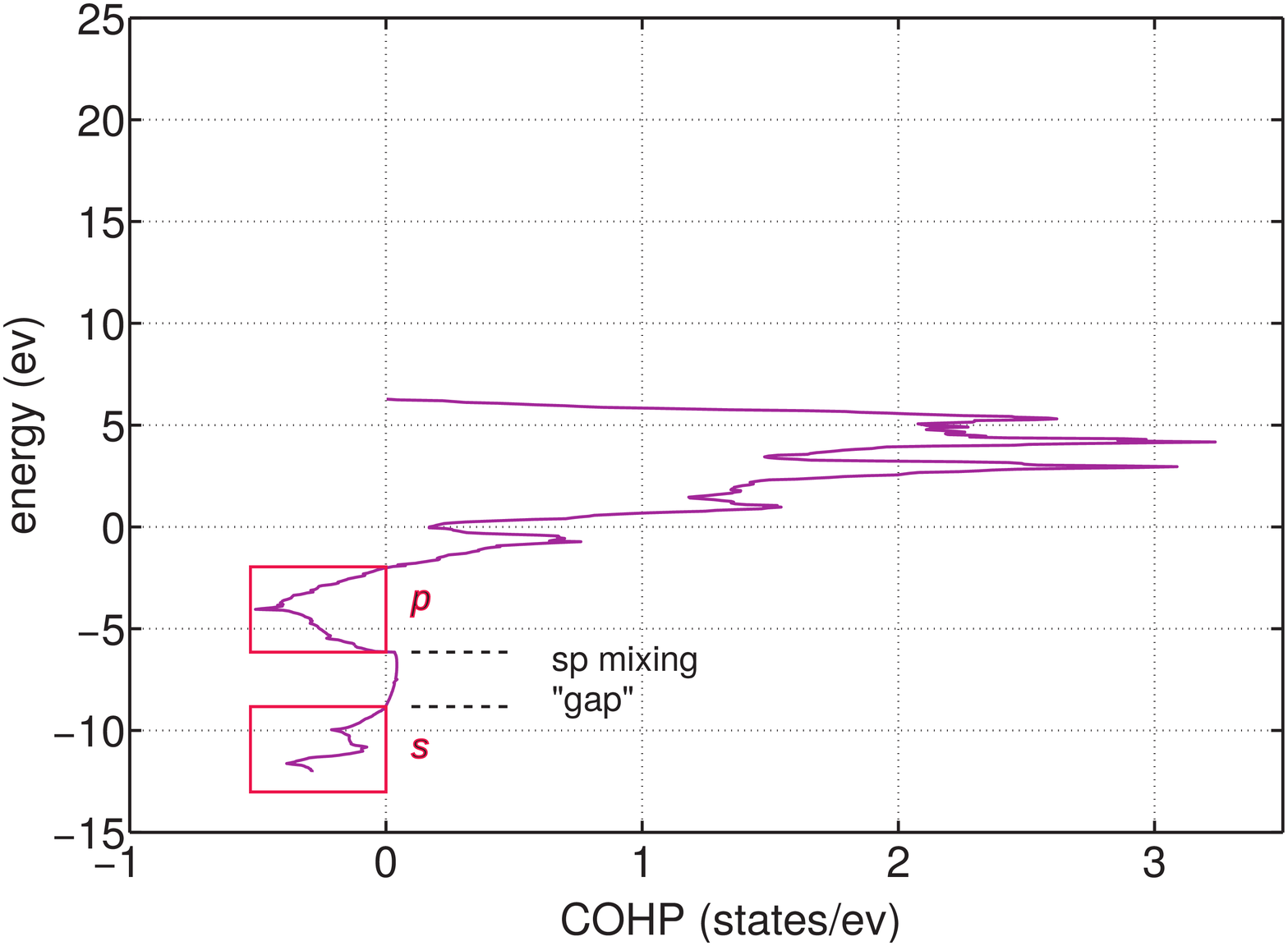}
  \caption{The dashed line indicate a gap between portions of the
    electronic density of states primarily associated with bonding $s$
    and $p$ electrons, in rhombohedral arsenic at normal
    conditions. The red boxes show $s$ and $p$ contributions to the
    density of bonding states.\label{As_COHP}}
\end{figure}

Density functional theory (DFT) calculations were performed in the
generalized gradient approximation (GGA) parameterization of Perdew,
Burke and Emzerhof~\cite{PBE38,CA45} using the Vienna ab-initio
Simulation Package (VASP) version 5.3~\cite{PhysRevB.47.558,KRESSE199615,PhysRevB.59.1758} with
Blochl's projector augmented-wave (PAW) method employed for the core
states.~\cite{PEB50} In the rhombohedral structure at ambient
pressure, there is a gap between the $s$ and $p$ contributions to the
overall density of bonding states, as shown in Fig.~\ref{As_COHP}.

We next compress the unit cell while simultaneously reducing the
rhombohedral distortion until we reach the density where the phase
transition to simple cubic is observed experimentally. The width of
the gap region shown in Fig.\ref{As_COHP} shrinks with density before
vanishing at the transition. We interpret this shrinking as a
decreased mixing between the $s$ and $p$ bonding sub-systems.

\begin{figure}[h]
  \centering
  \includegraphics[width = \figurewidth]{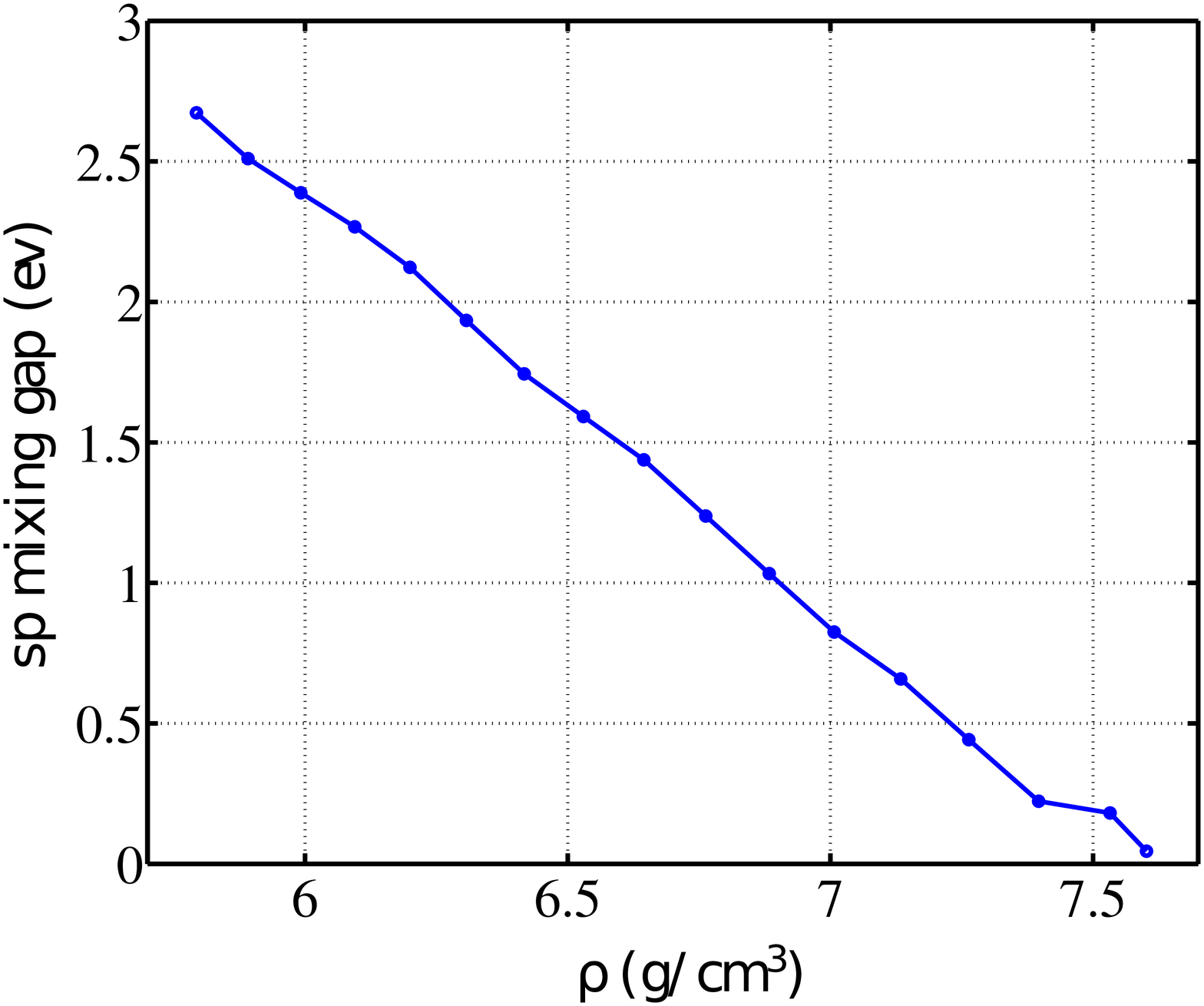}
  \caption{The dependence of the $sp$ from Fig.~\ref{As_COHP} as the
    sample is compressed and deformed toward the simple cubic
    structure.\label{sp_mixing_compressed}}
\end{figure}

\newpage

\section{Ab-initio verification of electronic symmetry breaking for
  symmetric AsH$_2$-AsH-AsH$_2$ trimer}

To verify the apparent change in the bond characteristics, as implied
by the MOPAC-based, LMO analysis, we have obtained variational HF
solutions for the symmetric 3/4 trimer AsH$_2$-AsH-AsH$_2$ using the
6-311G basis within Gaussian 09.\cite{g09} The latter Quantum Chemistry
package provides an alternative interpretational tool to analyse
bonding, viz., via the NBO 6.0 Natural Bond Orbital (NBO).~\cite{NBO6}
Natural Bond Orbitals (NBOs) are an orthonormal set of localized
orbitals in the form of core orbitals (CR) bonding orbitals (BD), lone
pairs (LP) and multi-center orbitals (nC).  The first three orbital
types are Lewis-type orbitals. The procedure attempts to maximize the
occupancy of these Lewis orbitals, the remaining electronic density
assigned to non-Lewis, multi-center orbitals. 

\begin{figure}[b]
  \centering
  \includegraphics[width=.9 \figurewidth]{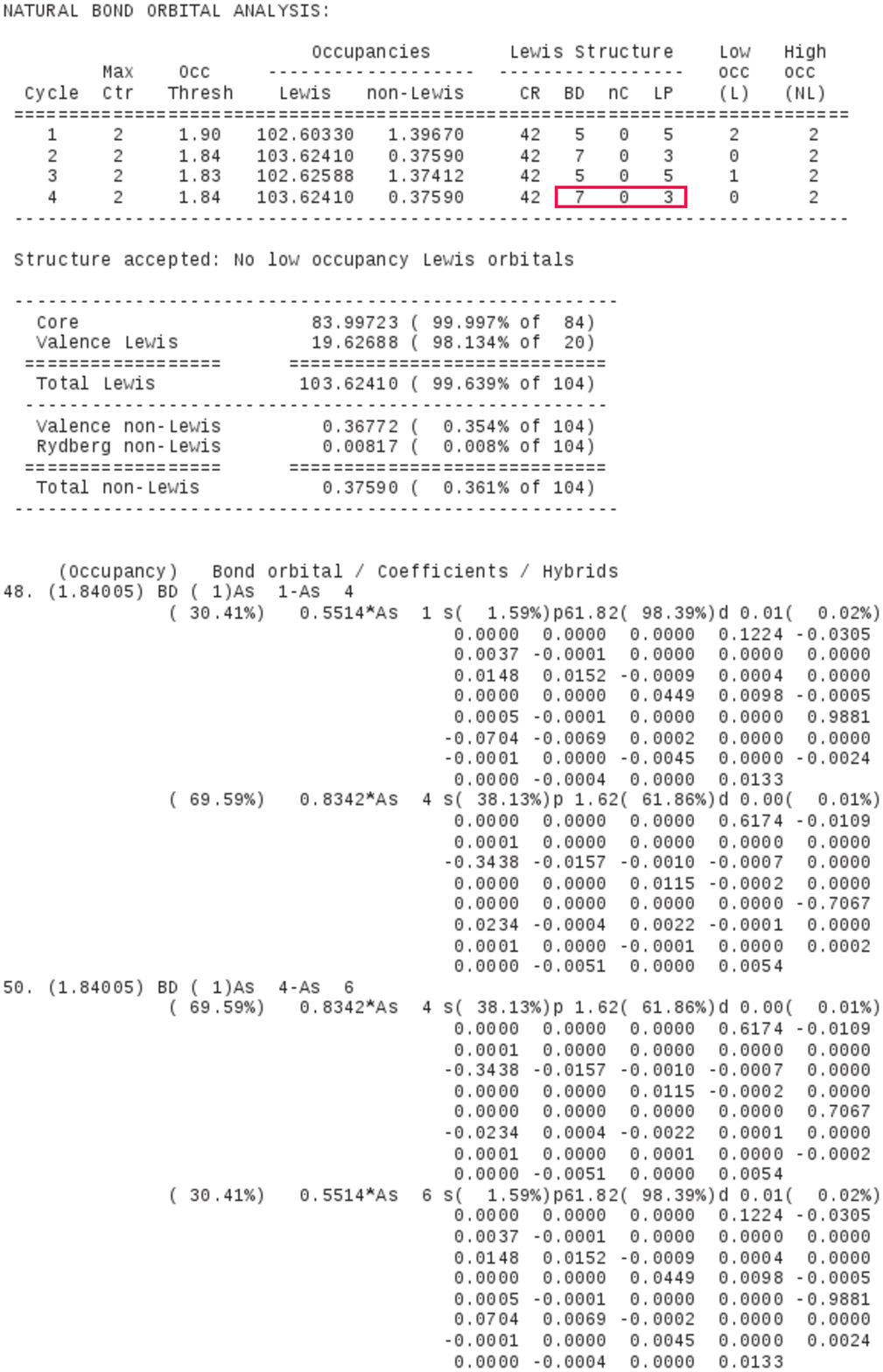}
  \caption{Fragment of the output of the NBO6 analysis within
    Gaussian09 for spatially-symmetric trimer at
    $b=3.2$~\AA.\label{320A_NBO6}}
\end{figure}

\begin{figure}[h]
  \centering
  \includegraphics[width=\figurewidth]{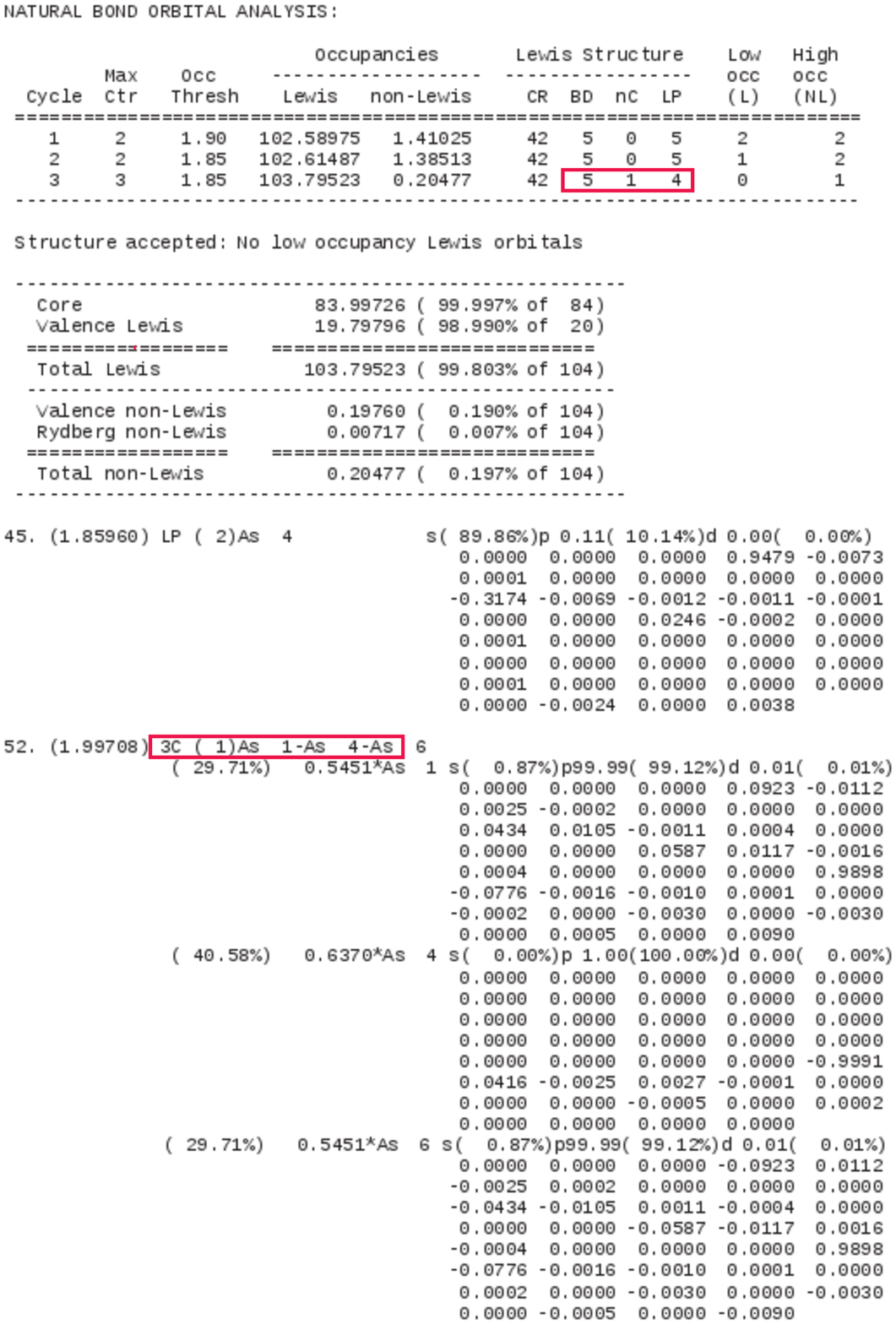}
  \caption{Fragment of the output of the NBO6 analysis within
    Gaussian09 for spatially-symmetric trimer at
    $b=3.3$~\AA.\label{330A_NBO6}}
\end{figure}

Similarly to the LMO-based conclusions in the main text, we have
detected a qualitative change in the NBOs by which a three-center
orbital and associated lone pair emerge for sufficiently long
trimers. Fragments of the corresponding output files just before and
after the transition are provided in Figs.~\ref{320A_NBO6} and
\ref{330A_NBO6}, respectively. For the reader's convenience, we framed
the results of the final iteration of the NBO procedure, which
indicate that following the transition, the number of two-center bonds
decreased by 2 while a three-center bond and a lone pair appeared. See
also the framed entry in Fig.~\ref{330A_NBO6} for orbital 52. The
configurations in Figs.~\ref{320A_NBO6} and \ref{330A_NBO6} correspond
to the As-As bond lengths $b = 3.2$ and $3.3$~\AA, respectively, which
is less than but reasonably close to MOPAC's figure $b \approx
3.6$~\AA. Importantly, this figure is still much longer than what is
expected for the covalent As-As bond, i.e., 2.4~\AA~or so.

\newpage

\section{Walsh diagram of the molecular terms of the trimer
  AsH$_2$-AsH-AsH$_2$ as functions of the trimer length}

\begin{figure}[H]
  \centering
  \includegraphics[width = \figurewidth]{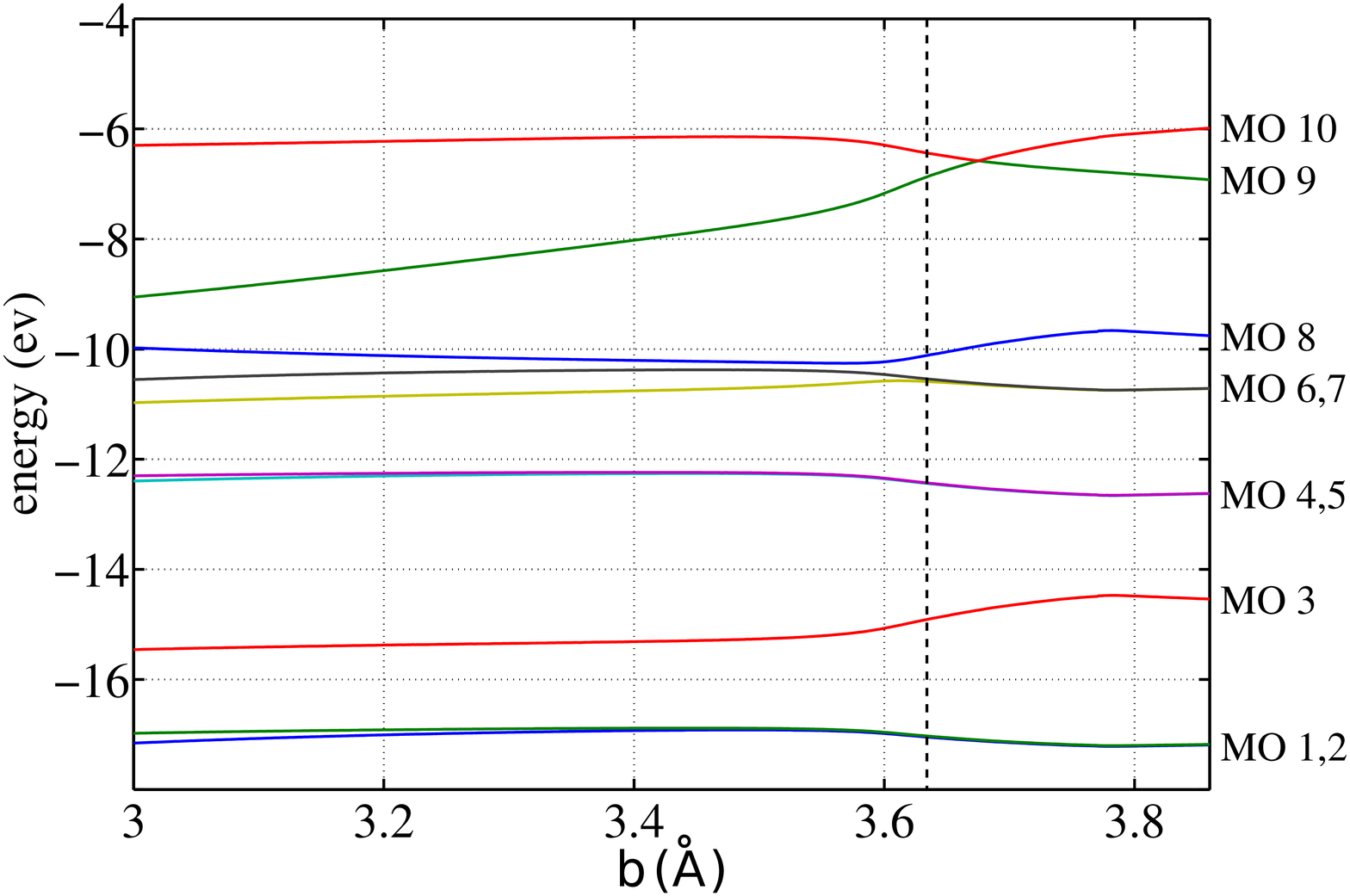}
  \caption{MOPAC-derived molecular orbital terms for the
    AsH$_2$-AsH-AsH$_2$ trimer as functions of the As-As bond length
    $b$. The vertical dashed line indicates the bond length at which
    the LMO transition takes place.\label{3/4 MO Walsh}}
\end{figure}

%\newpage

\section{Bonding from the viewpoint of localized molecular orbitals}

\label{LMO}

We will characterize bonding using the localized molecular orbital
(LMO) formalism. Originally inspired by symmetry adapted orbitals in
relatively symmetric molecules, the localized molecular orbitals can
be assigned for arbitrary geometries: Given a set of {\em occupied}
molecular orbitals $\psi_i$, one transforms to an alternative set of
orthonormal orbitals $\chi_j$ such that a certain quantity reflecting
the self-repulsion within individual new orbitals is
maximized.\cite{SB32, ER35, VN56, PS78} Specifically, here we adhere
to the Edmiston and Ruedenberg approach~\cite{ER35}, whereby one
maximizes the quantity $D_\chi \equiv \sum_j \int d^3 \br_1 d^3 \br_2
|\chi_j(\br_1)|^2 V(r_{12}) |\chi_j(\br_2)|^2$ or, equivalently,
minimizes the mutual repulsion among distinct orbitals in the form of
the quantity: $\sum_{j<k} \int d^3 \br_1 d^3 \br_2 |\chi_j(\br_1)|^2
V(r_{12}) |\chi_k(\br_2)|^2$. The function $V(r_{12})$ stands for a
purely repulsive potential energy function.  The LMOs are generally
not eigenfunctions of an effective one-particle Hamiltonian, such as
that arising in the Hartree-Fock (HF) approximation. Instead, they
represent an attempt by an interpreter to redistribute (the already
bound) electrons among orthonormal, maximally localized orbitals, each
of which thus binds together the smallest number of atoms. The latter
number serves as a lower bound on the number of centers in a bond; the
bound is not unique because it depends on the specific choice of the
% (The quantity $\sum_{mn} \int d^3 \br_1 d^3 \br_2 |\chi_m(\br_1)|^2
% V(r_{12}) |\chi_n(\br_2)|^2$ is independent of the basis choice, of
% course.)
test function $V(r_{12})$, which is made according to one's
convenience and computational means. For instance, $V(r_{12}) =
r_{12}^{-1}$ implies localization with respect to (non-screened)
Coulomb repulsion. A more computationally convenient choice is
$V(r_{12}) = \delta(\br_{12})$, which simply yields $D_\chi = \sum_j
\int d^3 \br |\chi_j(\br)|^4$.

\begin{figure}[t]
  \centering
  \includegraphics[width = .95 \figurewidth]{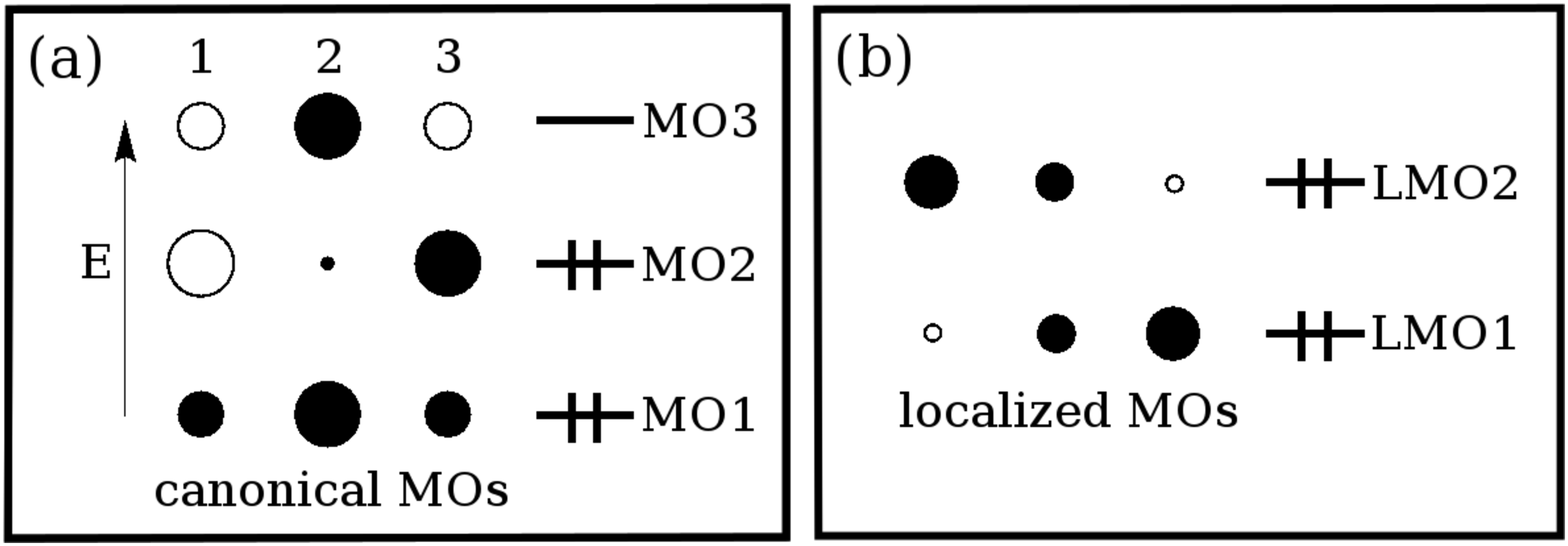}
  \caption{\label{H3} {\bf (a)} Graphical representation of the
    eigenvectors of the three-orbital, H\"{u}ckel energy function
    (\ref{Htrimer}). {\bf (b)} The localized molecular orbitals (LMOs)
    corresponding with panel (a). Note the two orbitals have the same
    energy, $(E_1 + E_2)/2$.}
\end{figure}

The above procedure is significantly simplified when the one-electron
energy function is diagonalized while neglecting the overlap between
distinct atomic orbitals, as is the case for certain semi-empirical
approximations.~\cite{PS78} Hereby, one can present the integral $\int
d^3 \br |\chi_j(\br)|^4$ as a sum over distinct atoms: $\sum_A \int_A
d^3 \br |\chi_j(\br)|^4$, where the individual integrations are over
regions occupied by distinct atoms. Each such integration yields a
quantity that scales inversely proportionally with the atom's
volume. Insofar as we are interested in the partitioning of the
electrons between atoms, irrespective of the extent of the atomic
orbitals, the above sum can be profitably replaced by a
(dimensionless) expression, viz.:
\begin{align} \label{partition1} \int d^3 \br |\chi_n(\br)|^4 &=
  \sum_A \int_A d^3 \br |\chi_n(\br)|^4 \to \\ \label{partition2} n_j
  &\equiv \sum_A \left( \sum_{\lambda_A} C^2_{\lambda_A j} \right)^2,
\end{align}
where we have expanded the localized orbital $\chi_j(\br)$ in terms of
the atomic orbitals $\phi_{\lambda_A}$:
\begin{equation} \label{chiExp} \chi_j(\br) = \sum_A \sum_{\lambda_A}
  C_{\lambda_A j} \phi_{\lambda_A}
\end{equation}
and the index $\lambda_A$ labels the atomic orbitals on atom $A$. As
already mentioned, $\int dV \phi_\lambda \phi_\mu = \delta_{\lambda
  \mu}$, implying $\sum_A \sum_{\lambda_A} C^2_{\lambda_A n} = 1$. The
quantity $n_j$ from Eq.~(\ref{partition2}) clearly provides the
participation number of distinct centers to the localized orbital $j$,
and is appropriately called the {\em bond-center number}.~\cite{PS78}
For a lone pair, one automatically gets $n = 1$. For a two-center
bond, the MO and LMO are equal: $\chi = \psi = \alpha \phi_1 + \beta
\phi_2$, ($\alpha^2 + \beta^2 = 1)$; the bond center number, $n =
[\alpha^{-4} + (1 - \alpha^2)^{-2}]^{-1}$, is maximized at 2 for a
covalent bond, $\alpha = 1/\sqrt{2}$, and is minimized at the lone
pair value 1 for a purely ionic bond, $\alpha = 0, 1$. For a linear
trimer with two equivalent bonds and one orbital per center, one may
use the H\"{u}ckel energy function:
\begin{equation}\label{Htrimer}
  \widehat{H} = - \left(  {\begin{array}{*{20}{c}}
        0 & \beta & 0 \\
        \beta & 0 & \beta\\
        0 & \beta &  0 \end{array}} \right).
\end{equation}
The corresponding MO's are shown graphically in Fig.~\ref{H3}(a).  The
lowest-energy MO is $\psi_1 = \phi_1/2 + \phi_2/\sqrt{2} + \phi_3/2$
and the (formally) non-bonding MO is $\psi_2 = (\phi_1 -
\phi_3)/\sqrt{2}$. If only the {\em ground} state MO contributes to
bonding, one gets $n = 8/3$. This value is less than three, in
reflection of the non-uniform distribution of charge among the three
centers, viz., $25\%$, $50\%$, and $25\%$ respectively.  The partially
ionic character of the three-center bond for a linear trimer made of
otherwise equivalent atoms was noted a while ago by
Musher.~\cite{MusherAngew1969} (For comparison, the bond-center number
for a three-center bond with three equivalent centers, $\chi = \psi =
(\phi_1 + \phi_2 + \phi_3)\sqrt{3}$ exactly equals $3$.) In the case
when MO2 is actually bonding, there are two LMOs. Direct computation
shows $\chi_{1, 2} = (\psi_1 \pm \psi_2)/\sqrt{2}$, the two shown
graphically in Fig.~\ref{H3}(b). The two LMOs are equivalent in that
they are mirror images of each other, up to the sign, and each can be
associated with a one-electron energy $(E_1 + E_2)/2$. (Generally, to
the $j$-th LMO $\chi_j = \sum_{l} c_{jl} \psi_l$, one may associate an
energy $E_j = \sum_{l} c_{jl}^2 E_l$, where the summations are over
the occupied MOs.) The bond number for each of the two LMOs is
$n_{1,2}= 32/19$, which is just short of 2. This (modest) deficit is,
again, due to the non-uniform distribution of charge among the three
centers. Still, insofar as the bond-center numbers for the two LMOs
are numerically close to 2, the 3 center/4 electron bond can be
thought of as two adjacent two-center bonds.

A useful pattern emerges from the above trimer example already at the
H\"{u}ckel level, as in Fig.~\ref{H3}: If the two-center bonds
comprising the three-center bond are equivalent and the two bonding
molecular orbitals have opposite parity, the two resulting LMOs are
necessarily mirror images of each other. (This pattern does not
necessarily hold when the number of MOs exceeds two, whether that
number is even or odd.) If only one LMO is present, on the other hand,
it must be either odd or even with respect to reflection in the
symmetry plane.

As a measure of the bond order for a two-center bond connecting atoms
A and B, one may evaluate the number of electrons shared by these
atoms according to:~\cite{APS627}
\begin{equation} \label{BOdef} B_{AB} \equiv \sum_{\lambda_A \!, \,
    \lambda_B} P^2_{\lambda_A \lambda_B},
\end{equation}
where $P_{\lambda \mu}$ is the density matrix:
\begin{equation} P_{\lambda \mu} \equiv 2 \sum_i
  \widetilde{C}_{\lambda i} \widetilde{C}_{\mu i},
\end{equation}
the summation being over the occupied MOs, and the $\widetilde{C}$s
are the expansion coefficients for the MOs in terms of the atomic
orbitals:
\begin{equation} \label{psiExpB} \psi_i(\br) = \sum_A \sum_{\lambda_A}
  \widetilde{C}_{\lambda_A i} \phi_{\lambda_A}.
\end{equation}
The quantity in Eq.~(\ref{BOdef}) is often called the ``Wiberg bond
index.''\cite{APS627} For a lone pair, $B = 0$ automatically. For
a two-center bond, $\psi = \alpha \phi_1 + \beta \phi_2$, $B_{AB} = 4
\alpha^2 (1-\alpha^2)$; it varies between 0 and 1, the two extremes
corresponding to a purely ionic and covalent bond, respectively. For
the trimer from Eq.~(\ref{Htrimer}), the bond orders are $B_{12} =
B_{12} = 1/2$, and $B_{13} = 1/4$, whether the number of the bonding
MO's is one or two.  We thus conclude that the bond order defined in
Eq.~(\ref{BOdef}) reflects, to some extent, the deviation of the
bonding in Fig.~\ref{H3} from a collection of two-center bonds, since
$1/2+1/2+1/4 = 1.25 > 1$.

To gain additional insight, one may quantify the bonding contribution
of an individual LMO, $j$, by evaluating the expectation value of the
density matrix for the orbital, see ``Localization Theory'' in the
online MOPAC Manual~\cite{MOPACmanual}
\begin{equation} \label{BcontDefb} C^\text{(bond)}_{jj} \equiv \la
  \chi_j | \widehat P | \chi_j \ra = 2 \sum_{\lambda \mu} C_{\lambda
    j} C_{\mu j} P_{\lambda \mu},
\end{equation}
where in the sum above, terms pertaining to the same atom are
excluded.  The bonding contribution for a lone pair is automatically
$0$. For a two-center bond $\chi = \psi = \alpha \phi_1 + \beta
\phi_2$, $C^\text{(bond)} = 8 \alpha^2 (1-\alpha^2)$, which is simply
the Wiberg bond index times two. Likewise, the trimer from
Eq.~(\ref{Htrimer}) in the one-LMO arrangement yields $C^\text{(bond)}
= 2.5$. The two-LMO arrangement leads to $C^\text{(bond)}_{11} =
C^\text{(bond)}_{22} = 1.25$, so that the total bond contribution is,
again, $2.5$. Thus based on the bond index and bond contribution
analysis, the bond order for the H\"uckel trimer is not sensitive to
whether MO2 is regarded as bonding or non-bonding. We shall see that
the two analyses give different results in more complicated
situations.

\newpage

\begin{widetext}
\section{The eventual dissociation of the trimer, from the LMO
  perspective}

\begin{figure}[h]
  \centering
  \includegraphics[width = 1.2
  \figurewidth]{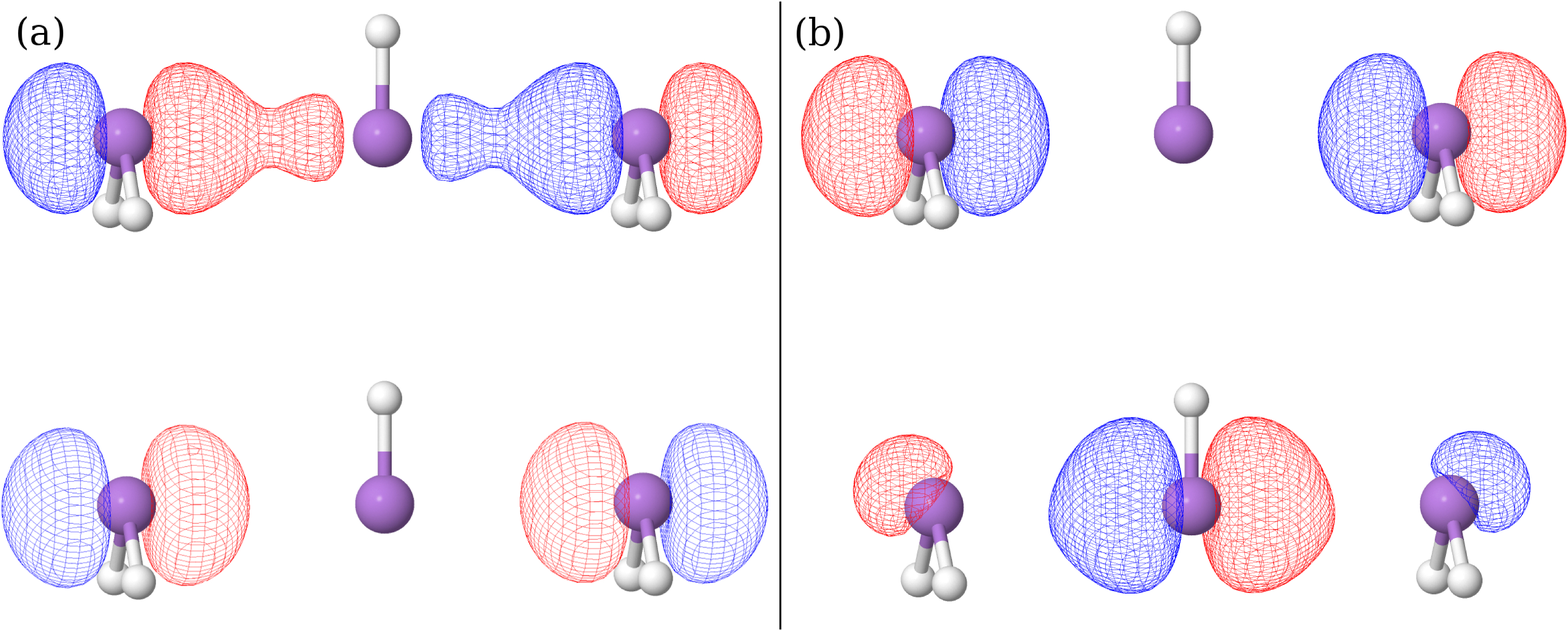}
  \caption{The bonding LMOs are shown for the AsH$_2$-AsH-AsH$_2$ for
    two values of the bond length, (a) $b = 3.74$~\AA~ and (b) $b =
    3.76$~\AA.}
\end{figure}

\section{A realization of a 5-center localized molecular orbital}

\begin{figure}[h]
  \centering

  \includegraphics[width = 1.4\figurewidth]{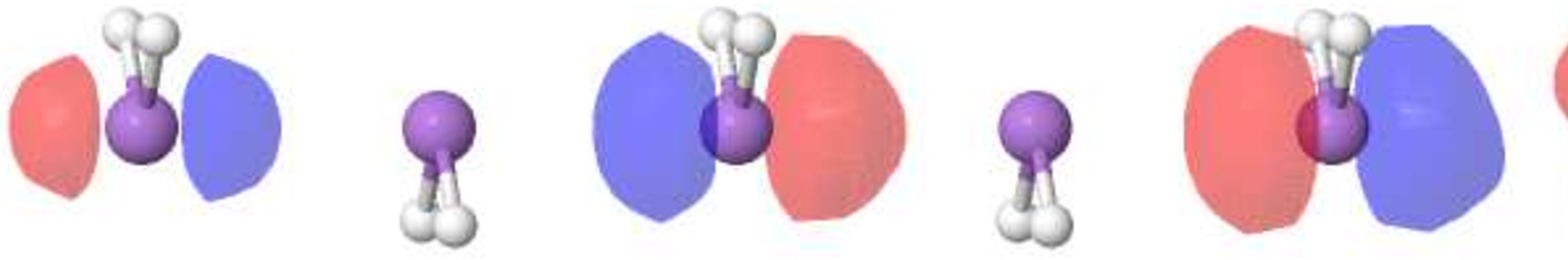}

  \includegraphics[width = 1.4 \figurewidth]{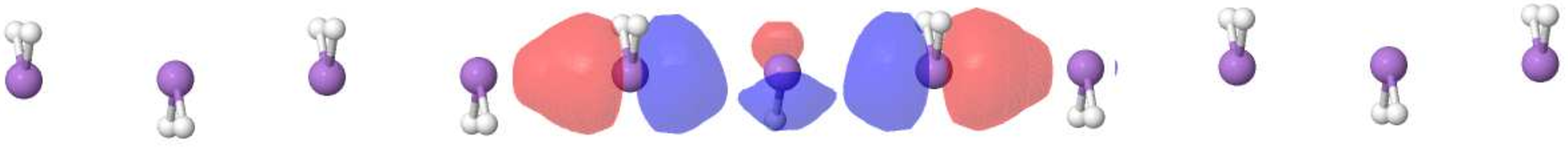}

  \caption{MOPAC-derived $pp\sigma$ molecular orbital (top) and
    corresponding localized molecular orbital (bottom). All nearest
    neighbor As-As bond lengths are equal to $3.39$~\AA. The hydrogens
    are constrained in the same fashion as for the 3/4 trimer in the
    main text.}
\end{figure}
\end{widetext}

\newpage

\section{The 20-member ring hosts a charge-density wave (CDW)}

\begin{figure}[H]
  \centering
  \includegraphics[width = 0.9
  \figurewidth]{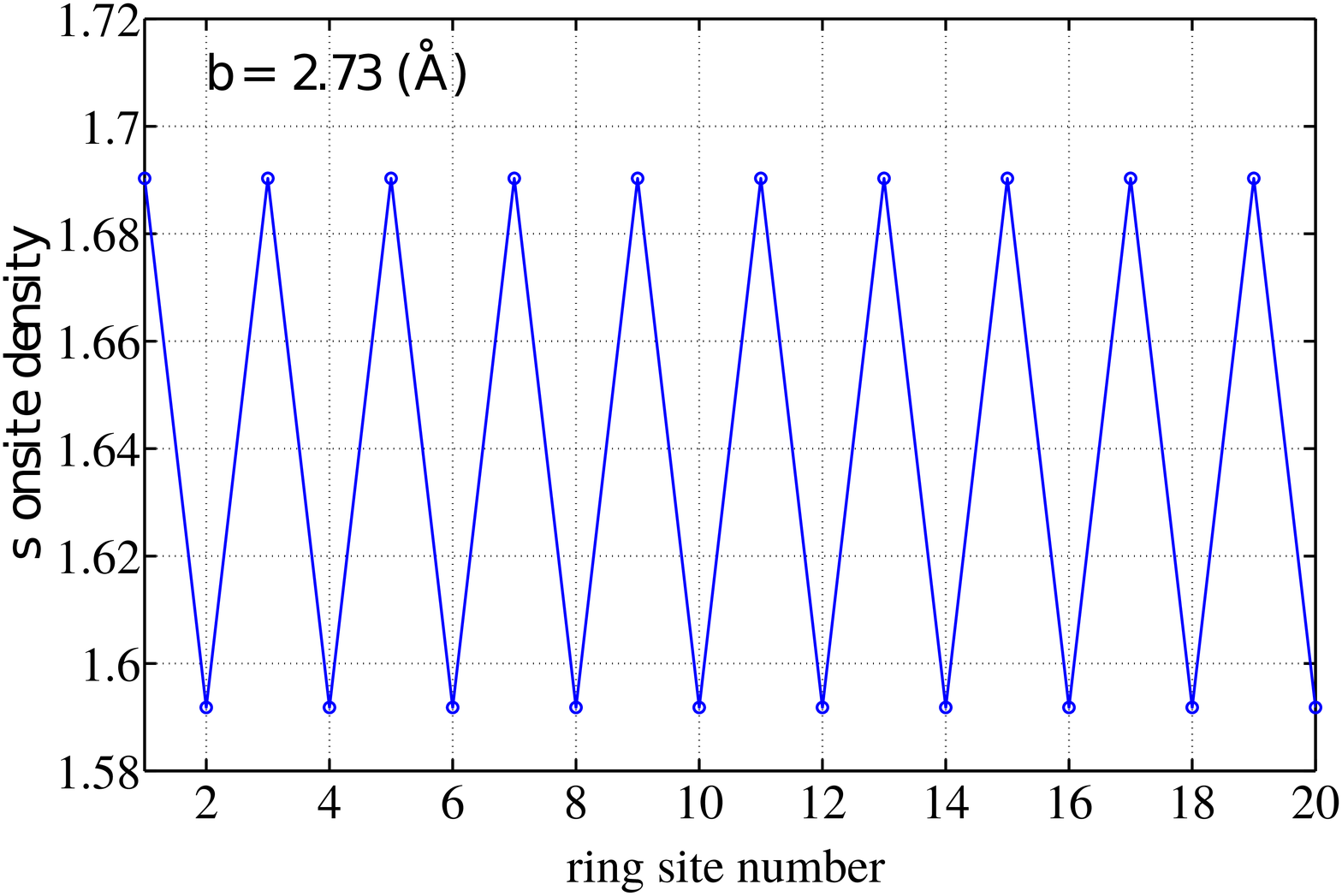}
  \caption{An example of a charge density wave in the 20-member ring
    (AsH$_{2}$)$_{20}$, as inferred from the diagonal entries of the
    density matrix, see main text. The specific bond length,
    $b=2.73$~\AA, corresponds with the ground state configuration of
    the non-dimerized 20 member ring at the ground state configuration
    indicated by the asterisk in Fig.~14 of main text.\label{CDWb}}
\end{figure}

\section{The 20-member ring becomes unstable toward dimerization for
  bond lengths exceeding 2.5~\AA}

\begin{figure}[H]
  \centering
  \includegraphics[width = 0.9
  \figurewidth]{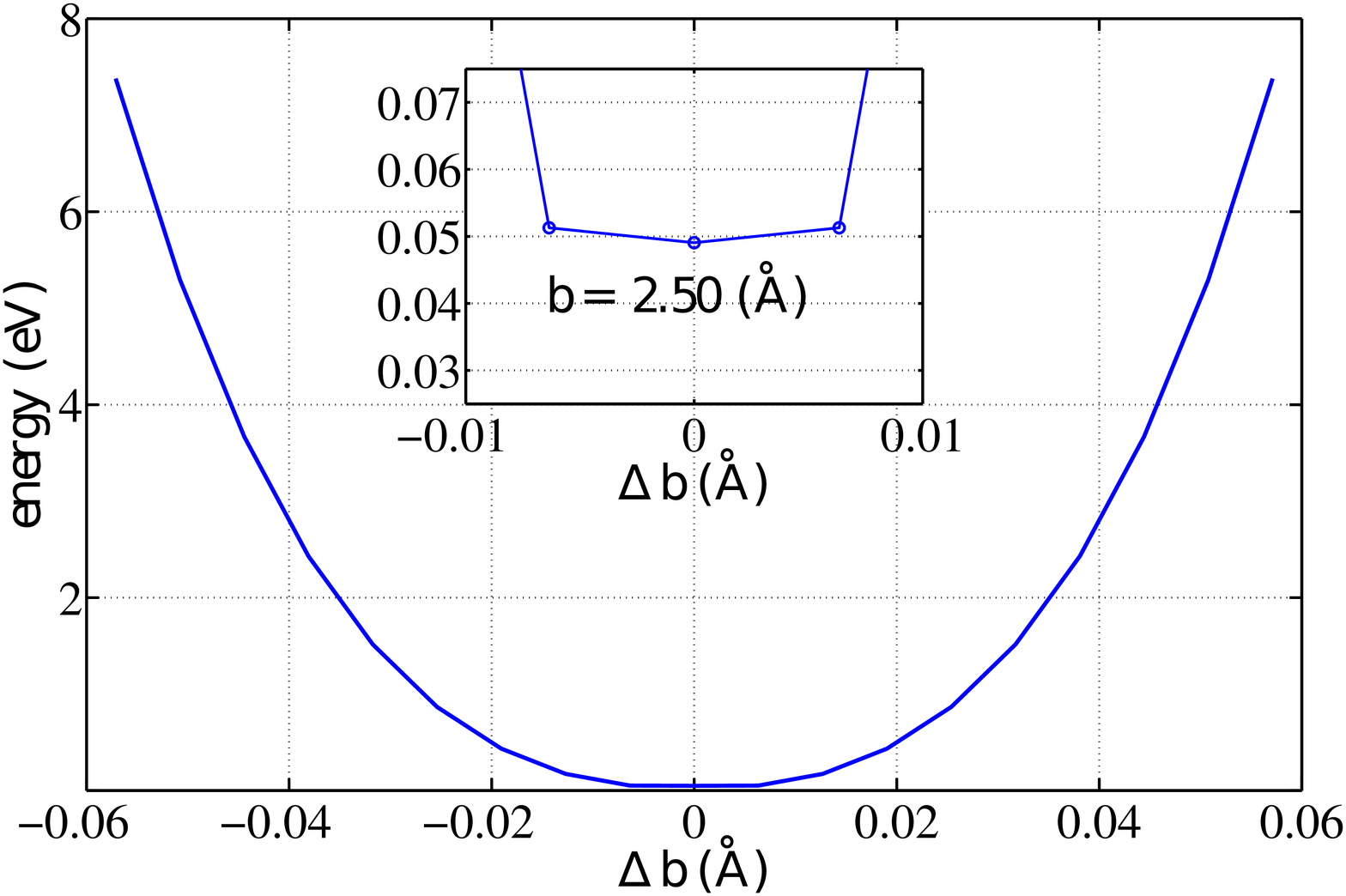}
  \caption{Electronic energy of the 20-member ring (AsH$_{2}$)$_{20}$
    as a function of the dimerization strength. The latter is
    indicated by the length difference between the longer and shorter
    bond. The average bond length is kept steady at
    $2.50$~\AA. \label{stable}}
\end{figure}

\begin{figure}[t]
  \centering
  \includegraphics[width = 0.9
  \figurewidth]{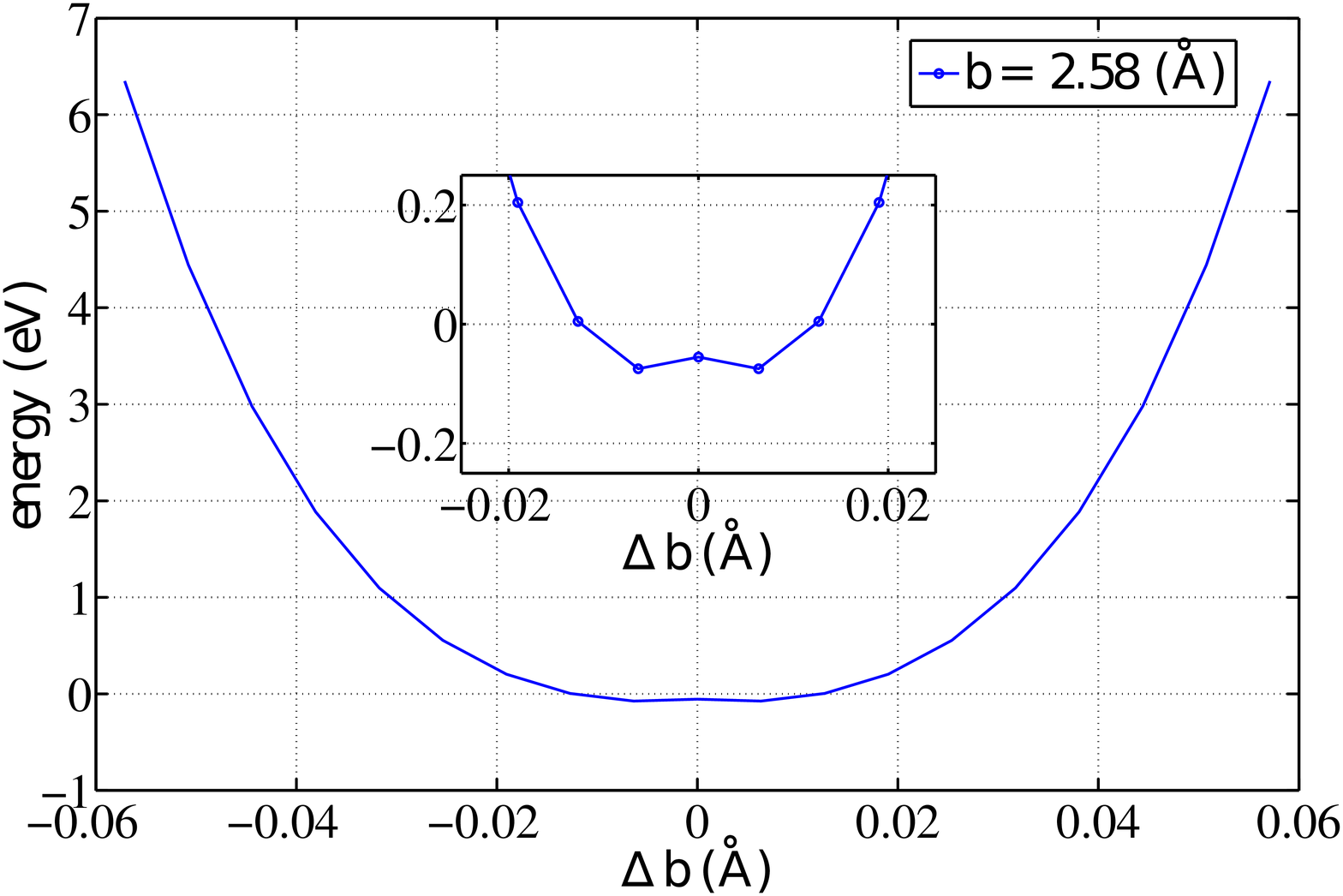}
  \caption{Electronic energy of the 20-member ring (AsH$_{2}$)$_{20}$
    as a function of the dimerization strength. The latter is
    indicated by the length difference between the longer and shorter
    bond. The average bond length is kept steady at
    $2.58$~\AA.  \label{bistable}}
\end{figure}

\begin{widetext}

\section{Additional examples of localized molecular orbitals near the
  ground state length of the non-dimerized 20-member ring}

\begin{figure}[h]
  \centering \subfigure{\includegraphics[width = 0.6
    \figurewidth]{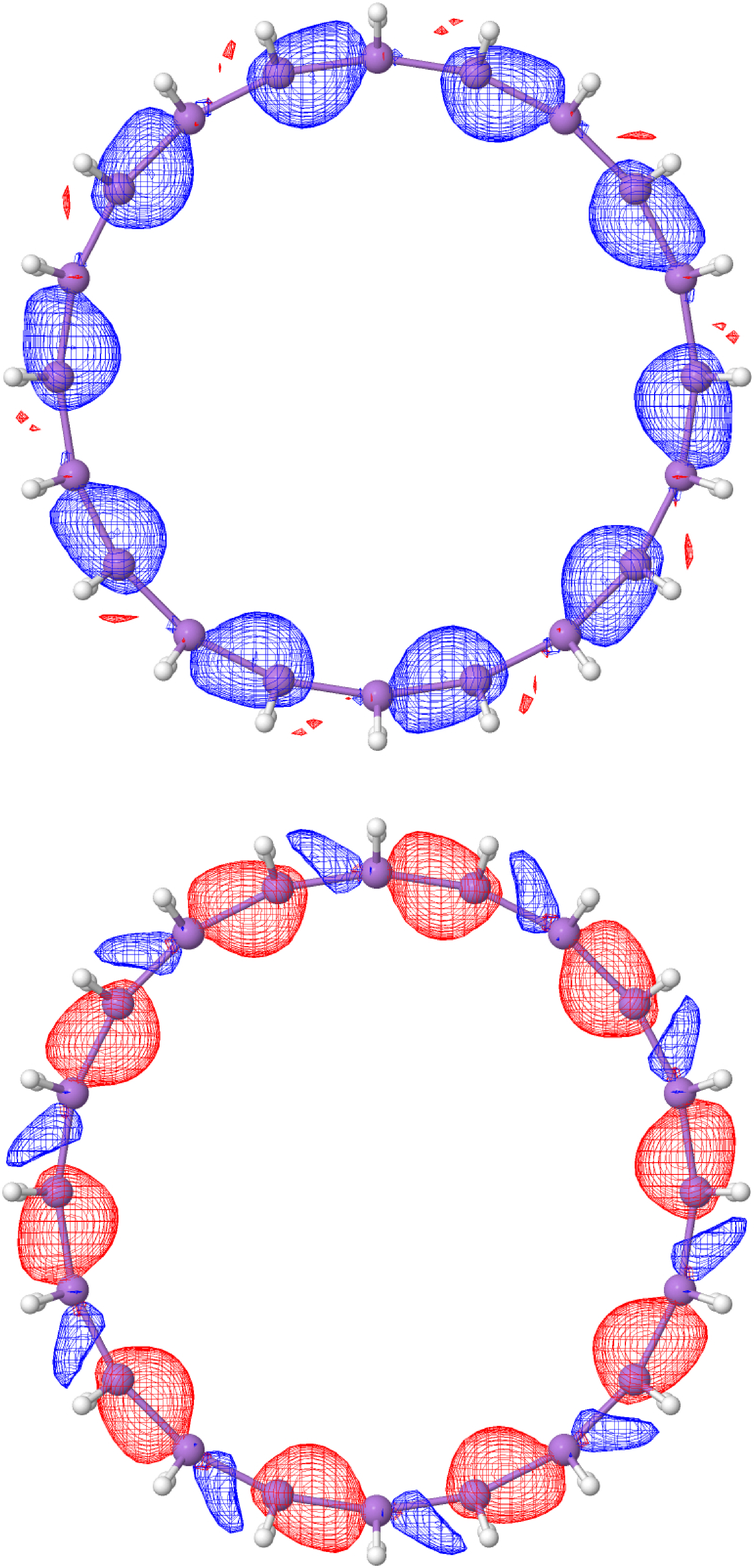}} \hspace{2cm}
  \subfigure{\includegraphics[width = 0.6
  \figurewidth]{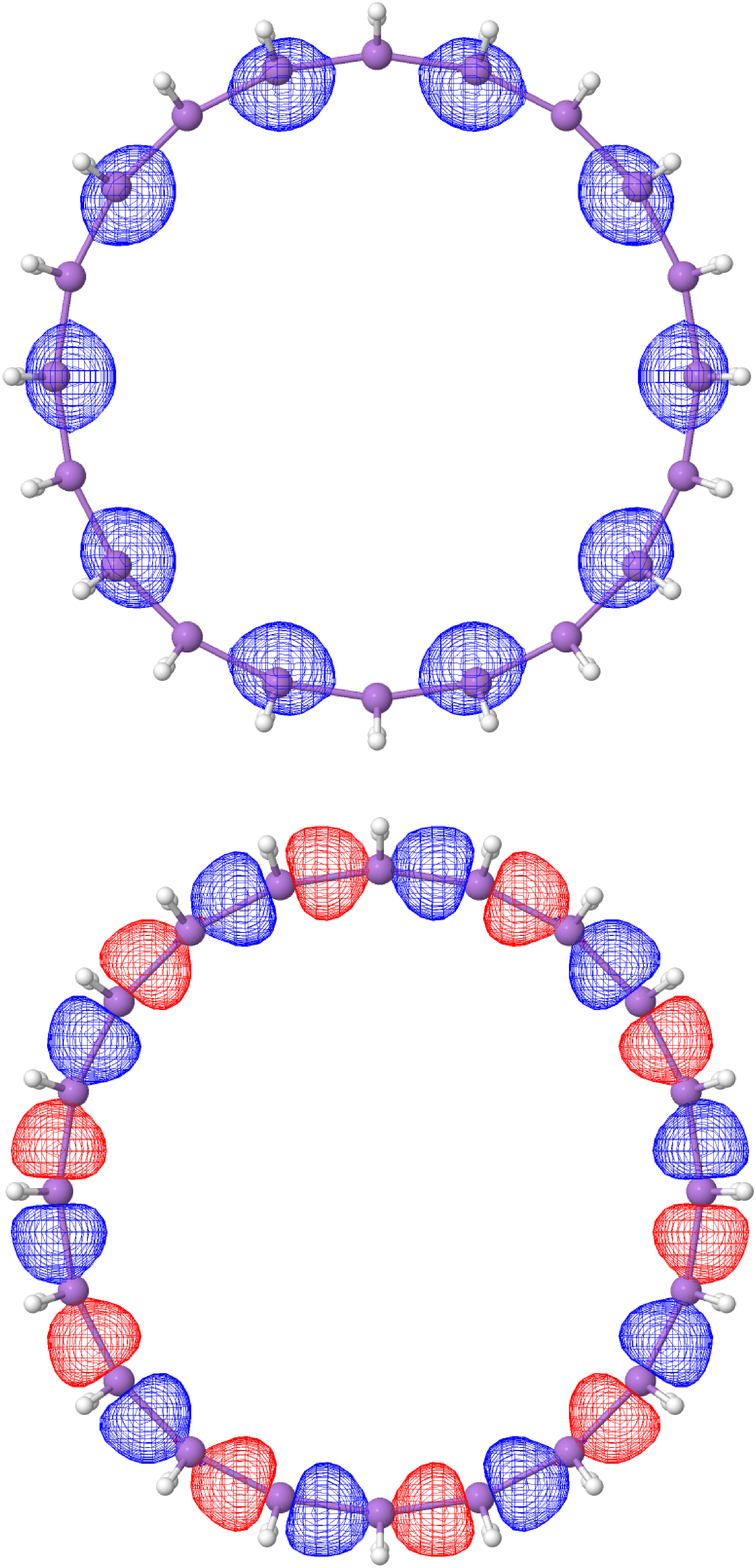}}
\caption{{\em Left}: The LMOs near the mechanical stability limit of
  the 20-member ring (AsH$_2$), a higher energy HF solution than the
  one visualized in Fig.~13 of main text, regime III; $b =
  2.58$~\AA. {\em Right}: The LMOs for the ground state configuration
  of the non-dimerized 20 member ring at the ground state
  configuration indicated by the asterisk in Fig.~14 of main text.}
\end{figure}
\end{widetext}

\section{Molecular geometry for the $AsH_{2}-AsH-AsH_{2}$ and $AsH_{2}-AsH_{2}-AsH_{2}$ trimers}

\begin{figure}[h]
  \centering
  \includegraphics[width = 0.7 \figurewidth]{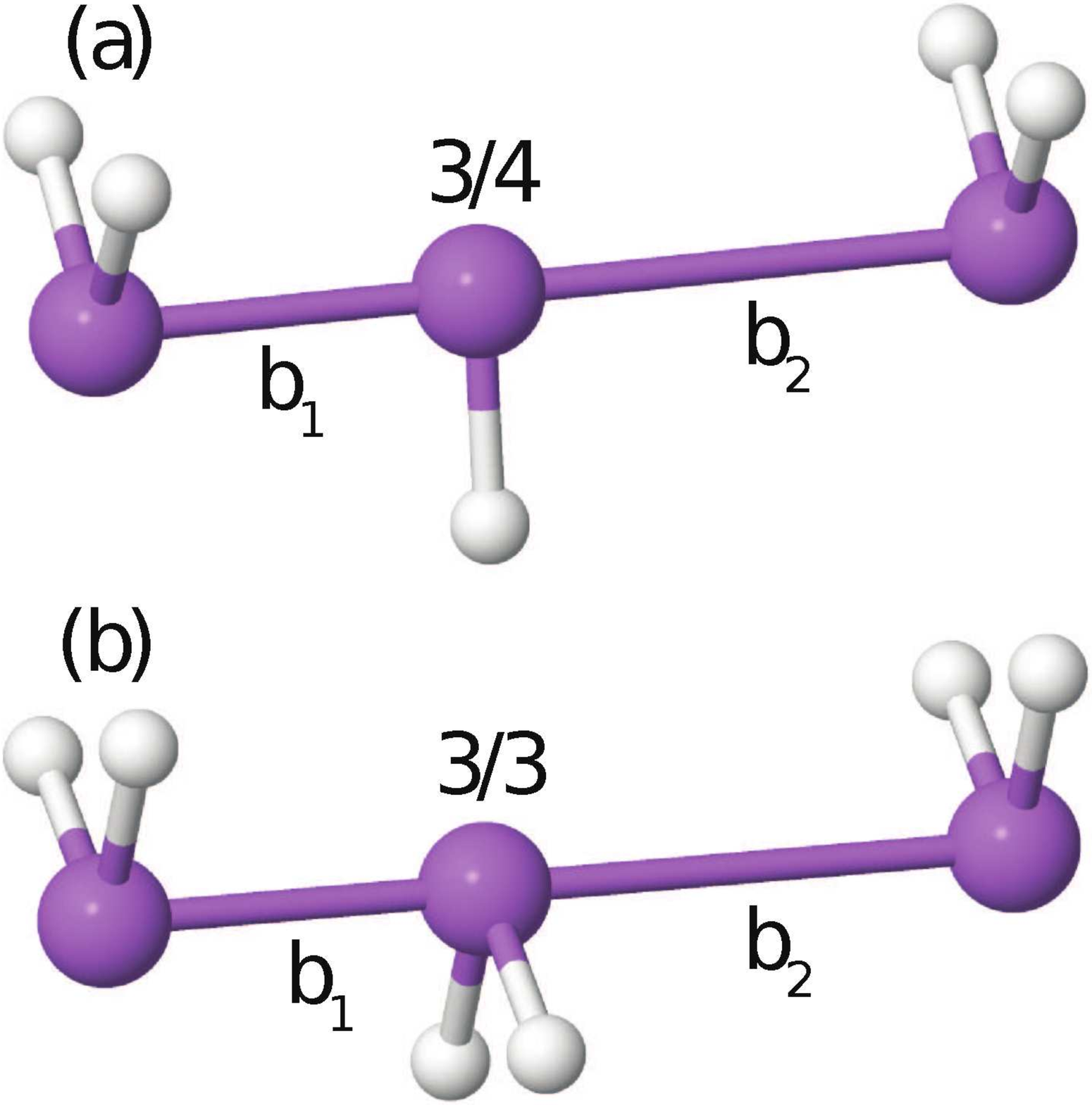}
  \caption{\label{As_trimers} The trimers AsH$_2$-AsH-AsH$_2$ (panel
    (a)) and AsH$_2$-AsH$_2$-AsH$_2$ (panel (b)), corresponding to
    3-center/4-electron and 3-center/3-electron $pp\sigma$ bond.}
\end{figure}

\section{Spatial symmetry breaking in the $AsH_{2}-AsH-AsH_{2}$ trimer}

\begin{figure}[h]
  \centering
  \includegraphics[width = \figurewidth]{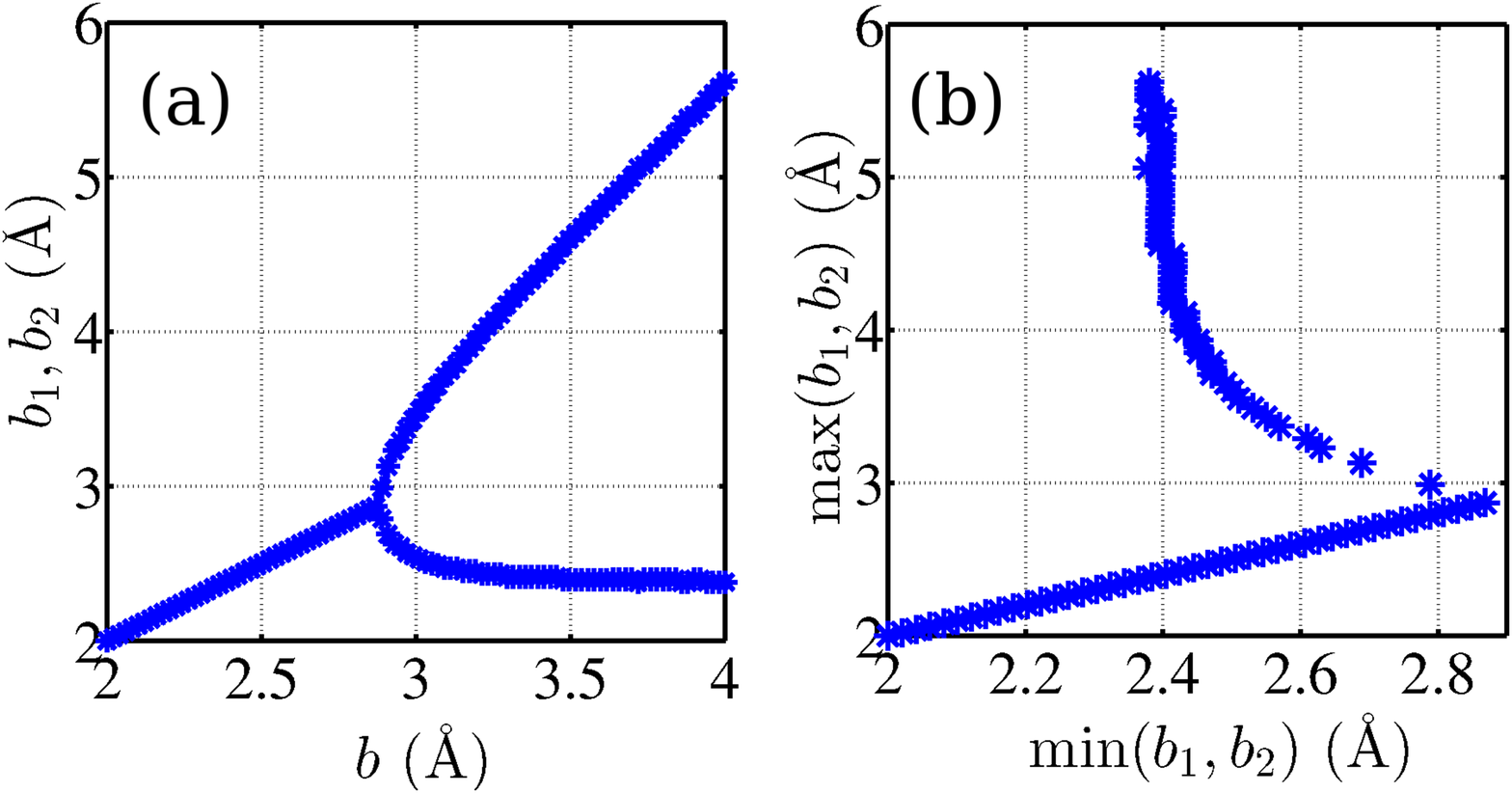}
  \caption{\label{b1b2A} The equilibrium values of the As-As bond
    lengths for a fixed value of the overall trimer length, according
    to the dashed line on the associated potential energy surface in the main text; {\bf (a)} as functions of
    the trimer length per bond, {\bf (b)} plotted parametrically, for
    the longer bond vs. the shorter bond.}
\end{figure}

\newpage

\section{Term crossing upstream of the LMO transition in the $AsH_{2}-AsH-AsH_{2}$ trimer}

\begin{figure}[h]
  \centering
  \includegraphics[width = 0.88
  \figurewidth]{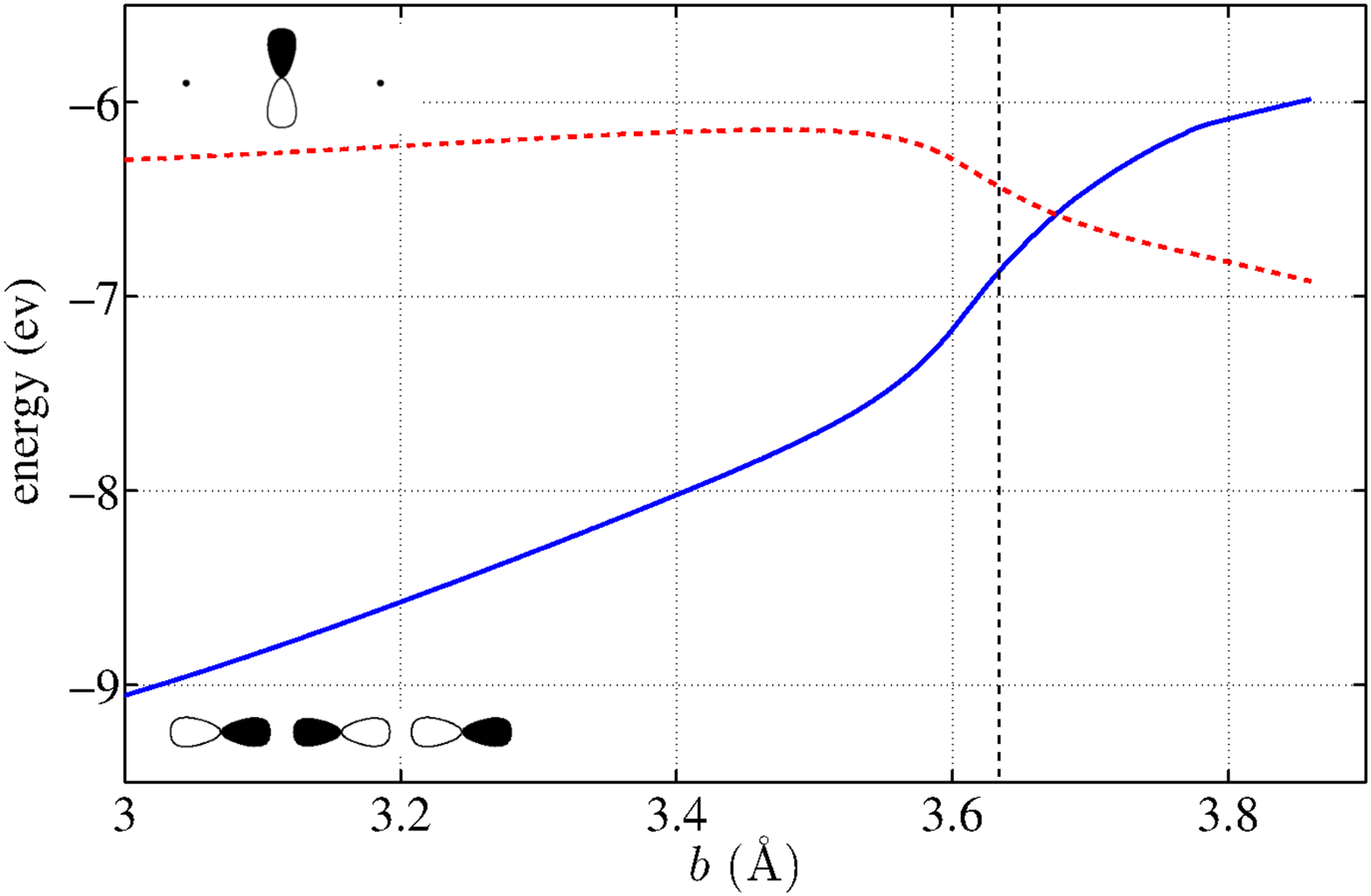}
  \caption{\label{MOexplanation} The energies of the MOs responsible
    for the $pp\sigma$ bond and the lone pair on the central arsenic,
    as functions of the As-As bond length $b = b_1 = b_2$, for the 3/4
    case. The vertical dashed line indicates the location of the LMO
    transition in Fig. 13 of the main text.}
\end{figure}

\section{Potential energy of the $AsH_{2}-AsH-AsH_{2}$ trimer}

\begin{figure}[h]
  \centering
  \includegraphics[scale=0.25]{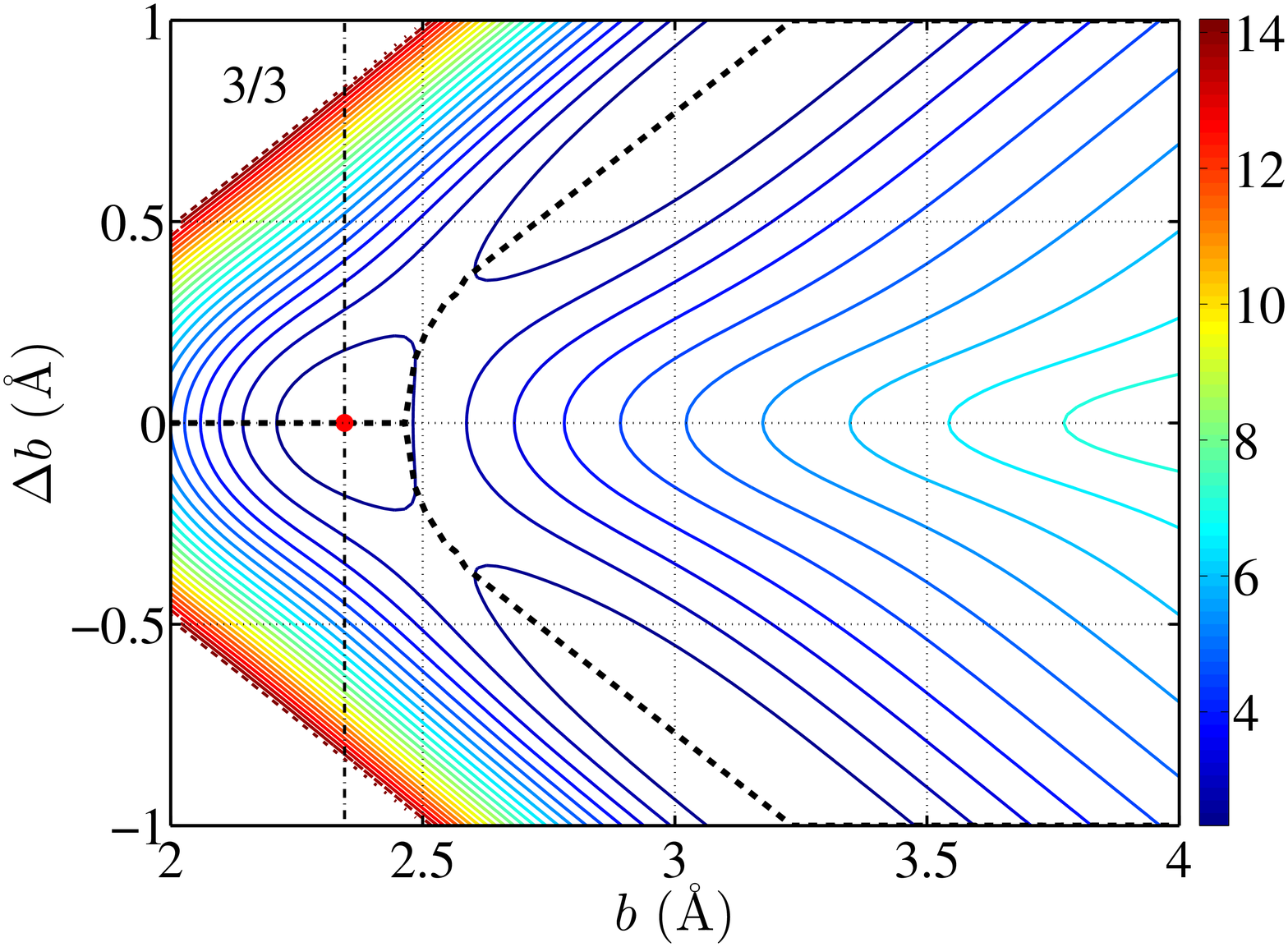}
  \caption{\label{33PES} The contour plot of the potential energy of
    the 3/3 trimer, Fig.~\ref{As_trimers}(a), as a function of the
    trimer length per bond $b \equiv (b_1 + b_2)/2$ and the distance
    of the center arsenic from the midpoint between the terminal
    arsenics, $\Delta \equiv (b_1 - b_2)/2$.  The metastable symmetric
    configuration is marked by the filled red circle. The dashed,
    bifurcating graph shows the minimum energy at fixed value of $b$.}
\end{figure}

\newpage

\section{Localized molecular orbital analysis of the 3 electron
multi-center bond}

\begin{figure}[h]
  \centering
  \includegraphics[width = \figurewidth]{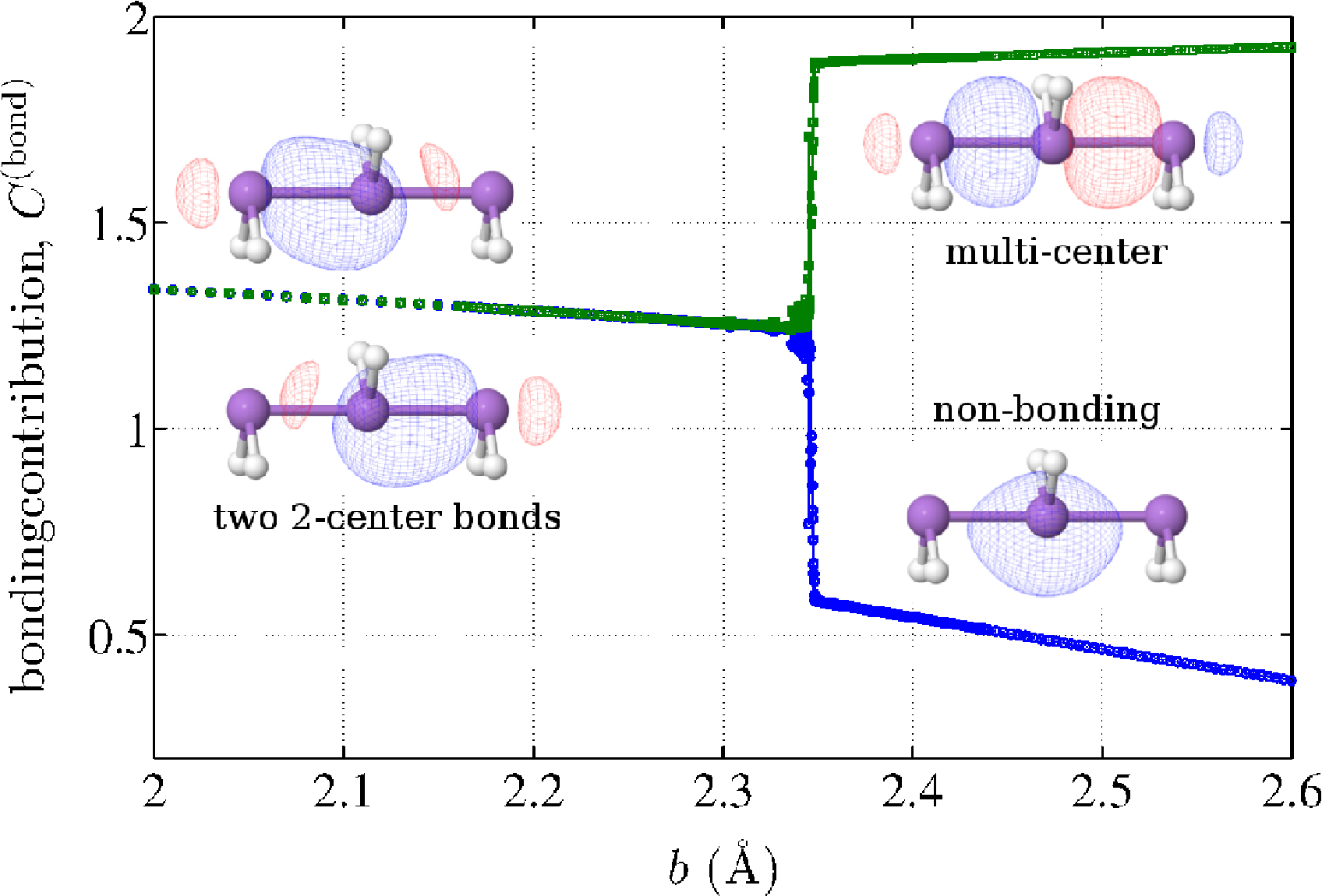}
  \caption{\label{33BOb} The bonding contribution of the LMOs as
    functions of the As-As bond length $b = b_1 = b_2$, for the 3/3
    molecule.}
\end{figure}

\begin{figure}[h]
  \centering
  \includegraphics[width=\figurewidth]{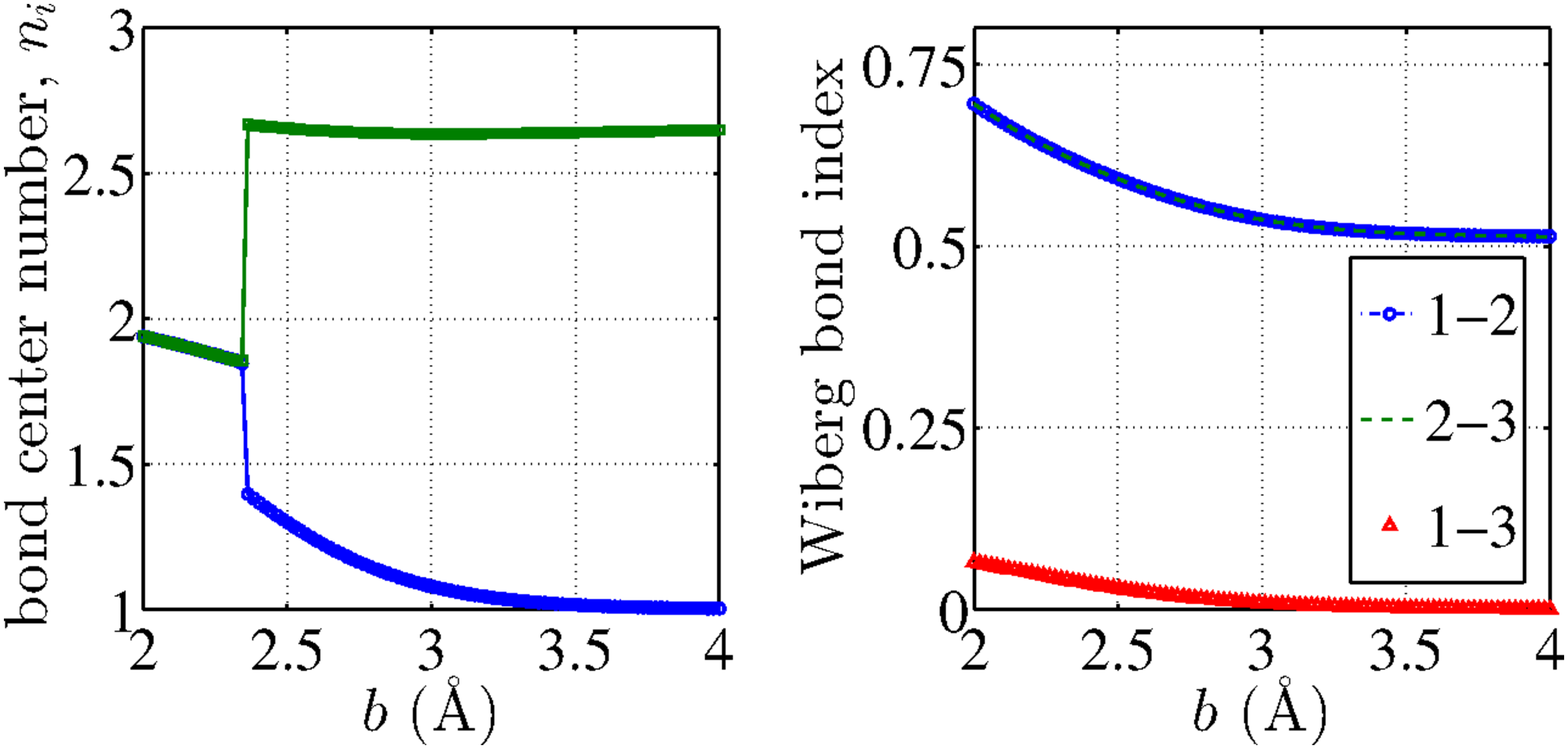}
  \caption{\label{33BI} Displayed as functions of the As-As bond
    length $b = b_1 = b_2$: {\bf (a)}, the bond center number for the
    LMOs from Fig.~\ref{33BOb}; {\bf (b)}, the Wiber bond index. Note
    the latter is computed using the density matrix and does not rely
    on the localization procedure.}
\end{figure}

\newpage

\section{Bond evolution in the $(AsH_{2})_{20}$ ring molecule}

\begin{figure}[h]
  \centering
  \includegraphics[width = .9 \figurewidth]{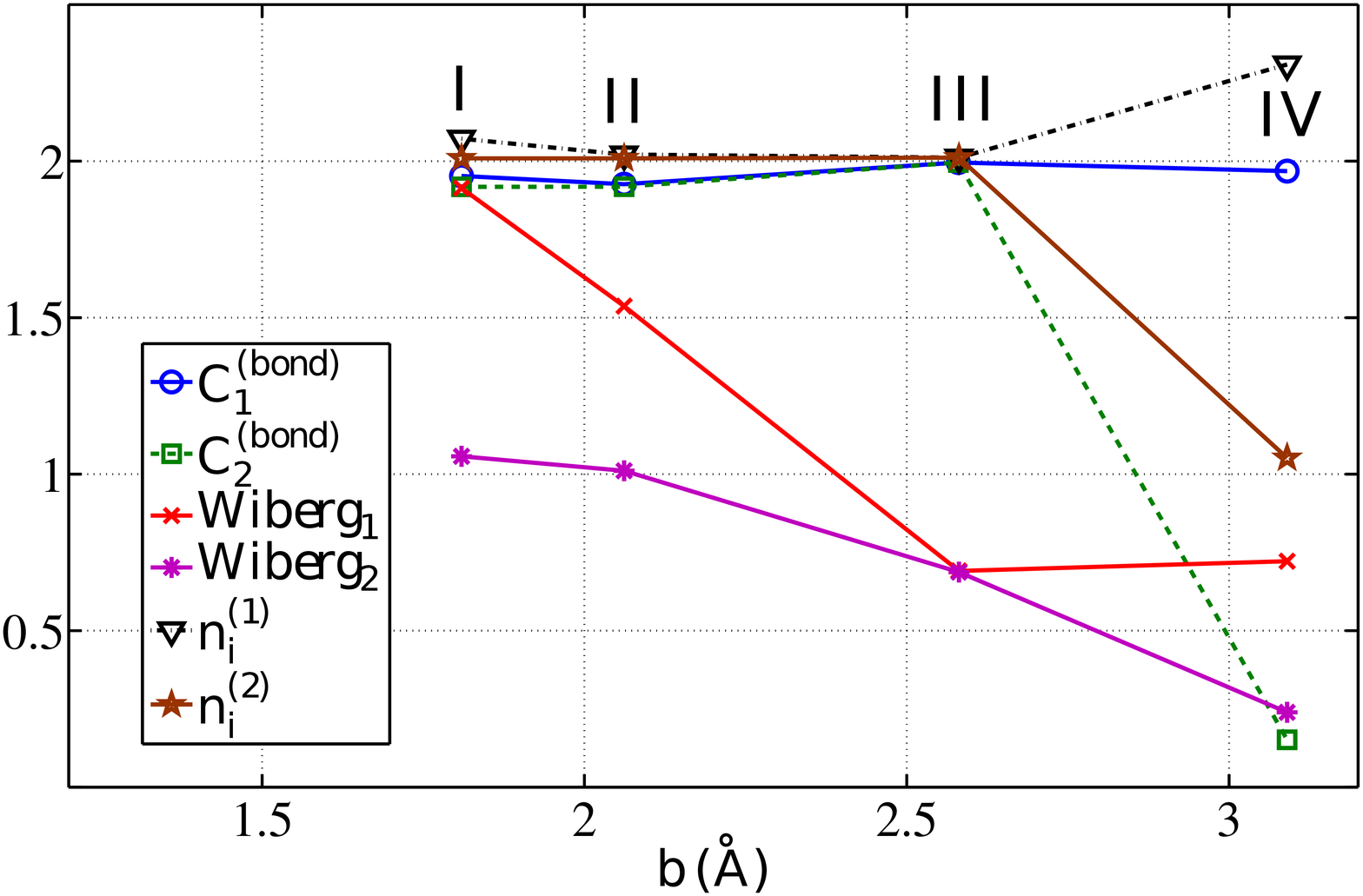}
  \caption{\label{BOring} The values the bonding contribution, Wiberg
    bond number, and the center number for the four configurations in
    Fig. 18 of the main text. The lines are guides to the eye.}
\end{figure}

\section{AIM based analysis of bond evolution in symmetric $AsH_{2}-AsH-AsH_{2}$ trimer}

\begin{figure}[h]
  \centering
  \includegraphics[width = 0.9\figurewidth]{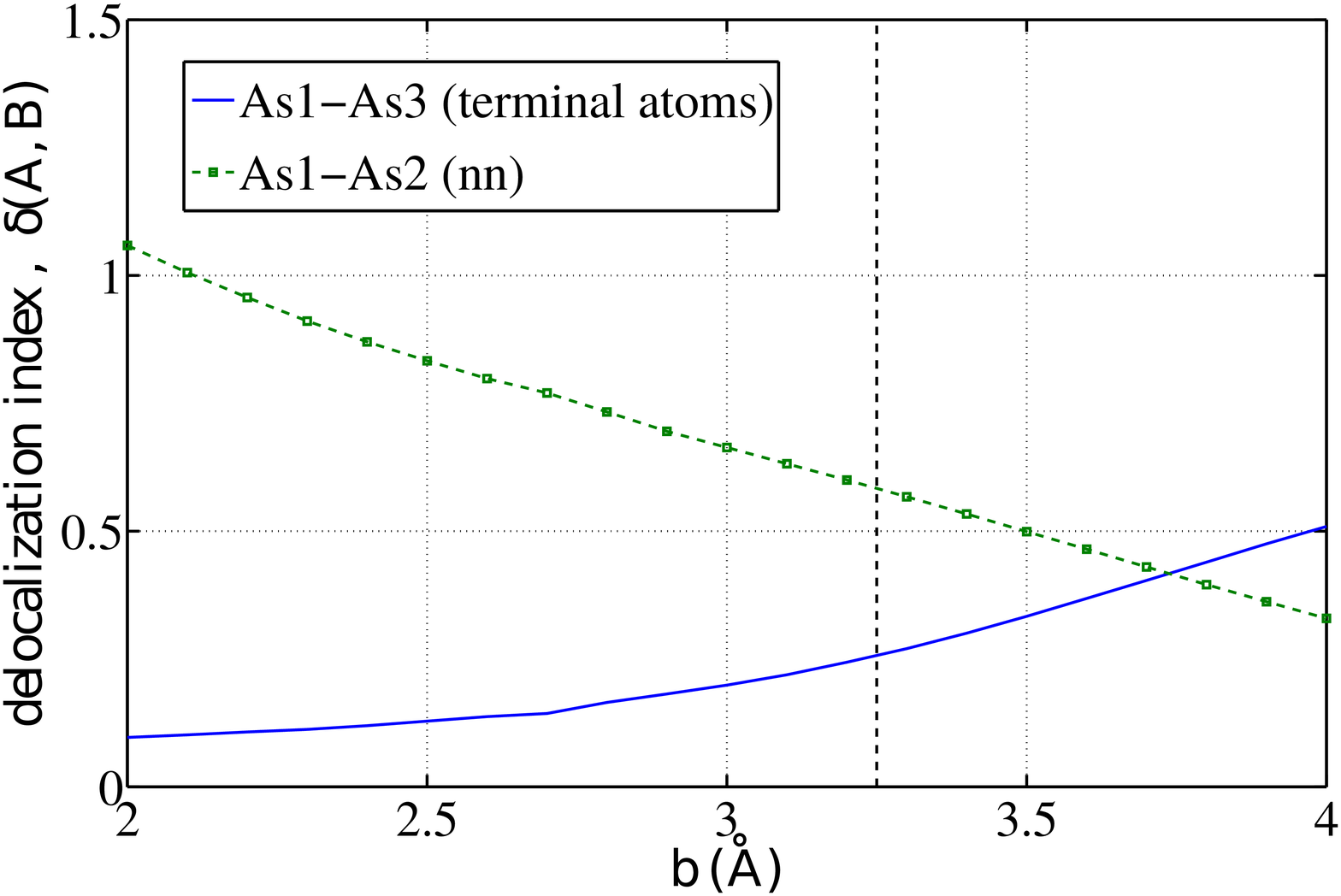}
  \caption{\label{DI} Evolution of the the delocalization indices for electron densities associated with nearest-neighbor and multi-center bonds in the 4-electron/3-center arsenic trimer.}
\end{figure}

\putbib[SI]
\end{bibunit}
  
\end{document}